\documentclass{aa}  

\usepackage{graphicx}
\usepackage{txfonts}
\usepackage{hyperref}
\usepackage{geometry}
\usepackage[utf8]{inputenc}
\usepackage[english]{babel}
\usepackage{hyperref}
\usepackage{graphics}
\usepackage{mathtools}
\usepackage[dvipsnames]{xcolor}
\usepackage{siunitx}
\usepackage{wrapfig}
\usepackage{listings}
\usepackage{enumitem}
\usepackage{booktabs}
\usepackage{subcaption}

\usepackage{natbib}

\makeatletter
\renewcommand*\aa@pageof{, page \thepage{} of \pageref*{LastPage}}
\makeatother
\numberwithin{equation}{section}
\let\oldsection\section  
\renewcommand{\section}{\setcounter{footnote}{0}\oldsection}

\setlist[itemize]{topsep=1pt, itemsep=0pt, parsep=0.5pt}  
\setlist[enumerate]{topsep=1pt, itemsep=0pt, parsep=0.5pt}  

\makeatletter
\renewcommand*{\@seccntformat}[1]{\csname the#1\endcsname\hspace{0.2cm}}
\makeatother

\setcitestyle{aysep={}}

\newcommand{\BracedIn}[1]{\left(#1\right)}

\newcommand{\BracedInCurl}[1]{\left\{#1\right\}}
\newcommand{\abs}[1]{\left|#1\right|}
\newcommand{\med}[1]{\text{med}\BracedIn{#1}}

\newcommand{\eromapper}{\texttt{eROMaPPer}}
\newcommand{\redmapper}{\texttt{redMaPPer}}
\newcommand{\erosita}{eROSITA}
\newcommand{\pointlike}{point-like}

\newcommand{\hsc}{HSC}
\newcommand{\hecate}{HECATE}
\newcommand{\efeds}{eFEDS}
\newcommand{\starex}{\texttt{STAREX}}
\newcommand{\mbproj}{\texttt{MBProj2D}}
\newcommand{\nway}{\texttt{NWAY}}

\newcommand{\circlez}{\texttt{CIRCLEZ}}

\defcitealias{Bulbul2024}{\scshape B24}
\defcitealias{Kluge2024}{\scshape K24}
\newcommand{\softbandRange}{\qtyrange[range-units = single]{0.2}{2.3}{keV}}
\newcommand{\softbandRangeMerloni}{\qtyrange[range-units = single]{0.2}{2}{keV}}

\newcommand{\rosatRange}{\qtyrange[range-units = single]{0.1}{2.4}{keV}}

\newcommand{\myeqref}[1]{Eq.~\eqref{#1}}
\newcommand{\pareqref}[1]{(Eq.~\ref{#1})}
\newcommand{\pareqrefsee}[1]{(see Eq.~\ref{#1})}

\newcommand{\secref}[1]{Sect.~\ref{#1}}
\newcommand{\Secref}[1]{Sect.~\ref{#1}}
\newcommand{\parsecref}[1]{(Sect.~\ref{#1})}
\newcommand{\parsecrefsee}[1]{(see Sect.~\ref{#1})}
\newcommand{\figref}[1]{Fig.~\ref{#1}}

\newcommand{\parfigrefsee}[1]{(see Fig.~\ref{#1})}
\newcommand{\tabref}[1]{Table~\ref{#1}}

\newcommand{\erass}[1]{eRASS#1}
\newcommand{\lsdr}[1]{LS DR #1}
\newcommand{\param}[1]{\textcolor{blue}{\lstinline[language=bash]{#1}}}

\newcommand{\sigtonoise}{\ensuremath{\rm S/N}}

\newcommand{\pAny}{\ensuremath{p_{\rm any}}}

\newcommand{\veldisp}{\ensuremath{\sigma_{\rm v}}}
\newcommand{\LMax}{\ensuremath{\mathcal{L}_{\rm max}}}

\newcommand{\zCtp}{\ensuremath{z_{\rm ctp}}}
\newcommand{\zEro}{\ensuremath{z_{\rm ero}}}
\newcommand{\zvlim}{\ensuremath{z_{\rm vlim}}}
\newcommand{\lambdaNorm}{\ensuremath{\lambda_{\rm norm}}}
\newcommand{\clusterClass}[1]{\textsf{Class~#1}}
\newcommand{\iauname}[1]{\textsf{1eRASS #1}}
\newcommand{\bcgXrayDist}{\ensuremath{d_{\rm X-ray - BCG}}}

\DeclareSIUnit{\deg}{deg}
\DeclareSIUnit{\Mpc}{Mpc}
\DeclareSIUnit{\kpc}{kpc}
\DeclareSIUnit{\erg}{erg}
\DeclareSIUnit{\mag}{^m}

\newcommand{\NumCTPAvailable}{\num{8134}}

\newcommand{\NumCTPInitialSpecZ}{\num{83646}}
\newcommand{\NumCTPSecure}{\num{5021}}

\newcommand{\NumCTPWithSecondaryMatch}{\num{1564}}

\newcommand{\NumClassFiveNoRedshift}{\num{25}}

\newcommand{\NumClassFiveStillMember}{\num{1111}}
\newcommand{\NumClassFive}{\num{3102}}
\newcommand{\NumClassFourCTPisBCG}{\num{572}}
\newcommand{\NumClassFourGalacticCTP}{\num{12}}
\newcommand{\NumClassFourMedianXRayBcgDist}{\qty{41}{kpc}}
\newcommand{\NumClassFourRedshiftHighOutliers}{\num{20}}
\newcommand{\NumClassFourRedshiftOutliers}{\num{188}}
\newcommand{\NumClassFour}{\num{1073}}
\newcommand{\NumClassOneInHecateRange}{\num{15}}

\newcommand{\NumClassOne}{\num{291}}
\newcommand{\NumClassThreeMedianXRayBcgDist}{\qty{324}{kpc}}
\newcommand{\NumClassThree}{\num{970}}
\newcommand{\NumClassTwoBehind}{\num{2305}}
\newcommand{\NumClassTwoInFront}{\num{390}}
\newcommand{\NumClassTwoMedianXRayBcgDist}{\qty{328}{kpc}}
\newcommand{\NumClassTwoUnreliableRedshift}{\num{512}}
\newcommand{\NumClassTwo}{\num{2698}}

\newcommand{\NumClassZero}{\num{213}}

\newcommand{\NumDupAssociationsAboveNine}{\num{41}}
\newcommand{\NumDupAssociationsAmbiguousNonExt}{\num{15}}
\newcommand{\NumDupAssociationsAmbiguous}{\num{54}}
\newcommand{\NumDupAssociationsWithExt}{\num{2148}}
\newcommand{\NumDupAssociationsWithoutExt}{\num{971}}
\newcommand{\NumDupAssociations}{\num{3119}}
\newcommand{\NumDupLargestAssociation}{\num{50}}
\newcommand{\NumDupNonPrimary}{\num{1228}}
\newcommand{\NumDupPlikeAssociatedWithExt}{\num{545}}
\newcommand{\NumDupPlikeInAssociationsDiscarded}{\num{5054}}
\newcommand{\NumDupPlikeInAssociations}{\num{6025}}
\newcommand{\NumDupPlikeSharingWithPlikeOrExt}{\num{9518}}
\newcommand{\NumDupPlikeSharingWithPlike}{\num{4118}}
\newcommand{\NumDupPlikeWithLowShareFrac}{\num{107}}
\newcommand{\NumEroAfterCloseExtCut}{\num{8366}}
\newcommand{\NumEroAfterDetlikeCut}{\num{576154}}
\newcommand{\NumEroAfterDuplicateCut}{\num{8382}}
\newcommand{\NumEroAfterLMaxCut}{\num{13436}}
\newcommand{\NumEroAfterMaskfracCut}{\num{151628}}
\newcommand{\NumEroAfterOpticalPropertiesCut}{\num{218613}}
\newcommand{\NumEroAfterRedshiftCut}{\num{215579}}
\newcommand{\NumEroAfterRichnessCut}{\num{23947}}
\newcommand{\NumEroAfterVisualContCut}{\num{8347}}
\newcommand{\NumEroAfterXGoodCut}{\num{574281}}
\newcommand{\NumEroAfterZVLimCut}{\num{158774}}
\newcommand{\NumEroBeforeDupClean}{\num{13436}}
\newcommand{\NumEroCloseExtendedDifferentRedshift}{\num{121}}
\newcommand{\NumEroCloseExtendedSameRedshift}{\num{16}}
\newcommand{\NumEroCloseExtendedSecondaryRedshiftAvailable}{\num{9}}
\newcommand{\NumEroCloseExtendedSecondaryRedshiftMatch}{\num{7}}

\newcommand{\NumEroExtFiltered}{\num{5721}}
\newcommand{\NumEroFinal}{\num{8347}}
\newcommand{\NumEroInitial}{\num{1250795}}
\newcommand{\NumEroSpecZNotConverged}{\num{83}}
\newcommand{\NumEroVisualContaminants}{\num{19}}
\newcommand{\NumExtFilteredMedianRichness}{\num{38.8}}
\newcommand{\NumFullSampleMedianXRayBcgDist}{\qty{214}{kpc}}
\newcommand{\NumKlugeExtClusters}{\num{12247}}
\newcommand{\NumPlikeFilteredMedianRichness}{\num{28.4}}

\hypersetup{
  colorlinks=true,
  allcolors=[rgb]{0,0,0.8},
  pdftitle={Misclassified eRASS1 clusters in the point source catalog},
  pdfauthor={Balzer et al.},
}
\newcommand{\extlike}{\ensuremath{\mathcal{L}_{\rm ext}}}
\newcommand{\detlike}{\ensuremath{\mathcal{L}_{\rm det}}}
\newcommand{\ext}{\texttt{EXT}}

\begin{document} 
\title{The First SRG/eROSITA All-Sky Survey}
\subtitle{Characterization of clusters of galaxies misclassified in the eRASS1 point source catalog}
\titlerunning{Misclassified galaxy clusters in the eRASS1 point source catalog}

\author{F. Balzer
            \inst{1}
        \and
        E. Bulbul
            \inst{1}
        \and
        M. Kluge
            \inst{1}
        \and
        A. Liu
            \inst{1}
        \and
        M. Salvato
            \inst{1}
         \and
        M. Fabricius
            \inst{1},
            R.~Seppi\inst{1},
E.~Artis\inst{1}, Y.~E.~Bahar\inst{1},  R.~Bender\inst{1}, N.~Clerc\inst{2}, J.~Comparat\inst{1}, V.~Ghirardini\inst{1}, S.~Grandis\inst{3}, S.~Krippendorf\inst{4}, G.~Lamer\inst{5}, N.~Malavasi\inst{1}, A.~Merloni\inst{1}, K.~Nandra\inst{1}, M.~E.~Ramos-Ceja\inst{1}, J.~S.~Sanders\inst{1}, X.~Zhang\inst{1} \and S.~Zelmer\inst{1}
}

\institute{
Max-Planck-Institut für Extraterrestrische Physik, Gießenbachstraße, D-85748 Garching, Germany\\            
\email{fbalzer@mpe.mpg.de}
\and
IRAP, Universite de Toulouse, CNRS, UPS, CNES, F-31028 Toulouse, France
\and
Universit\"at Innsbruck,  Institut f\"ur Astro- und Teilchenphysik, Technikerstr. 25/8, 6020 Innsbruck, Austria
\and
Arnold Sommerfeld Center for Theoretical Physics, LMU Munich, Theresienstr. 37, 80333 M\"unchen, Germany
\and
Leibniz-Institut f\"ur Astrophysik Potsdam (AIP), An der Sternwarte 16, 14482 Potsdam, Germany
}

\date{\today{}}

\abstract{
The detection of the extended X-ray-emission of the intracluster medium by the first SRG/eROSITA All-Sky Survey (\erass{1}), combined with optical and near-infrared follow-up, resulted in the identification of more than \num{12000} galaxy clusters, yielding precise constraints on cosmological parameters. However, some clusters of galaxies can be misclassified as point sources by \erosita{}'s source detection algorithm due to the interplay between the point-spread function, the shallow depth of the survey, compact (cool core) X-ray emission, and bright active galactic nuclei hosted in their centers or their vicinity. To identify such misclassified galaxy clusters and groups, we apply optical follow-up to the \erass{1} X-ray point sources analogously to the treatment of the extent-selected catalog. After rigorous filtering to ensure purity, we find a total of \NumEroFinal{} clusters of galaxies, of which \num{5819} are novel detections, in a redshift range $0.05 < z \lesssim 1.1$. This corresponds to a \qty{70}{\percent} discovery rate, a fraction similar to that of the extent-selected sample. To facilitate finding new exceptional clusters such as the Phoenix cluster (which is recovered in our sample), we divide the clusters into five classes based on the optical properties of likely single-source counterparts to the X-ray emission. We further investigate potential biases in our selection process by analyzing the optical and X-ray data. With this work, we provide a catalog of galaxy clusters and groups in the \erass{1} point source catalog, including their optical and X-ray properties along with a meaningful classification. 
}


\maketitle

\section{Introduction}

Understanding the astrophysical properties and the statistical signatures of the population of clusters of galaxies plays a crucial role in constraining the fundamental parameters of our Universe.
Most baryons in clusters are encapsulated in the intracluster medium (ICM), which, due to their large gravitational potential wells, emits mainly in the X-ray band through thermal Bremsstrahlung. However, minor contributions to the overall baryonic budget in clusters are provided by galaxies, whose stellar light allows for optical and near-infrared detection. Additionally, the energetic electrons in the ICM interact with the Cosmic Microwave Background (CMB) through inverse Compton scattering, which is known as the Sunyaev-Zel'dovich (SZ) effect \citep{Sunyaev1972}. These observational signatures of the various components of clusters of galaxies in the electromagnetic spectrum are widely used to facilitate their detection. This is the basis for statistical analyses, where a thorough study of cluster count statistics can tightly constrain cosmological parameters \citep[see for example the review by][and the references therein]{Clerc2022}.

Surveys based on the various observational tracers have been conducted, each with their strengths and weaknesses. Large-area photometric surveys in the optical band, performed mainly via ground-based telescopes, usually identify galaxy clusters through the detection of the characteristic overdensity of early-type galaxies, the so-called red sequence \citep{Dressler1980}. The red sequence is exhibited through a tight linear relationship in color-magnitude space \citep{Bower1992}. These surveys use photometric data to select the clusters \cite[e.g.][]{Gladders2000, Oguri2014, Rykoff2016}, which can effected by projection effects \citep{Zu2017, Busch2017}. 
Cluster surveys that rely on detecting signatures of the SZ effect are less affected by projection effects, enable cluster detection at high redshifts, and catalogs based on this selection method have high completeness. However, they are limited by the angular resolution of the instruments. For overviews of SZ cluster surveys conducted using various telescopes, see, e.g., \citet{PlanckCollaboration2014} for the Planck Space Telescope, \citet{Bleem2015} for the South Pole Telescope (SPT), or \citet{Hilton2021} for the Atacama Cosmology Telescope. 

Cluster surveys in the X-ray band are also sensitive to the emission from the hot ICM. The spatial resolution of X-ray telescopes is typically higher than that of those used for SZ surveys, and projection effects are far less severe than in the optical; X-ray surveys are thus well-suited to detect large samples of clusters of galaxies and study their statistical properties. The first wide-area X-ray cluster surveys include but are not limited to the NORAS survey in the Northern Hemisphere \citep{Boehringer2000}, and the REFLEX survey in the Southern Hemisphere \citep{Boehringer2004}, both derived from the ROSAT All-Sky Survey \citep[RASS,][]{Voges1999}. Cluster surveys with limited area coverage but deeper observations with XMM-Newton bridge this first All-Sky X-ray survey and the next generation X-ray surveys \citep{Adami2018, Liu2022a}.

The extended ROentgen Survey with an Imaging Telescope Array (\erosita{}) on board the Spektrum Roentgen Gamma (SRG) mission, which is a wide-field X-ray instrument, completed its first All-Sky Survey (\erass{1}, hereafter) six months after its launch in June 2020 \citep[][]{Sunyaeve2021, Predehl2021}. The first catalog of $\sim1.1$ million X-ray sources in the Western Galactic Hemisphere is presented in \cite{Merloni2024}.
Based on this catalog, \citet[\citetalias{Bulbul2024} hereafter]{Bulbul2024} and \citet[\citetalias{Kluge2024} hereafter]{Kluge2024} selected \NumKlugeExtClusters{} clusters of galaxies via their extended X-ray emission, compiling a sample with redshifts up to $z = 1.3$. The details of the optical identification scheme and the X-ray properties of these clusters and sample properties are presented in \citetalias{Kluge2024} and \citetalias{Bulbul2024}, respectively.

Clusters of galaxies in the \erosita{} source catalogs are characterized based on their extent (\ext), extent likelihood (\extlike), and detection likelihoods (\detlike) \citep[\citetalias{Bulbul2024}]{Brunner2022, Liu2022a}. Due to the sizable point-spread function \citep[PSF; survey averaged \qty{\sim30}{\arcsec}~HEW][]{Merloni2024} of \erosita{}, galaxy clusters with peaked cool-core emission at higher redshifts or clusters hosting Active Galactic Nuclei (AGN) in their central regions are sometimes misclassified as point sources. This can lead to their omission from the extent-selected sample.
One infamous example of such a misclassification is the Phoenix cluster, which was first detected in the RASS. Due to its emission's extent comparable to the PSF of ROSAT and an association with an AGN in the optical, it was classified as a point source and thus discarded from the cluster catalogs. Only more than ten years later it was revealed to be a massive galaxy cluster at $z = 0.597$ \citep{McDonald2012}, uncovered by an observation in the millimeter-wave band with the SPT \citep{Carlstroem2011}. 
The realization that such clusters are missed in the extent-selected searches sparked thorough investigations on the RASS bright source catalogs \citep{Green2017, Donahue2020, Somboonpanyakul2021b}, including studies focused on the interaction of the central AGN and Brightest Cluster Galaxies (BCGs), and X-ray and optical follow-up observations. One challenging aspect of employing X-ray surveys in population studies involves modeling the selection function, which assesses the completeness of the extent-selected sample \citep[see][for further details]{Clerc2018, Clerc2024}. Searching for misclassified clusters can help to understand such selection effects. 

Since \erass{1} with a median flux limit of \qty{\sim5e-14}{\erg\per\s\per\cm\squared} in the \softbandRangeMerloni{} band \citep[see Fig. 9 in][]{Merloni2024} is already deeper than the RASS, which had a limit of $\qty{\sim e-13}{\erg\per\s\per\cm\squared}$ in the \rosatRange{} band \citep{Boller2016}, a first assessment of the abundance and nature of misclassified clusters was conducted on the \erosita{} Final Equatorial Depth Survey (\efeds{}) data by \cite{Salvato2022} and \cite{Bulbul2022}. Studies on the \efeds{} field showed a non-negligible number of galaxy clusters and groups (346, compared to the 542 extent-selected clusters) to remain in the \efeds{} point source catalog. This catalog was obtained via the application of an extent likelihood cut of $\extlike\geq6$ \citep{Liu2022a}. Of these misclassified clusters, only a small fraction of $\qty{\sim10}{\percent}$ seemed to host AGN in their centers. The majority of clusters was missed by the extent-selection due to having fluxes below the flux limit. With the same separation scheme, \citet{Bulbul2022} predicted \num{\sim6000} misclassified clusters to be found in \erass{1}, with \num{\sim400} candidates hosting AGN in their BCGs. The authors recommended lowering the \extlike{} cut (with the downside of increasing contamination of the extent-selected sample) to recover more of such clusters. As a result, to keep the majority of high redshift clusters with bright cores and compact groups, a cut of $\extlike\geq3$ was adopted to construct the primary \erass{1} cluster catalog \citepalias{Bulbul2024}. Even with this reduced \extlike{} cut, underluminous clusters and clusters hosting AGN in their cores can be included in the remaining X-ray point sources. Furthermore, bright AGN point source emission can outshine and thus disguise the fainter extended emission of clusters. 

The goal of this paper is to characterize galaxy clusters found at the positions of \erass{1} X-ray point sources. We base this on an identification of a red sequence in the optical and infrared using the \eromapper{} algorithm \citep{IderChitham2020, Kluge2024}, and on the properties of single-source counterparts identified via the \nway{} algorithm \citep{Salvato2018, Salvato2024}. With this information, we assemble a comprehensive sample of interesting clusters of galaxies that had otherwise been missed by the extent-selection cuts in the main \erass{1} cluster catalog. This sample can be used as a foundation for various astrophysical studies: The cool-core clusters with prominent \pointlike{} X-ray flux will enable extensive analysis of the AGN feedback cycle, while optical clusters associated with background quasars will allow for absorption line studies of the ICM. Numerous high-redshift clusters, which can be used in evolutionary studies of clusters of galaxies, are also part of this sample.

This work first introduces the underlying X-ray and optical data in \secref{sec:main_sec_two}, followed by a description of the filtering and classification procedure in \secref{sec:main_sec_three}. A comparison to a selection of optical, X-ray, and SZ surveys is performed in \secref{sec:main_sec_four}. We then characterize the sample by the optical properties of the clusters and the single-source counterparts in \secref{sec:main_sec_five}, followed by a discussion of the X-ray properties in \secref{sec:main_sec_six}.
Finally, our results are summarized in \secref{sec:main_sec_summary}.
Throughout this paper, we assume a flat $\Lambda$CDM cosmology with $H_0=\qty{70}{km\per\second\per\Mpc}$, $ \Omega_{\rm m} = 0.3$, and $\sigma_8=0.8$. 
Redshifts are provided in the heliocentric frame of reference, and no corrections are applied for Virgo infall or the CMB dipole moment.

\section{X-ray, Optical, and Near-infrared Data}
\label{sec:main_sec_two}

In this section, we introduce the data used throughout this paper. We describe the X-ray catalogs used in \secref{ssec:data_intro_xray}. We then introduce the photometric optical and near-infrared data in \secref{ssec:data_intro_optical} as it is used both for optical cluster follow-up and the single source counterpart identification.
Next, in \Secref{ssec:data_intro_eromapper}, we briefly describe the data products obtained from the red-sequence-based cluster finding algorithm \eromapper{}, followed by a description of the \nway{} single-source counterparts to the X-ray point sources in \secref{ssec:data_intro_nway}, and a brief summary of how photometric redshifts for these counterparts are determined.

\subsection{X-ray data}
\label{ssec:data_intro_xray}

The first \erosita{} All-Sky Survey \citep[\erass{1}, ][]{Merloni2024} catalogs include X-ray-detected \pointlike{} and extended sources in the Western Galactic half of the \erosita{} sky, i.e., at a Galactic longitude $179.9442\,\deg<l<359.9442\,\deg$.
The calibration and source detection algorithms used in creating the catalogs are described in detail in Sect.~4.3.2 of \citet{Merloni2024}. In short, the \erosita{} Science Analysis Software System \citep[eSASS,][]{Brunner2022}\footnote{further details are available at \url{https://erosita.mpe.mpg.de/dr1/eSASS4DR1/}} is used in pipeline configuration 010, which, in comparison to the \efeds{} processing c001, performs a stronger telescope module specific noise suppression for pattern events, more accurately computes the subpixel position, has improved flagging of pixels neighboring bad pixels, and has an improved projection accuracy.

The ICM of galaxy clusters has an extended appearance in the X-rays, setting it apart from AGN and stars that appear consistent with \erosita{}'s PSF. This offers a natural selection criterion for galaxy clusters given the \erass{1} X-ray data. The sample of extended X-ray objects, i.e., objects with an extent likelihood $\extlike\geq\num{3}$, is described in \citetalias{Bulbul2024}, including all sources above an X-ray detection likelihood of $\detlike\geq6$. Further cleaning of the sample was performed via optical identification using the \eromapper{} red sequence-based cluster finding algorithm \parsecref{ssec:data_intro_eromapper}, selecting \NumKlugeExtClusters{} galaxy clusters in the common Legacy Survey DR9 (north) and DR10 (south) area \citetalias{Kluge2024}. We use this sample for comparisons throughout this work. At the same time, more detailed analysis and further experiments have been conducted elsewhere, including a characterization of the selection function \citep{Clerc2024}, constraints on cosmology and general relativity \citep{Ghirardini2024, Artis2024, Artis2024b, Seppi2024, Grandis2024, Kleinebreil2024}, a catalog of large scale structure and evolutionary studies \citep{Liu2024, Liu2024b}, and studies of AGN feedback in galaxy groups \citep{Bahar2024}.

In this work, we complement the extent-selected catalog by identifying clusters hiding in plain sight, i.e., misclassified clusters which ended up in the point source catalog. The \erass{1} main catalog presented in \citet{Merloni2024} includes point sources selected from the \softbandRange{} band. This catalog, comprising \NumEroInitial{} sources selected by incorporating the detections with an extent of \num{0} and a detection likelihood of at least \num{5}, is the basis of our analysis.

%
\subsection{Optical and near-infrared data}
\label{ssec:data_intro_optical}

For the optical and near-infrared galaxy cluster identification, we use the DESI Legacy Imaging Surveys \citep[LS,][]{Dey2019} akin to \citetalias{Kluge2024}. These are a collection of sky surveys obtained with four telescopes. They consist of the Dark Energy Camera Legacy Survey \citep[DECaLS,][]{Dey2019} in the $g,r,i,$ and $z$ bands in the southern sky at ${\rm Dec}\lesssim32.375\deg$. We refer to this dataset as \lsdr{10} south because it is taken from the $10^{\rm th}$ data release of the LS. In the northern sky at ${\rm Dec}\gtrsim32.375\degr$, the Beijing-Arizona Sky Survey \citep[BASS,][]{Zou2017aa} is used to cover the $g$ and $r$ bands while the Mayall $z$-band Legacy Survey \citep[MzLS,][]{Silva2016aa} is used to cover the $z$ band. We refer to these two datasets as \lsdr{9} north because, although they are included in data release 10, the data remains unchanged since data release 9. The overlap of the \lsdr{10} south with \erass{1} is \qty{12791}{\deg^2} (\qty{12205}{\deg^2}) without (with) the $i$-band included, while \lsdr{9} north has an overlap of \qty{462}{\deg^2}. The calculation of these areas is described in \cite{Kluge2024}, Appendix~B.

The \lsdr{9} data are only used for the cluster identification \parsecref{ssec:data_intro_eromapper}, but not for the counterpart identification \parsecref{ssec:data_intro_nway}.
Additionally, for the full sky, near-to-mid-infrared data from the Near-Earth Object Wide-field Infrared Survey Explorer \citep[NEOWISE,][]{Lang2014,Mainzer2014,Meisner2017a,Meisner2017b} is included to increase the wavelength range. Of these data, only the near-infrared $W1$ band at \qty{3.4}{\micro\meter} is used for identifying the cluster counterparts of \erass{1} \parsecrefsee{ssec:data_intro_eromapper}, and all bands are considered (when available) for the photometric redshift measurements of single-source counterparts \parsecrefsee{ssec:data_intro_nway}.

Photometric measurements are consistently performed using \textit{The Tractor} algorithm \citep{Lang2016} based on seeing-convolved analytic models. Galaxy luminosities are given in AB magnitudes and were corrected for galactic extinction.


%
\subsection{\eromapper{} optical cluster follow-up for the X-ray point sources}
\label{ssec:data_intro_eromapper}
Galaxy clusters are identified using the red-sequence-based optical cluster finding algorithm \eromapper{} \citep{IderChitham2020,Kluge2024}, adopted to the \erosita{} source confirmation process based on the \redmapper\ algorithm \citep{Rykoff2012,Rykoff2014,Rykoff2016}. The optical identification is employed in scanning mode on the full \erass{1} sample, consisting of \num{26682} extended and \NumEroInitial{} \pointlike{} X-ray detections with a detection likelihood \detlike$\geq 5$ \citep{Merloni2024}. Results for the extended sources (\extlike$\geq3$) are published \citepalias{Kluge2024, Bulbul2024}. In this work, we study the results for clusters at positions of the point sources detected in \erass{1}. This initial catalog contains many non-unique and spurious detections. Hence, we will filter it further in \secref{ssec:data_intro_cat_cleaning}. While the \eromapper{} algorithm is described in detail in \citetalias{Kluge2024}, we will briefly outline its most important outputs here:

When run in \textit{scanning} mode (i.e., using positional priors), \eromapper{} iteratively searches for cluster member galaxies within a radius $R_\lambda$ defined as
\begin{equation}\label{eq:richness_radius}
    R_\lambda=R_0\BracedIn{\frac{\lambda}{100}}^\beta 
\end{equation}
where $\lambda$ is the richness \pareqrefsee{eq:richness_definition}, and where the parameters $R_0=1.0h^{-1}$\,Mpc and $\beta=0.2$ were found to minimize the scatter within the redshift-richness relation \citep{Rykoff2012, Rozo2011}. 
The conversion from angular to physical scales is performed at the photometric redshift.
Each of the galaxies within that radius is assigned a membership probability $p_{\rm mem}$ that depends on the color distance from the red sequence model, the radial distance from the search location, and global galaxy background probability \citep{Rykoff2014, Rykoff2016}. Akin to \redmapper{}, \eromapper{} constructs and evaluates a likelihood function $\mathcal{L}_\lambda$ on a redshift grid,
\begin{equation}\label{eq:cluster_likelihood_function_definition}
    \ln{\mathcal{L}_\lambda}(z)=-\frac{\lambda(z)}{S(z)}-\sum_i\ln\BracedIn{1-p_{{\rm mem}, i}(z)}.
\end{equation}
The richness $\lambda$ of a cluster is defined as
\begin{equation}\label{eq:richness_definition}
    \lambda:=S\sum_{i}p_{{\rm mem},i},
\end{equation}
where $S$ is a scaling factor that depends on the fraction masked due to the flagged photometry of galaxies and the limited depth of the images.
To be considered for the richness estimate, galaxies are required to have a minimum luminosity of $L>fL_*$, where $L_*$ is the break of the Schechter luminosity function \citep{Schechter1976}, and $f$ a factor that varies depending on the available photometry for a given \eromapper{} run. Following \citet{Rykoff2012}, $f=0.2$ was employed for the $grz$ and $griz$ runs to minimize the scatter of the X-ray luminosity at fixed richness, and $f=0.4$ was employed for the $grzW1$ and $grizW1$ runs to have an unbiased richness estimate for higher redshifts $z>0.8$. To correct for the fact that the richness systematically decreases for a higher luminosity threshold applied to the member galaxies, a normalized richness $\lambdaNorm{}$ is defined via
\begin{equation}
   \lambdaNorm{} = S_{\rm norm}\lambda,
\end{equation}
with the normalization factor $S_{\rm norm}$ varying for each of the \eromapper{} runs ($S_{\rm norm}\approx1$ for the runs with $f=0.2$, and $S_{\rm norm}\approx2.4$ for the runs with $f=0.4$, see Table~2 of \citetalias{Kluge2024}).
For each positional input, \eromapper{} determines the maximum of the likelihood function $\ln\mathcal{L}_\lambda(z_\lambda)$, using the $z_\lambda$ value that maximizes it as the photometric redshift. The associated richness $\lambda$ and the maximum value \LMax{} are registered. While both only depend on the membership probabilities, a lower \LMax{} value for two clusters with identical richness can indicate a lower $p_{\rm mem}$ for more members, which in turn can hint at an offset of the identified cluster to the input position or spurious member detections.

For each cluster and \eromapper{} run, a limiting redshift \zvlim{} is calculated \citepalias[see Appendix~B in][]{Kluge2024}. \zvlim{} is defined as the redshift at which member galaxies at the low-luminosity limit $fL_*$ can still be detected with a Signal-to-Noise ratio of \num{10} in the $z$-band. \citetalias{Kluge2024} caution (e.g., shown in their Fig.~13) that clusters with photometric redshifts $z_\lambda> \zvlim{}$ are prone to contain false-positive detections, and therefore provide the \param{IN_ZVLIM}-flag defined as $z_\lambda\leq\zvlim{}$ for each cluster.

In addition to the quantities introduced above, \eromapper{} defines the BCG as the brightest member galaxy in the $z$ band within $R_\lambda$. This choice has been shown to agree with other selection criteria based on extended stellar envelopes in \qty{\sim80}{\percent} of the cases \cite[see also][]{VonDerLinden2007,Kluge2020}.

\subsection{\nway{} single source counterparts for the X-ray point sources}\label{ssec:data_intro_nway}
For each of the \citet{Merloni2024} X-ray point sources, \citet{Salvato2024} present single source counterpart associations in the optical and near-infrared, making use of the \lsdr{10} data described in \secref{ssec:data_intro_optical}.
For the task of identifying them, they employ the \nway{} algorithm \citep{Salvato2018} combined with a Machine Learning (ML) algorithm based on a Random Forest classifier. This ML algorithm, trained on more than \num{40000} secure X-ray emitters (stars, compact objects, AGN, and QSOs), determined a model that, to each source in the \lsdr{10} survey within one arcmin from the X-ray position, assigns the probability to be an X-ray emitter. The model takes into account the fluxes in the $g,r,i,z$ and the 4 WISE bands, the constructed colors, and signal-to-noise.
This probability is then combined with the probability determined on the basis of astrometry (i.e., separation, positional errors, and number density in the \erosita{} and \lsdr{10} catalogs) to rank each counterpart candidate.\\
To assess the reliability of these counterparts, \citet{Salvato2024} randomized the coordinates of the \erass{1} point sources and re-applied the same model as for the real position. This allowed them to measure the distribution of $\pAny$ (i.e., the probability that an \erass{1} source is correctly associated with a counterpart) for chance associations.
\citet{Salvato2024} show that in the area covered by \lsdr{10}, the accuracy and purity is higher than \qty{95}{\percent} already at very low $\pAny{}$.

\subsubsection{Distinction between Galactic and extragalactic counterparts}\label{sssec:data_description_galactic_classification}
In addition to providing the counterpart identification, \citet{Salvato2024} also assess the Galactic/extragalactic nature of these counterparts, providing a \param{class_gal_exgal} parameter for each source. For this purpose, another ML algorithm, \starex{}, was developed. A training sample of more than \num{22000} elements (\num{7735} stars and \num{14517} extragalactic sources) was constructed, and the model was trained using $g,r,z,$ WISE/$W1$ and X-ray fluxes, morphological information (expressed with TYPE from \lsdr{10}). For sources brighter than $\sim20$ magnitudes in the $g$ band, the proper motion and related error was gathered from Gaia DR3 \citep{Gaia2016, Gaia2023}.
On a validation sample, \starex{} correctly identifies the Galactic/extragalactic nature of the sources in more than \qty{98}{\percent} of the cases.
Furthermore, other criteria introduced in \citet{Salvato2018, Salvato2022} are adopted, together with the classification from SIMBAD, when available. In this work, we opted to utilize a binary classification between the Galactic and extragalactic nature of counterparts securely associated with X-ray emission. For this, we take into account both the assigned \param{class_gal_exgal} and the reliability of the counterpart photometry; if the photometry of a counterpart is considered unreliable, it will only be labeled as Galactic if $\param{class_gal_exgal}=-5$; otherwise, we label it as Galactic if $\param{class_gal_exgal}<3$. This is encoded in the \param{CTP_IS_GALACTIC} flag.

\subsubsection{Counterpart redshifts}\label{sssec:data_description_photo_z}
For each \lsdr{10} south extragalactic counterpart to an \erass{1} source, \citet{Salvato2024} also compute photometric redshifts. These are determined using \circlez{}, an algorithm based on a Fully Connected Neural Network (FCNN) \citep{Saxena2024}. Unlike the standard approach, \circlez{} uses fluxes and colors within apertures in addition to total fluxes. The information on the light profile of the sources is indirectly related to the redshifts via the size.
In terms of accuracy and the fraction of outliers \circlez{}, provides the same accuracy (\qty{5}{\percent}) and a smaller fraction of outliers (\qty{<10}{\percent}) than the photometric redshift computed in \efeds{}, where SED fitting was applied to photometry from the UV, optical, near- and mid-infrared \citep{Salvato2022}.

In addition to these photometric redshifts, \citet{Salvato2024} provide spectroscopic redshifts and associated errors for \NumCTPInitialSpecZ{} sources collected from the literature.
We define the best counterpart redshift $\zCtp$ and associated $1\sigma$ errors as the spectroscopic redshift when available and photometric redshift otherwise.

\section{Filtering and classifying the sample}\label{sec:main_sec_three}
Here, we describe how the final sample of misclassified galaxy clusters is obtained from the data described in the previous section. In \secref{ssec:data_intro_cat_cleaning}, we outline the steps conducted to clean the initial catalog, followed by a description of the cluster classification scheme in \secref{ssec:data_intro_classification_explanation}.
\subsection{Catalog cleaning}\label{ssec:data_intro_cat_cleaning}
\begin{figure*}
    \centering
    \includegraphics[width=\linewidth]{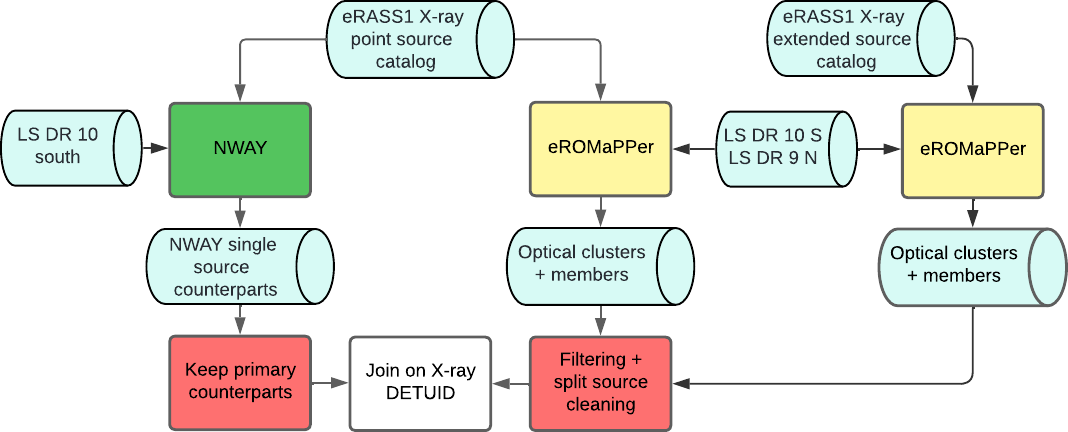}
    \caption{Illustration of the data flow discussed in \secref{ssec:data_intro_cat_cleaning} for assembling the cleaned and merged sample.}
    \label{fig:2_5_sample_assembly}
\end{figure*}

We apply various filters on the X-ray point source catalog by \citet{Merloni2024} and the resulting cluster catalog obtained by \eromapper{} \parsecref{ssec:data_intro_eromapper} to ensure a high level of purity in the final catalog. The details of the filtering are laid out in \secref{sssec:data_intro_sample_cleaning_eromapper}. We further focus on a novel split source cleaning scheme in \secref{sssec:data_intro_split_source_cleaning} and describe the contributions of different \eromapper{} run configurations in \secref{sssec:data_intro_contributions}. We finally outline how the \nway{} counterparts \citep[\secref{ssec:data_intro_nway}]{Salvato2024, Saxena2024} are merged with this sample in \secref{sssec:data_intro_sample_cleaning_and_merging_ctps}.
The data flow for the catalog assembly and cleaning is shown in \figref{fig:2_5_sample_assembly}.

%
\subsubsection{Filtering steps}\label{sssec:data_intro_sample_cleaning_eromapper}
As described in \secref{ssec:data_intro_eromapper}, \eromapper{} is run in \textit{scanning} mode on the full \erass{1} sample consisting of \num{26682} extended and \NumEroInitial{} \pointlike{} X-ray detections with a detection likelihood \detlike$ \geq 5$. This process is similar to the optical identification process of the extent-selected catalog \citepalias{Kluge2024}.
The sample of optical cluster counterparts identified at the positions of the X-ray point sources via \eromapper{} is expected to be highly contaminated by projections and chance associations. We expect that only a low fraction truly has cluster ICM contribution to the observed X-ray flux of the point source. In contrast to the extent-selected sample cleaning, we cannot rely on the \textit{Mixture Model} \citep[covered in detail in][]{Ghirardini2024} to provide contamination probabilities. This is because the model uses redshift and richness information of \eromapper{} detections both at positions of point sources and of random points to infer said contamination probability. While we tried to devise a similar scheme by comparing just to the run at random positions, the run on the point sources is too degenerate with it in redshift-richness space \citepalias[see Fig.~14 in][]{Kluge2024}. Therefore, we have shifted our aim to assembling a catalog of misclassified clusters with high purity. To achieve this, we apply the following conservative cuts on the sample (see \tabref{tab:eromapper_filtering_stages} for a summary):

\begin{enumerate}
    \item As a first step, we clean the initial \NumEroInitial{} X-ray point sources by employing a cut on the detection likelihood \param{DET_LIKE_0} of $\mathcal{L}_{\rm det} \geq 8$, reducing the sample to \NumEroAfterDetlikeCut{} point sources. For this cut, contamination from spurious sources of less than \qty{4}{\percent} is to be expected according to Table~3 of \citet{Seppi2022}, which we deem to be a good trade-off to reach high purity on the input.
    
    \item In addition to that, we mask regions of known non-cluster extended X-ray sources such as supernova remnants, positions of known X-ray binaries and globular clusters, and the region of the Virgo cluster (denoted by the flag \param{IN_XGOOD}), which was analogously performed for the extent-selected cluster catalog \citepalias{Bulbul2024, Kluge2024}. This removes an area equivalent to \qty{62}{\deg\squared}, reducing the total area to \qty{13116}{\deg\squared}, and the number of considered point sources to \NumEroAfterXGoodCut{}.

    \item For \NumEroAfterOpticalPropertiesCut{} of those X-ray point sources, \eromapper{} could identify a cluster candidate with optical properties (redshifts, richnesses, optical centers, and BCG positions). No optical overdensity of red-sequence galaxies could be found at the positions of the remaining point sources.
    
    \item The minimum redshift for the red sequence model employed by \eromapper{} is $z_{\rm min}=0.05$ \citepalias[see appendix~C in ][]{Kluge2024}. The final photometric redshift inferred by \eromapper{} can still be lower but is highly unreliable as the model was not calibrated in that regime. We, therefore, exclude clusters with $z_{\rm best}\leq z_{\rm min}$ for further analysis, leaving \NumEroAfterRedshiftCut{} cluster candidates.
    
    \item As described in \secref{ssec:data_intro_eromapper}, \eromapper{} calculates the limiting redshift \zvlim{} for each identified cluster based on the $z$-band survey depth and a limiting luminosity $L_{\rm thresh}$, which varies based on whether the $W1$ band was used in the run. To limit the amount of false-positive detections, in contrast to the extent-selected catalog, we from here on only consider detections that are \param{IN_ZVLIM} (i.e., $z_\lambda<\zvlim{}$), further limiting the sample to \NumEroAfterZVLimCut{} cluster candidates.
    
    \item The \eromapper{} algorithm discards sources that have specific MASKBITS assigned in the LS (e.g., due to bright nearby objects) and accounts for this via the scaling factor $S$ in the richness estimate \pareqrefsee{eq:richness_definition}. The \param{MASKFRAC} parameter $f_{\rm mask}$ provided for each optical cluster describes the fraction of potential members masked this way. We apply a cut at $f_{\rm mask}\leq0.3$ to exclude clusters with high masking fractions, reducing the sample to \NumEroAfterMaskfracCut{} cluster candidates. Since \eromapper{} was tiled into heal pixels to allow for parallelization, some clusters near the edges of these heal pixels were unintentionally cropped, especially affecting clusters at low redshift due to their larger extent on the sky. As noted in Appendix~A of \citetalias{Kluge2024}, this concerns a few percent of the removed clusters, especially those with $f_{\rm mask}\lesssim0.6$. We thus expect to lose a few hundred clusters (in the final sample) due to this effect by applying our specific \param{MASKFRAC} cut. This is acceptable since we strive for a pure sample, which this cut warrants by removing candidates heavily masked due to nearby bright stars.
    
    \item Since the relative amount of contamination and chance alignments is especially high for the low-richness detections \citepalias[Fig.~14 in][]{Kluge2024}, we employ a richness cut of $\lambda_{\rm norm}\geq \lambda_{\rm min}=16$. \citetalias{Kluge2024} found this value to roughly correspond to a DES~Y1 \citep[optical clusters found using \redmapper{},][]{Abbott2020} richness of $\lambda_{\rm DES}\approx20$, i.e., clusters with $M_{200}\gtrsim10^{14}M_\odot$.
    This filtering step leaves \NumEroAfterRichnessCut{} cluster candidates.

    \item Through visual inspection of the sample obtained at this point, we noticed that the richness cut is sometimes not enough to exclude contaminants, i.e., \eromapper{} cluster candidates found far from the X-ray detection with a clearly different center. We have found the \LMax{} value to be a good indicator of such contaminants, and after extensive visual inspection and comparison to the catalog run on random positions, we decided to discard all candidates with $\LMax{}<20$. This reduces the sample to \NumEroAfterLMaxCut{} cluster candidates.

    \item In the next step, we identify and discard duplicate sources, i.e., separate point sources for which \eromapper{} has matched virtually the same optical clusters.
    The details of this step are further laid out in \secref{sssec:data_intro_split_source_cleaning}. In short, we use a joint catalog of member galaxies of the \NumEroBeforeDupClean{} remaining misclassified candidates and the extent-selected clusters \citepalias{Bulbul2024, Kluge2024}. This enables us to identify \NumDupPlikeSharingWithPlikeOrExt{} misclassified cluster candidates that share members with others or with extent-selected clusters. We calculate the fraction of members they share, forming transitive \emph{associations} of candidates with shared member fractions above \qty{70}{\percent}. Using these, we can discard \NumDupPlikeAssociatedWithExt{} cluster candidates associated with extent-selected clusters. We identify the primary misclassified cluster for the other associations by comparing \LMax{} values within them and discard the \NumDupNonPrimary{} non-primary ones.
    After rejecting all candidates in associations with extended clusters or in an association with a higher \LMax{} candidate, we are thus left with \NumEroAfterDuplicateCut{} cluster candidates.

    \item In a related step, we remove \NumEroCloseExtendedSameRedshift{} further candidates that were found to be a duplicate to an existing extent-selected cluster but missed by the shared member analysis, reducing the sample to \NumEroAfterCloseExtCut{} cluster candidates. The details and reasons for this are also discussed at the end of  \secref{sssec:data_intro_split_source_cleaning}, omitted here for brevity. 

    \item Finally, we manually remove \NumEroVisualContaminants{} sources found to be contaminants due to phantom source detections in the \lsdr{10} caused by foreground dust. In these cases, \eromapper{} identified contaminants as apparent galaxies with similar (red) colors and assigns them to a cluster close to the point source. We found these contaminants through a brief visual inspection of the full sample. After removing them, \NumEroAfterVisualContCut{} cluster candidates remain in the final sample.
\end{enumerate}
\begin{table}[]
    \centering
    \caption{The number of cluster candidates remaining after each filtering step described in \secref{sssec:data_intro_sample_cleaning_eromapper}.}
    \begin{tabular}{ll}
        \toprule
        Filtering Stage & Amount \\
        \midrule
        Initial sample & \NumEroInitial{}\\
        \param{DET_LIKE_0} cut ($\detlike{}\geq8$) & \NumEroAfterDetlikeCut{}\\
        \param{IN_XGOOD} cut & \NumEroAfterXGoodCut{}\\
        Optical cluster identified & \NumEroAfterOpticalPropertiesCut{}\\
        Redshift cut ($z>0.05$) & \NumEroAfterRedshiftCut{}\\
        \param{IN_ZVLIM} cut ($z_\lambda<\zvlim{}$) & \NumEroAfterZVLimCut{}\\
        \param{MASKFRAC} cut ($f_{\rm mask}\leq0.3$) & \NumEroAfterMaskfracCut{}\\
        Richness cut ($\lambdaNorm{}\geq 16$) & \NumEroAfterRichnessCut{}\\
        \param{LMAX} cut ($\LMax{}\geq 20$) & \NumEroAfterLMaxCut{} \\
        Shared member cleaning & \NumEroAfterDuplicateCut{}\\
        Remaining duplicates removed & \NumEroAfterCloseExtCut{}\\
        Visual contaminants removed & \NumEroAfterVisualContCut{}\\
        \bottomrule
    \end{tabular}
    \label{tab:eromapper_filtering_stages}
\end{table}

We thus narrowed down the initial catalog of \NumEroInitial{} X-ray point sources to a sample of \NumEroFinal{} clusters confirmed by \eromapper{}. To compare its properties to those of the extent-selected clusters, we apply a series of the same cuts to the \NumKlugeExtClusters{} sources presented in \citetalias{Kluge2024}. We rely on their \param{SHARED_MEMBERS} parameter for duplicate removal. This results in a subset of \NumEroExtFiltered{} filtered extent-selected clusters.

%
\subsubsection{Identification of duplicates via shared member analysis}\label{sssec:data_intro_split_source_cleaning}
A crucial part of identifying the clusters is assessing how many appear multiple times (i.e., the same clusters which \eromapper{} identified for different point sources or even re-identified from the extent-selected source catalog). The source detection algorithm {\tt{ermldet}} sometimes divides the extended emission into multiple points or extended source detections \citepalias[discussed in detail in ][]{Bulbul2024}, which can lead to extended sources being classified as multiple point sources. These cases naturally would be included in our sample. Another possibility is the incidence of a point source close to a cluster, which can lead to \eromapper{} identifying the cluster twice.
To disentangle these cases, we developed a method to identify duplicates. For this, we use the catalogs of member galaxies that \eromapper{} provides for each cluster, allowing us to find candidates that share the same red-sequence galaxy members. 

\begin{figure}
    \centering
    \includegraphics[width=\linewidth]{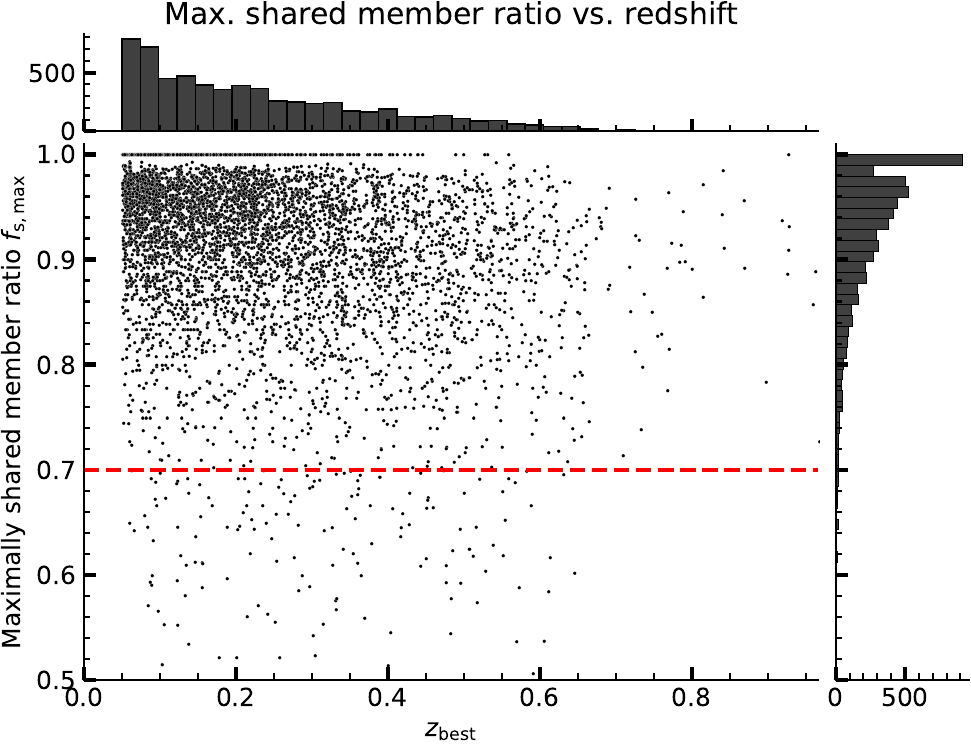}
    \caption{Distribution of the maximally shared member fraction $f_{\rm s, max}$ \pareqref{eq:maximally_shared_member_fraction} as a function of redshift for \NumDupPlikeSharingWithPlikeOrExt{} misclassified clusters that share at least one member. $f_{\rm s, max}$ is calculated w.r.t. the joint member sample of other misclassified clusters and the extent-selected clusters presented by \citetalias{Kluge2024}. The histograms in the top and right panels show projections of the distributions. The dashed red line indicates the threshold at $f_{\rm s, max}=0.7$ that is adopted to mark candidates as part of associations with others.}
    \label{fig:2_5_2_shared_member_frac_vs_z}
\end{figure}

We start with the \NumEroBeforeDupClean{} filtered misclassified cluster candidates identified in the previous cleaning steps \parsecref{sssec:data_intro_sample_cleaning_eromapper}. Of these, \NumDupPlikeSharingWithPlike{} share at least one member with another point source candidate.
Additionally, we consider the \NumKlugeExtClusters{} extent-selected clusters and their member galaxies provided in \citetalias{Kluge2024} to identify those cluster candidates from our point source catalog that mirror their detection. This can happen if true point sources, such as an AGN, are close to the cluster. Including the extent-selected catalog, we find that in total \NumDupPlikeSharingWithPlikeOrExt{} of our misclassified cluster candidates share at least one member with another cluster or cluster candidate. 
For all of these candidates and additionally for all extent-selected clusters that share members, we identify the number of members shared with others in the sample and define a candidate's fraction of maximally shared members with a single different candidate $f^i_{\rm s,max}$ as 
\begin{equation}\label{eq:maximally_shared_member_fraction}
    f^i_{\rm s,max}
        =\max_{i\neq j}\BracedInCurl{N^j_{\rm s}}\frac{1}{N^i_{\rm tot}},
\end{equation}
where $N^j_{\rm s}$ are the shared members of candidate $i$ with candidate $j$, and ${N^i_{\rm tot}}$ the number of members of candidate $i$.
The distribution of this fraction as a function of redshift among the misclassified cluster candidates is displayed in \figref{fig:2_5_2_shared_member_frac_vs_z}. The distribution is skewed towards lower redshifts as the area covered in the sky is larger for low-redshift clusters, so true X-ray point sources such as AGN are more likely to be in the vicinity of them. The distribution of $f_{\rm s, max}$ (projected in the panel on the right) peaks at $\sim0.97$ and displays an additional peak at  $f_{\rm s, max}=1$, reflecting those cluster candidates that share all of their members with another one.

\begin{figure}
    \centering
    \includegraphics[width=\linewidth]{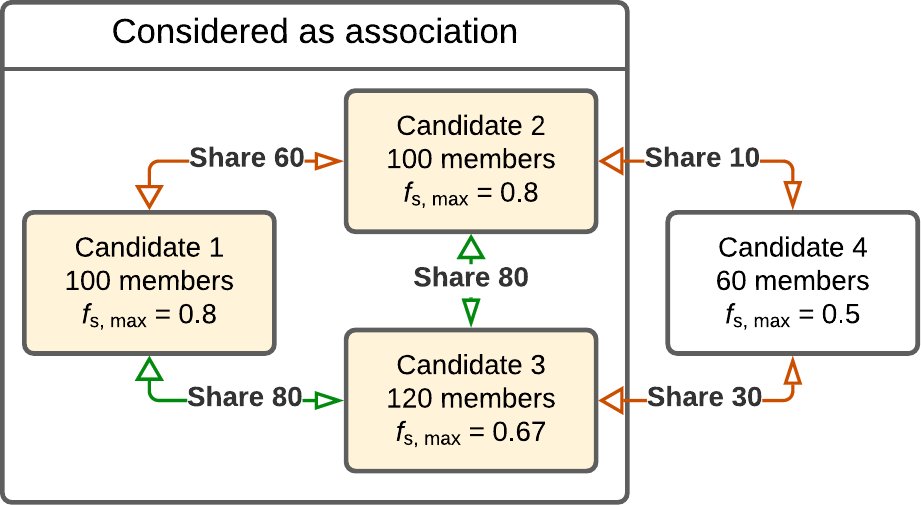}
    \caption{Illustration of the transitive association process.
    The numbers of members and shared members are arbitrarily chosen.
    The labels on the arrows indicate the number of shared members between the respective candidates; their colors denote whether their shared numbers would lead to the two candidates being associated (green) or not (orange).}
    \label{fig:2_5_2_association_example}
\end{figure}

\begin{figure}
    \centering
    \includegraphics[width=\linewidth]{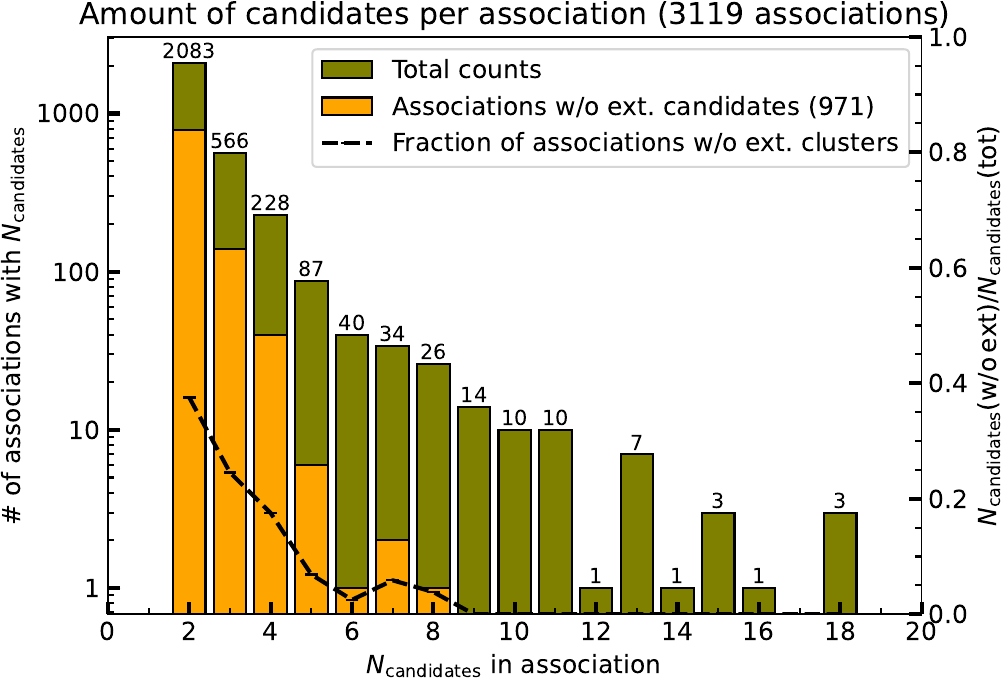}
    \caption{Distribution of the number of cluster candidates in each transitive association sharing members above the threshold.
    The total height of the green bars and the numbers annotated show the total amount of associations that contain $N_{\rm candidate}$. The yellow bars display the number of associations containing $N_{\rm candidate}$ where none of the included candidates in the association is an extent-selected cluster. The dashed line shows the fraction of these associations as a function of the total amount for each $N_{\rm candidate}$. For better visibility, we omit the \NumDupAssociationsAboveNine{} associations where $N_{\rm candidate}$ exceeded \num{20}, most notably one association containing \NumDupLargestAssociation{} overlapping candidates.}
    \label{fig:2_5_2_association_count_distribution}
\end{figure}

Akin to \citetalias{Kluge2024}, who flagged (and discarded) all extended clusters that shared more than \qty{70}{\percent} of their members with another extended cluster with higher $\mathcal{L}_{\rm max}$, we apply a similar threshold at $f_{\rm s,min}=0.7$ (indicated by the red dashed line in \figref{fig:2_5_2_shared_member_frac_vs_z}) to consider a candidate to be of a shared member-source\footnote{\citetalias{Kluge2024} found that only \qty{3}{\percent} of their extended clusters exceeded that threshold; note that, different to our point source sample, the extended X-ray sources had already been split-cleaned based on their extent before the application of \eromapper{}. While a similar procedure (i.e., position-based split cleaning) is possible here via assuming a minimum radial extent for each source, after some experiments, we found that approaching this problem via the shared member fractions is more successful at removing duplicates.}.
For each cluster candidate, we identify all other cluster candidates where $f_{\rm s}$ is above the threshold and assemble them in \textit{associations}. In a second step, we iteratively combine all associations of such cluster candidates that transitively share member fractions above that threshold. This can lead to two candidates being placed into the same association while not having a respective membership fraction $f_{\rm s} \geq 0.7$ but with fractions exceeding the threshold with a third candidate. An illustration of this scenario is provided in \figref{fig:2_5_2_association_example}. Here, candidates 1, 2, and 3 form an association because the $f_{\rm s, max}$ of candidates 1 and 2 links to candidate 3 (additionally indicated by arrow color). Candidate three has only $f_{\rm s, max}=80/120=0.67$ with candidates 1 and 2. Since both have their shared member fraction with 3 exceeding the threshold, they are linked to it and are thus also transitively linked with each other. The example also contains another candidate (4), which is not associated with any of the other candidates as neither of its shared member fractions exceeds $f_{\rm s}=0.7$, independent of the side they are being considered from\footnote{Note that with, e.g., 50 shared members with either one of the other clusters, candidate four would become part of the association since it would then have $f_{\rm s}=0.83$.}. In this case, we would count candidate four as an individual candidate.
We note that for this analysis, the membership probabilities assigned by \eromapper{} are not considered (except for determining \LMax{}), as the association alone provides a handle on the overlap.
We employ this transitive association process to identify \NumDupAssociations{} of such unique associations with shared membership fractions exceeding the threshold among respective candidates, in total finding \NumDupPlikeInAssociations{} misclassified cluster candidates to be part of them. A brief visual inspection confirms that the association method works as anticipated, i.e., the identified candidates in associations indeed share member galaxies appearing to be part of the same optical cluster. The distribution of the number of candidates in those associations is displayed in \figref{fig:2_5_2_association_count_distribution}. We distinguish between associations that contain clusters already present in the extent-selected catalog (\NumDupAssociationsWithExt{}) and ones that only contain candidates identified at the positions of the clusters in the point source catalog (\NumDupAssociationsWithoutExt{}). We note that the fraction of clusters in the point source catalog (shown with the dashed line in \figref{fig:2_5_2_shared_member_frac_vs_z}) decreases with the $N_{\rm candidate}$. All \NumDupAssociationsAboveNine{} associations with \num{10} or more candidates contain at least one extended cluster. When looking at their redshifts, we find that almost all of those are at $z<0.1$, where the larger extent increases the chance of the clusters overlapping with random background X-ray point sources.
For each association, we assess whether the maximum physical distance between each pair of clusters is feasible. For this, we consider their position in the sky and photometric redshifts $z_\lambda$. The limits are set by considering the uncertainties on the photometric redshifts: We flag all associations with any candidates $i, j$ that have $d_{ij,\rm max}> 2 \Delta_d$, where we define $\Delta_d$ as the uncertainty on the comoving distance derived from the maximum photometric redshift uncertainty. We find \NumDupAssociationsAmbiguous{} of such associations in total, of which only \NumDupAssociationsAmbiguousNonExt{} are purely point source associations. Visual inspection of these cases showed that even for these, the member galaxies could still belong to the same (possibly disturbed) optical cluster. Therefore, we decided to keep them in the sample as part of the same association.
We also note that, as also illustrated in \figref{fig:2_5_2_association_example}, our association procedure can lead to cases where even $f_{\rm s, max}$ is below the threshold for some of the candidates in an association. This is the case for \NumDupPlikeWithLowShareFrac{} candidates. Most associations only consist of just two or three candidates, reflecting the cases where only the cluster with the lower amount of members has a shared member fraction above the threshold. 

To summarize, in this cleaning step, we discard \NumDupPlikeInAssociationsDiscarded{} of the \NumEroBeforeDupClean{} misclassified cluster candidates due to sharing high member fractions either with close extent-selected clusters or higher \LMax{} misclassified cluster candidates. We keep the highest \LMax{} candidate for each of the \NumDupAssociationsWithoutExt{} associations without extent-selected clusters. The \param{SPLIT_NEIGHBOR_NUM} column of the final catalog denotes the number of other point sources that point toward the same cluster.

To verify that the shared member analysis described above was successful and avoid duplicates, we cross-match the cluster sample remaining with the extent-selected catalog \citetalias{Kluge2024}. For each cluster, we calculate the closest distance to a match, $d_{\rm match, min}$, and define a \param{HAS_CLOSE_EXT_NEIGHBOR} flag via the condition $d_{\rm match, min}\leq R_{\lambda,{\rm ext}}^*$. Here, the extent-selected cluster's richness radius $R_{\lambda,{\rm ext}}$ is being calculated as introduced in \myeqref{eq:richness_radius}, and converted to an angular distance $R_{\lambda,{\rm ext}}^*$ using its redshift $z_{\rm ext}$. We also calculate the relative redshift offset between misclassified cluster redshift \zEro{} and extent-selected cluster redshift $z_{\rm ext}$, $\Delta z_{\rm rel}:=\abs{\zEro-z_{\rm ext}}/(1+z_{\rm ext})$. We then consider all misclassified cluster candidates for which both \param{HAS_CLOSE_EXT_NEIGHBOR} and $\Delta z_{\rm rel}\leq0.05$ to be duplicates of extent-selected clusters. This concerns only \NumEroCloseExtendedSameRedshift{} candidates, as the shared member analysis has successfully removed most duplicates at this point; we identify two reasons for those being missed by the shared member analysis:
\begin{enumerate}
    \item The optical cluster is very extended on the sky, causing the shared member fraction between the two to be just below the threshold of $f_{\rm s, max}=0.7$.
    \item The extent-selected and corresponding misclassified cluster candidate identified by \eromapper{} are located on different heal pixel tiles. Thus, they do not share any members even though they correspond to the same cluster. This effect is rare since our \param{MASKFRAC} cut already excludes most candidates that could be affected by it.
\end{enumerate}
There remain \NumEroCloseExtendedDifferentRedshift{} misclassified cluster candidates that are located close to an extent-selected cluster (with \param{CLOSEST_EXT_IS_CLOSE}) but are at different redshifts. For \NumEroCloseExtendedSecondaryRedshiftAvailable{} of these, a secondary redshift estimate \param{Z_LAMBDA_SECOND} of the extent-selected cluster is available, which for \NumEroCloseExtendedSecondaryRedshiftMatch{} sources matches \zEro{} of the misclassified cluster. Most of the other extent-selected cluster matches are at rather low redshifts ($z_{\rm ext}\lesssim0.1$), and we believe those to be genuinely different clusters that just happen to overlap in projection with the misclassified cluster matches.

\subsubsection{Filter band contributions}\label{sssec:data_intro_contributions}
As presented in \citetalias{Kluge2024}, \eromapper{} conducted runs using different filter band combinations ($grz$, $griz$, $grzW1$, and $grizW1$). The final catalog is selected from these runs based on a priority scheme. For low and intermediate redshifts ($z_\lambda\leq0.8$), the $grz$ combination performs best, while for higher redshifts ($z_\lambda >0.8$), the $grizW1$ combination had the best performance. The separate catalogs were merged using a priority scheme (e.g., a source with a cluster identified at $z_\lambda\leq0.8$ would only be taken from the $griz$ run if not available in the $grz$ run). In total, there were six runs, four of them done in the \lsdr{10} south area, and two of them (without $i$-band photometry) in the \lsdr{9} north area. The final catalog of misclassified cluster candidates, therefore, consists of sources identified using the combinations shown in \tabref{tab:run_distribution_table}.
\begin{table}
\caption{The contribution of clusters from \eromapper{} runs using the filter band combinations in the \emph{Bands} column, as described in \secref{sssec:data_intro_contributions}.}
\label{tab:run_distribution_table}
\begin{tabular}{llllr}
\toprule
Survey & Region& Bands & Condition & Amount\\
\midrule
\lsdr{10} & south & $grz$ & if $z_\lambda\leq0.8$ & 7901 \\
\lsdr{10} & south & $griz$ & if $z_\lambda\leq0.8$ & 12 \\
\lsdr{9} & north & $grz$ & if $z_\lambda\leq0.8$ & 213 \\
\lsdr{10} & south & $grzW1$ & if $z_\lambda>0.8$ & 3 \\
\lsdr{10} & south & $grizW1$ & if $z_\lambda>0.8$ & 216 \\
\lsdr{9} & north & $grzW1$ & if $z_\lambda>0.8$ & 2 \\
\bottomrule
\end{tabular}

\end{table}

\subsubsection{Merging the sample with the \nway{} single source counterpart catalog}\label{sssec:data_intro_sample_cleaning_and_merging_ctps}
To infer further characteristics of the galaxy clusters identified by \eromapper{}, we consider the \lsdr{10} single source counterparts in the optical and infrared \parsecrefsee{ssec:data_intro_nway}.
Such counterparts have been found for \num{\sim750000} \erass{1} X-ray point sources \citep{Salvato2024} in the \lsdr{10} footprint. By merging these to the \NumEroFinal{} \eromapper{} clusters, we find that \NumCTPAvailable{} have at least one counterpart available. The remaining \NumClassZero{} sources do not have a counterpart since they fall in the northern \lsdr{9} region without $i$-band photometry, for which no \nway{} counterpart cross-matching was performed. We keep those in the sample, labeling them as \clusterClass{0}.
For each X-ray source in the point source catalog, \nway{} provides a general probability \pAny{} that any of the matches within \qty{1}{\arcmin} represent a correct match, and an individual probability $p_i$ for each match within that radius. In some cases, this leads to multiple possible counterparts being identified. We discard all higher-order matches beyond the primary one (i.e., the one with the highest $p_i$). Such secondary matches are discarded for \NumCTPWithSecondaryMatch{} of the counterparts.

%
\subsection{Classification}\label{ssec:data_intro_classification_explanation}
In this section, we first describe which properties of the cluster candidates are considered for the classification scheme \parsecref{sssec:classification_parameters} and then describe the classes we derive from those. We provide an overview of redshifts and colors of counterparts.

We adopt a classification scheme reflecting the method established for the X-ray misclassified cluster candidates in the \efeds{} field by \citet{Salvato2022} but apply modifications to account for additional information and different data available. \citet{Salvato2022} ranked their misclassified clusters into 5 classes, where \clusterClass{1} denoted true X-ray point sources with no cluster association. In their scheme, the classes were defined so that their order roughly reflected the likelihood of the cluster being associated with the X-ray point source emission.
Although we follow their general scheme, the modifications we apply lead to some major differences due to, e.g., the shallower X-ray data and varying exposures of \erass{1}, the updated version of the LS catalogs, and the different cluster-finding algorithms. Also, we introduce a distinction based on assessing the Galactic nature of secure counterparts.

\subsubsection{Classification parameters}\label{sssec:classification_parameters}
\begin{figure*}
    \centering
    \includegraphics[width=\linewidth]{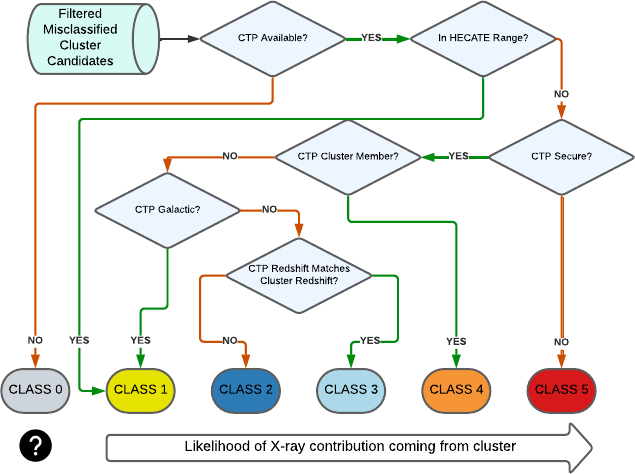}
    \caption{Illustration of the classification process for the cluster classes introduced in \secref{ssec:data_intro_cluster_class_definition}. We abbreviate the \emph{counterpart} associated via \nway{} to \emph{CTP} for better readability. The decision points are described in detail in \secref{sssec:classification_parameters}.}
    \label{fig:2_6_classification}
\end{figure*}

Before introducing the cluster classes, we describe the features we consider for the classification. Abbreviations (which shorten \textit{counterpart} to CTP) of these are used in the decision diagram \figref{fig:2_6_classification}. The following distinctions are useful for characterizing the nature of the cluster candidates:

    \noindent \textbf{Is counterpart information available?} As discussed above, there are \NumClassZero{} cluster candidates in the \lsdr{9} north region for which no \nway{} counterpart information is available at all.
    This is the \param{CTP Available?} decision point in \figref{fig:2_6_classification}.
    
    \noindent \textbf{Is the X-ray source in range of a \hecate{} galaxy?} The \hecate{} galaxy catalog\footnote{\url{https://hecate.ia.forth.gr/}} \citep{Kovlakas2021} lists nearby ($z\lesssim0.047$) bright galaxies that might be responsible for the detected X-ray point source emission. We, therefore, introduce the \param{In HECATE Range?} decision point in \figref{fig:2_6_classification} to distinguish all sources that are inside the area covered by a \hecate{} galaxy.
    
    \noindent \textbf{Are we certain the counterpart association is secure?} For each optical counterpart to an X-ray point source, \citet{Salvato2024} provide a purity value based on the association likelihood \pAny{} and the surrounding $\detlike{}\geq8$ sources (by tile). We assume the \nway{} single source counterpart to be a secure association if the \param{PURITY8} value is above \qty{90}{\percent}. Of the \NumCTPAvailable{} counterparts, \NumCTPSecure{} are above this threshold.
    This is the \param{CTP Secure?} decision point in \figref{fig:2_6_classification}.

    \noindent \textbf{Is the counterpart associated with the cluster?} If the counterpart is identified as a member of the cluster by \eromapper{}, it is likely that no single AGN is responsible for the X-ray emission, even in the few cases where its redshift estimate is off, see \secref{sssec:3_1_ctp_redshift_comparison}. We, therefore, use that membership, indicated by the \param{CTP Cluster Member?} decision point in \figref{fig:2_6_classification}.
    
    \noindent \textbf{Is the counterpart Galactic} As described in \secref{sssec:data_description_galactic_classification}, we assign the \param{CTP_IS_GALACTIC} flag based on the \citep{Salvato2024} classification to distinguish between Galactic and extragalactic \nway{} counterparts. Counterparts flagged as Galactic are likelier to have wrong redshift measurements as the \circlez{} algorithm is not calibrated for those. If these counterparts are securely associated with the X-ray detection, it is likely that any extended X-ray emission from the optically identified cluster is contaminated by the foreground source. A similar thing can happen if the counterpart is within the radius of an object in the \hecate{} catalog \citep{Kovlakas2021}, i.e., a bright nearby ($z\lesssim0.47$) galaxy. We bundle this categorization in the \param{CTP Galactic?} decision point. 
    
    \noindent \textbf{Do counterpart and cluster redshifts match?} The photometric redshifts derived via \circlez{} come with a probability distribution function that reflects the uncertainties and degeneracies \citep{Saxena2024}. We use this to calculate an associated symmetric $1\sigma_{\zCtp}$ error. We flag all counterparts with \param{CTP_FLAG_BEST_Z_UNRELIABLE} where this error relative to the redshift is high, i.e., a low precision is evident as $\sigma_{\zCtp}/(1+\zCtp)>0.15$. We choose this particular value as estimated redshifts deviating by more than that are conventionally considered catastrophic outliers in the literature. Additionally, we require the photometry of the counterpart to be reliable when no spectroscopic redshift is available. For each counterpart, \citet{Salvato2024} provides a flag that marks whether the nominal depth has been reached in all of the $griz$ bands (\param{IN_ALL_LS10}). We both use this flag and enforce a stricter criterion by requiring a minimum signal-to-noise ratio of $\sigtonoise{}_{\rm min}=3$ in all of the $grzW1$ filter bands.
    We define the \param{CTP_FLAG_BAD_PHOTOMETRY} as either \param{IN_ALL_LS10} or $\sigtonoise{}_{\rm min}\geq3$ not being met in all bands.\\
    For all secure counterparts, we thus assume the redshifts of counterpart and cluster to coincide if the redshift is reliable, the photometry is reliable if no spectroscopic redshift is available, and the criterion
    \begin{equation}\label{eq:same_redshift_criterion}
        \abs{\zEro-\zCtp}\leq \sqrt{\sigma_{\zEro}^2+\sigma_{\zCtp}^2+(0.15\zEro)^2}
    \end{equation}
    is fulfilled. Here, $\sigma_{\zEro}$ is the error of the \eromapper{} cluster photometric redshift, and the $0.15\zEro$ term takes into account the commonly employed outlier boundary. Secure counterparts whose redshift estimate is unreliable (\param{CTP_FLAG_BEST_Z_UNRELIABLE}, see above) are considered at different redshifts by default. This is the \param{CTP Redshift Matches Cluster Redshift?} decision point.
\subsubsection{Cluster Class definitions}\label{ssec:data_intro_cluster_class_definition}

\begin{figure*}
\centering
\includegraphics[width=\linewidth]{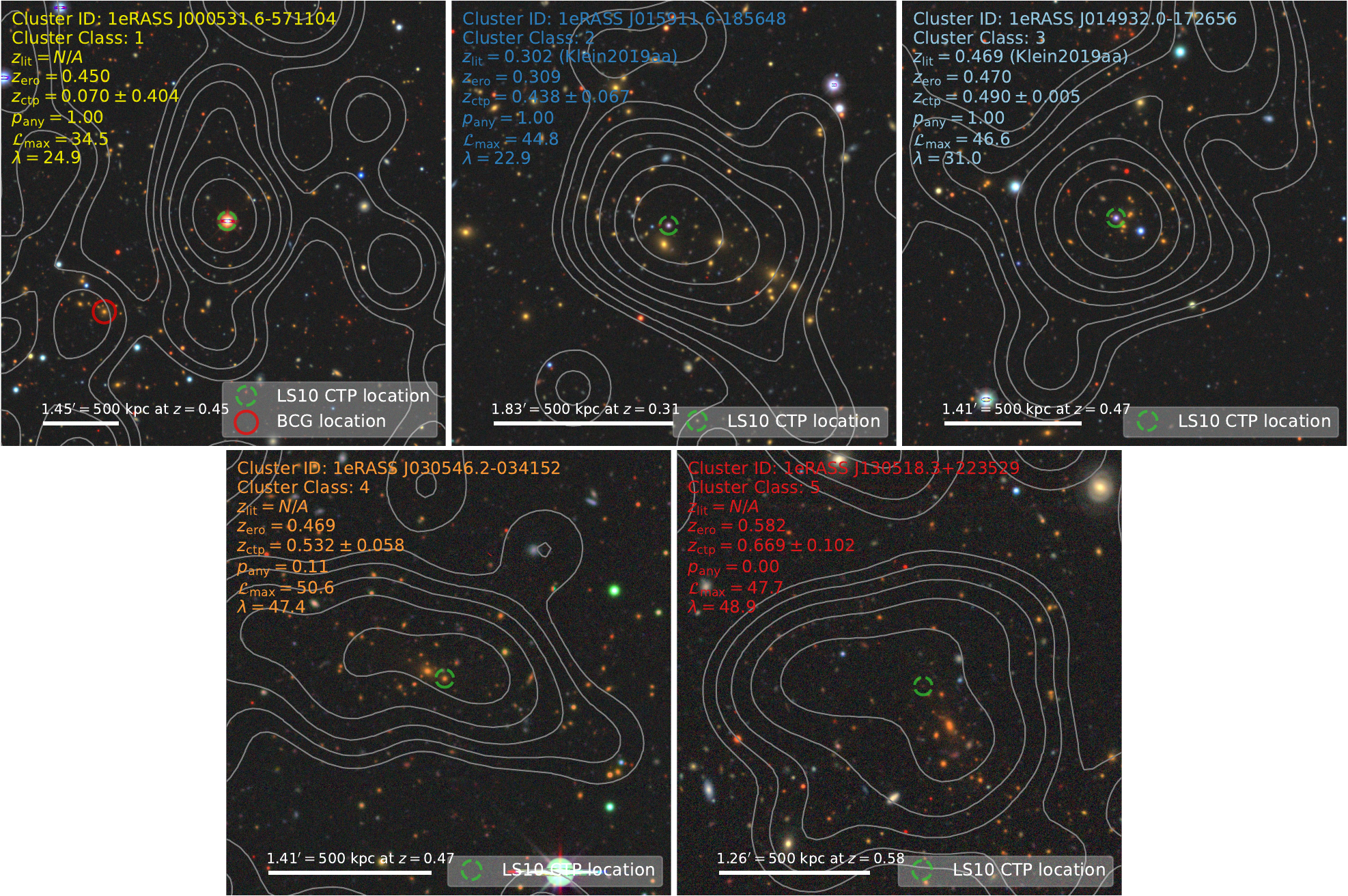}
    \caption{Example \lsdr{10} $grz$ images with overlaid X-ray contours for one misclassified cluster of each \clusterClass{}. The green-dotted circle marks the location the X-ray detection's most likely single-source counterpart determined via \nway{}. Important parameters of cluster and counterpart association are annotated. \textit{Upper left:} A \clusterClass{1} cluster. Here, the X-ray point source is clearly associated with the star, while \eromapper{} identified a nearby cluster in the lower left quadrant, where we marked the BCG to highlight the location of the cluster. \textit{Upper center} A \clusterClass{2} cluster with an AGN clearly associated with the X-ray emission. The foreground cluster emission might contribute partially as seen by the elongated shape. \textit{Upper right:} A \clusterClass{3} cluster, having an AGN located at the same redshift $z\sim0.48$. The contours resemble those of a point source, with the AGN likely outshining any ICM emission. \textit{Lower left:} A \clusterClass{4} cluster, for which the most likely single counterpart is an early type galaxy that is identified as a member. There is no AGN discernible, and the contours are a little more extended. \textit{Lower right:} A \clusterClass{5} cluster, for which the \nway{} counterpart is not secure ($\pAny{}\approx0$). The X-ray contours have a more extended appearance than e.g. the point source of the \clusterClass{3} cluster. North is up; east is left, and the images are centered around the initial X-ray point source detections.}
    \label{fig:6_1_1_examples_one}
\end{figure*}

In the context of this paper, together with a catalog, we provide classifications of clusters based on their optical counterpart properties to identify extreme and interesting clusters and facilitate follow-up studies. A detailed study of the counterpart colors is provided in \secref{sec:main_sec_five}, and their X-ray properties are presented in \secref{sec:main_sec_six}. The parameters and decision points described above allow us to define the following cluster classes for any \eromapper{} cluster we associate with an X-ray point source, with examples provided in \figref{fig:6_1_1_examples_one}:

{\bf \clusterClass{0}:} This class contains the \NumClassZero{} cluster candidates for which no \nway{} counterpart is available but which still reflect X-ray point sources for which \eromapper{} has identified an optical cluster counterpart in the \lsdr{9} north data. \clusterClass{0} candidates could thus belong to any of the other classes, but we cannot distinguish them. We leave these in the sample for completeness but note that they are not as well characterized as the rest of the sample that contains single-source counterpart information.

{\bf \clusterClass{1}:} This class contains cluster candidates for which the X-ray source is in range of a \hecate{} galaxy or for which the secure counterpart association is likely to be Galactic and not identified as a member of a cluster (see also the definition of \clusterClass{4}). 
For the point sources of \clusterClass{1} clusters, any X-ray emission from the optically identified cluster is, therefore, likely to be contaminated. We provide an example for such a cluster in the upper left corner of \figref{fig:6_1_1_examples_one}, where the optical cluster detection is clearly offset from the Galactic point source detection. The contours show some faint extended emission at the location of the cluster, but that was not enough to warrant a detection.
In total, this class contains \NumClassOne{} cluster candidates, of which only \NumClassOneInHecateRange{} are in the range of a \hecate{} galaxy.

{\bf \clusterClass{2}:} This class is used for cluster candidates for which the \nway{} counterpart is securely associated, is extragalactic, is \textit{not} a member galaxy of the cluster, and has good photometry. In addition to that it is required to have either a different redshift than the cluster \pareqrefsee{eq:same_redshift_criterion} or an unreliable redshift estimate.
The criteria lead to a total of \NumClassTwo{} \clusterClass{2} sources, of which \NumClassTwoUnreliableRedshift{} have an unreliable redshift estimate or bad photometry with only photometric redshift available. Physically, these objects correspond to X-ray point sources for which the majority of the measured X-ray emission is expected to stem from AGN along the line of sight to the optically identified clusters; there are \NumClassTwoBehind{} sources with a counterpart redshift higher than the cluster, and \NumClassTwoInFront{} sources where the counterpart appears to be in the foreground the cluster. An example of such a cluster is shown in the upper central panel of \figref{fig:6_1_1_examples_one}. Here the blue counterpart is very close to the peak of the X-ray emission and seems to be behind the cluster given \zCtp{}.

{\bf \clusterClass{3}:} This class indicates sources for which the \nway{} counterpart is secure, is extragalactic, \eromapper{} has not identified it as a member galaxy of the cluster, and the counterpart redshift and cluster redshift match according to \myeqref{eq:same_redshift_criterion}.
We find \NumClassThree{} sources that fulfill the criteria for \clusterClass{3}. These objects are likely to represent AGNs that reside in clusters, and the detected X-ray signal might be the cluster's emission boosted by the point source or the faint extended emission from the ICM gas dominated by the X-ray emission from the central AGN. An example of such a cluster is shown in the upper right panel of  \figref{fig:6_1_1_examples_one}; the counterpart looks similar to the one of the \clusterClass{2} example, but its spectroscopic redshift pinpoints it to be part of the cluster.

{\bf \clusterClass{4}:} This class represents sources for which the \nway{} counterpart is secure and is a member of the cluster as assigned by \eromapper{} and the colors of the counterpart are consistent with an early type red galaxy (see the details in \secref{ssec:properties_color_distribution}). We find a total of \NumClassFour{} sources fulfilling these criteria. We expect the sources in this class to be galaxy clusters which might be more compact than the extent-selected ones, usually having a cool core. An example of such a cluster is shown in the lower left panel of \figref{fig:6_1_1_examples_one}. There is clearly no AGN candidate that could be attributed to the emission, and \nway{} selects a member of the cluster to be the most likely single-source association. We decided to prioritize the membership decision point both over the Galactic/extragalactic and the matching redshift criteria after realizing in visual inspection that those clusters are a better fit for this class than \clusterClass{1} or \clusterClass{2}. We note that there are \NumClassFourGalacticCTP{} counterparts classified as Galactic that have also been identified as cluster members. We include these in \clusterClass{4} after a visual inspection showed that their Galactic classification might be inaccurate. Also, there are \NumClassFourRedshiftOutliers{} counterparts that seemingly conflict with the cluster redshift estimate, although they are found to be members. We discuss the latter cases in \secref{sssec:3_1_ctp_redshift_comparison}.

{\bf \clusterClass{5}:} This class represents sources for which no reliable \nway{} counterpart to the X-ray point source could be identified. For sources in \clusterClass{5}, the measured X-ray flux likely stems from the clusters. We find a total of \NumClassFive{} clusters in this class. 
These clusters have been misclassified in part due to having emission comparable to the PSF of \erosita{}, in part due to having another point source close by, and in part due to the nature of the $\extlike=3$ cut, where sources with only a slight extent can end up being classified as a point source. The latter might be the case for the example shown in \figref{fig:6_1_1_examples_one}. Close to the reported coordinates of the X-ray emission, there seems to be no single source that could account for it, so \nway{} assigns the closest (faint) galaxy with $\pAny\approx0$. The cluster is slightly offset.

\begin{figure}
    \centering
    \includegraphics[width=0.9\linewidth]{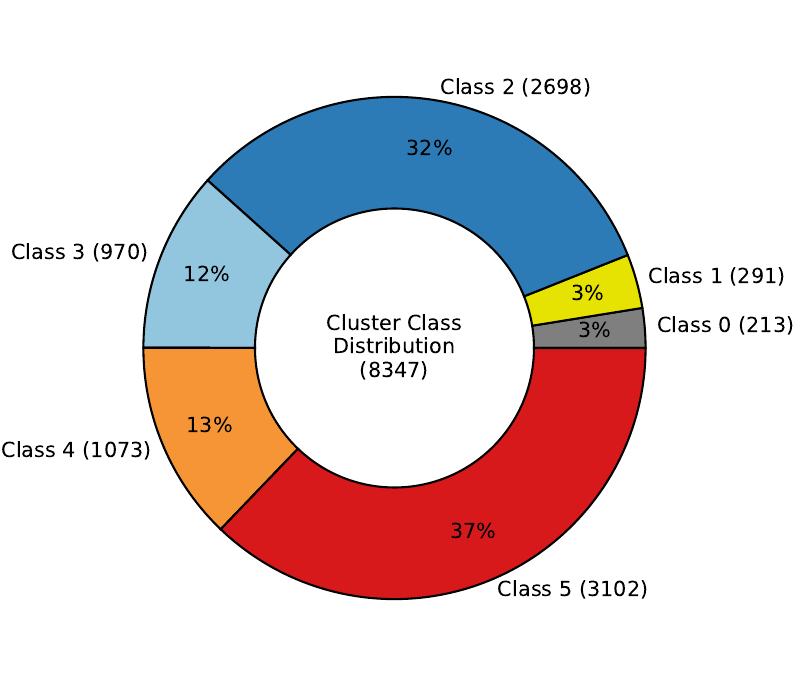}
    \caption{Distribution of the number of clusters in each \clusterClass{}. About a third of the clusters belong to \clusterClass{2}, where an AGN is located in projection to the cluster. The clusters hosting bright AGN are categorized in \clusterClass{3}. Misclassified clusters without apparent AGN contribution (\clusterClass{4} and \clusterClass{5}) comprise about half of the sample, and Galactic contaminants (\clusterClass{1}) and those sources without counterpart information (\clusterClass{0}) contribute only a small part.}
    \label{fig:2_6_cluster_class_distribution}
\end{figure}

We have provided an overview of how each class is derived decision-wise in \figref{fig:2_6_classification}. The distribution of candidates with respect to these cluster classes is presented in \figref{fig:2_6_cluster_class_distribution}.
About half of the identified cluster candidates lack a secure AGN counterpart (\clusterClass{4} or \clusterClass{5}). Most of the other optically identified galaxy clusters seem to be in the vicinity of the projection of an AGN (\clusterClass{3} and \clusterClass{4}). Only a small proportion of our sources seem to have securely Galactic counterparts or are in the range of a \hecate{} galaxy (\clusterClass{1}), or are lacking counterpart information (\clusterClass{0}).

\subsubsection{Counterpart redshifts}\label{sssec:3_1_ctp_redshift_comparison}

\begin{figure}
    \centering
    \includegraphics[width=\linewidth]{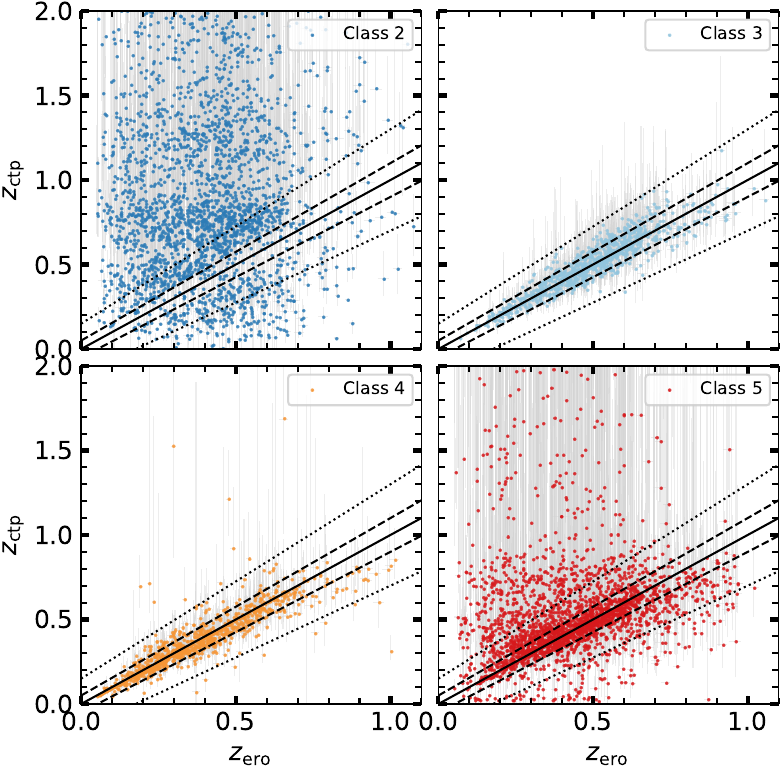}
    \caption{Relation between the most likely counterpart best redshifts $\zCtp$ and the corresponding \eromapper{} cluster redshifts $\zEro$, split by cluster class. $1\sigma$ error bars corresponding to these best redshifts are indicated in gray.
    The black lines show $\zCtp=\zEro\pm m(1+\zEro)$ with $m=0$ for the straight, $m=0.05$ for the dashed lines, and $m=0.15$ for the dotted lines. Since photometric redshift estimates for Galactic counterparts are not reliable, we omit to show \clusterClass{1}.}
    \label{fig:3_1_ctp_redshift_comparison}
\end{figure}

Since the redshifts of the counterparts are a vital part of the classification, we provide a comparison to the cluster redshifts here. More details on the cluster redshifts are discussed in \secref{ssec:cluster_redshift_distribution}; we are using the best redshift estimate \zEro{}.
As described in \secref{sssec:data_description_photo_z}, the best counterpart redshift \zCtp{} is spectroscopic if available, and photometric determined by \circlez{} otherwise. The \circlez{} photometric redshifts are associated with realistic errors, which can be high in a relative sense.
The relation of \zCtp{} to \zEro{} is shown in \figref{fig:3_1_ctp_redshift_comparison} for classes \num{2} to \num{5}, as \clusterClass{0} clusters do not have \nway{} counterparts, and \clusterClass{1} clusters have counterparts that are likely to be Galactic, which results in flawed redshift estimates.

For \clusterClass{2} (upper left panel in \figref{fig:3_1_ctp_redshift_comparison}), the counterpart redshifts are by construction offset from the \eromapper{} redshifts. There are almost no counterparts where $\zCtp\approx\zEro$; if they are, it is due to the counterpart being associated with bad photometry or an unreliable redshift estimate (\param{CTP_FLAG_BEST_Z_UNRELIABLE}, see \secref{sssec:classification_parameters}). The counterpart redshifts follow no clear trend, having a median at a higher redshift than any of the other classes and being usually associated with high uncertainties. There are more clusters with counterparts at higher redshifts (in total \NumClassTwoBehind{}), while \NumClassTwoInFront{} are associated with counterparts at lower redshifts.

For \clusterClass{3}, the counterpart redshifts by construction are close to the cluster redshifts, with deviations only due to the uncertainties in \zCtp{}. This is shown in the upper right panel of \figref{fig:3_1_ctp_redshift_comparison}. The $1\sigma$ error bars shown indicate that some \clusterClass{3} clusters should be expected to belong to \clusterClass{2}; without spectroscopic redshift information of the counterparts, it is currently not possible to disentangle these cases.

For \clusterClass{4}, the correlation of counterpart and cluster redshifts is generally tight (see lower left panel on \figref{fig:3_1_ctp_redshift_comparison}). This confirms our method of selecting these clusters via their counterpart's membership assignment through \eromapper{}, as the counterpart spectroscopic and photometric redshift estimates agree independently. There are a few estimates that disagree. We find moderate offset (i.e., $z_{\rm diff}:=\abs{\Delta z}/(1+\zEro)>0.05$) for \NumClassFourRedshiftOutliers{} clusters, and stark disagreement ($z_{\rm diff} > 0.15$) for \NumClassFourRedshiftHighOutliers{} of those. There are two explanations for these disagreements: Either the \circlez{} photometric redshift estimate of the counterpart is uncertain (with \zCtp{} mostly being in agreement with \zEro{} within the $1\sigma$ uncertainty estimate, indicated by the gray uncertainty bars in \figref{fig:3_1_ctp_redshift_comparison}), or the \nway{} counterpart that \eromapper{} has identified as a member galaxy is not a true member galaxy. We visually inspected the \NumClassFourRedshiftHighOutliers{} clusters with stark disagreement to disentangle these cases. For most of them, the former seems to be the reason. 

In \clusterClass{4}, we find two cases for which the \eromapper{} membership association has an issue. In first case, an AGN (\iauname{J123745.1+240542}) at $\zCtp=0.716$ is incorrectly identified as a member of a $\zEro=0.450$ cluster, most likely due to high uncertainties in the photometry. Therefore, this cluster should instead be categorized as \clusterClass{2}. In the second case (\iauname{J020705.3-272657}), \eromapper{} adopted the spectroscopic redshift of the central galaxy (in this case coinciding with the BCG) at $z_{\rm spec,cg}=1.00$. In contrast, its photometric redshift of $z_\lambda=0.365$ is consistent with $\zCtp=0.32\pm0.05$. $z_{\rm spec,cg}$ does not correspond to the BCG but to a gravitationally lensed background quasar whose arc surrounding the BCG is visible. 
Other than for those outliers, we find that the photometric redshift estimation with \circlez{} seems to be precise for most elliptical galaxies. All of the \NumClassFourGalacticCTP{} \clusterClass{4} clusters with counterparts classified as Galactic still have a counterpart redshift assigned that matches the cluster's redshift.

For \clusterClass{5} (lower right panel of \figref{fig:3_1_ctp_redshift_comparison}), \NumClassFiveNoRedshift{} \clusterClass{5} sources with no counterpart redshift estimate are not shown. For many of those where it is available, \zCtp{} and \zEro{} seem to agree. This is surprising, as the counterpart found by \nway{} has a very low association probability with the X-ray source and could therefore have a random redshift. Their agreement (and later also the counterpart's colors, see \secref{ssec:properties_color_distribution}) indicates that many of the sources identified as the low-probability counterpart are galaxies belonging to the cluster, not unlike \clusterClass{4} clusters. This is reflected by the fact that for \NumClassFiveStillMember{} of the \NumClassFive{} \clusterClass{5} clusters, the most likely counterpart from which \zCtp{} is taken is considered by \eromapper{} to be a member of the cluster.

%
\section{Cross-matching with public cluster catalogs}\label{sec:main_sec_four}

\begin{figure}
\includegraphics[width=0.5\textwidth,trim={0 0 0 0}, clip]{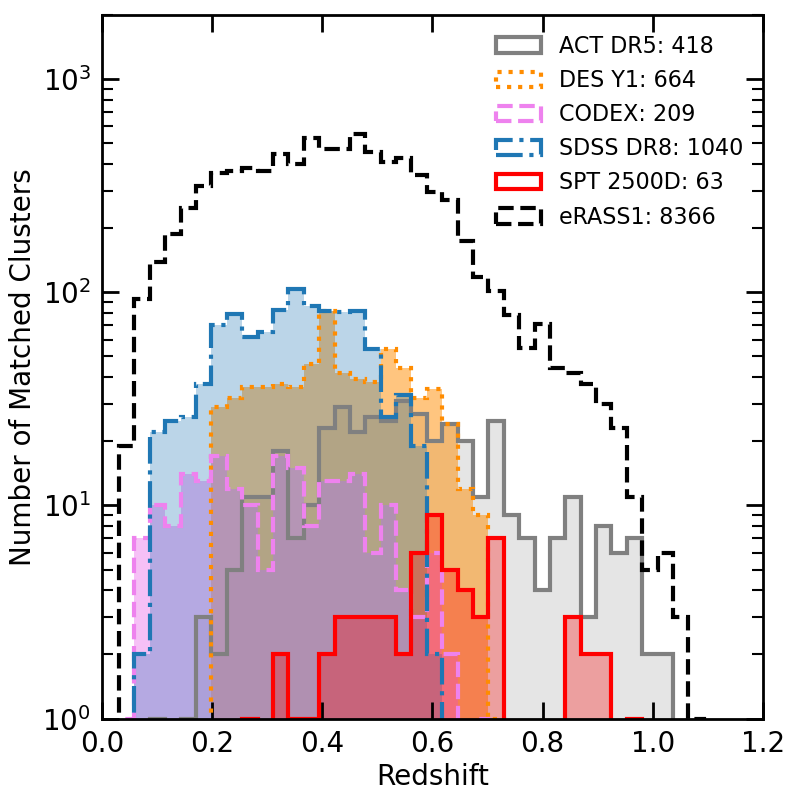} 
\caption{Redshift histogram of the \erass{1} misclassified clusters that have matches in the ACT~DR5, DES~Y1, CODEX, SDSS~DR8, SPT~2500D Surveys, and the redshift histogram ($z_{\rm best}$) of all misclassified clusters. The largest overlap of clusters is between optically selected samples and \erosita{}.}
\label{fig:hist_matched}
\end{figure}

The clusters of galaxies detected in the point source catalog carry significant implications for understanding the selection effects and the completeness of the extent-selected samples compiled from the same data. The cross-matches with public optically, X-ray, and SZ-selected clusters also help to evaluate the cluster populations these surveys detect. This section provides counterparts of \erass{1} clusters in the point source catalog to those found in several previously published cluster catalogs compiled from optical, X-ray, and SZ surveys. A compilation of cross-matches with these surveys, their median redshift, and offset distributions are provided in \tabref{table:match}. We use a matching radius of \qty{2}{\arcmin} applied to the \erass{1} for a fair comparison. The identifiers of the matched clusters are given in the catalog under the column \param{MATCH_NAME}. Out of the \NumEroFinal{} clusters of galaxies detected in this sample, we find that \num{2528} of them have known counterparts, suggesting a \qty{70}{\percent} discovery rate. This percentage falls below the rate of the extent-selected sample with the reported rate of \qty{80}{\percent} \citepalias[see][for details]{Bulbul2024}. Figure~\ref{fig:hist_matched} shows a histogram of cross-matched clusters with selected cluster samples.

\subsection{Comparison with optically selected cluster samples}\label{ssec:comparison_optical}

In the literature, extensive work has been done to detect optically selected samples from data of ground-based observatories, such as the Sloan Digital Survey (SDSS), Dark Energy Survey (DES), and Abell Catalogs, as well as a dedicated infrared/optical surveys for high-redshift clusters \citep{Rykoff2014, McClintock2019, Andernach1991, Gonzalez2019}. Those samples, such as CODEX, which uses information in optical and X-ray bands, have one of the largest matches as the cluster sample was sampled similarly from the point source catalog in this work. We find \num{209} out of \num{3106} clusters at a median redshift of $0.33$ in the common footprint between \erosita{} and CODEX. The median separation is rather large, $57\farcs6$, similar to the counterparts of the \erass{1} extent-selected sample. The selection of this sample is based on the optical properties of the X-ray-selected parent sample, with the expectation of a higher proportion of matches with the \erosita{} identified samples.

One of the most extensive optically selected nearby rich galaxy cluster catalogs was compiled by \citet{Abell1989}. As the Abell catalogs are mainly composed of nearby clusters, they are expected to have a higher matched rate with the extent-selected sample, and indeed, many cross-matches (\num{152}) have already been identified in the catalog of \citetalias{Bulbul2024, Kluge2024}. In this work, we match another \num{23} of them at a median redshift of \num{0.1}; we note that some are not recovered due to the $z\geq0.05$ cut we apply to our sample.

Another wide area survey with a large common footprint with \erosita{} is the Dark Energy Survey (DES) \citep{McClintock2019}. The cluster catalogs based on DES-Y1 data are compiled through the \redmapper{} algorithm, which has \num{664} close associations in our catalog, a number comparable to the extent-selected sample. The median sample redshift of \num{0.42} and a median separation of $28\farcs9$ are larger than the extent-selected sample reported in \citet{Bulbul2024}. We find \num{82} close matches clusters with the surveys dedicated to finding high redshift clusters at high redshifts $z>0.8$, e.g., MaDCoWS \citep{Gonzalez2019}, NEURALENS \citep{Huang2021}, GOGREEN-GCLASS \citep{Balogh2021}, and WENHAN-HIGHZ-SPECZ \citep{Wen2018}. The shallow depth of the WISE data included in the \lsdr{10} survey makes the confirmation challenging in our catalogs; however, we still can identify the most massive of these high redshift clusters in our sample at $z>0.5$ through cross-matching.

In general, the number of the cross-matched clusters in this sample is as large as for the extent-selected clusters but with a redshift distribution skewed to slightly higher values. We observe a slightly larger median offset in this sample. This may be attributed to the fact that the clusters in our sample are either below the survey's flux limit or predominantly dominated by X-ray emission from a point source. As a result, the X-ray centroids of these clusters are less accurately determined than those in the extended source catalog.

\subsection{Comparison with X-ray Catalogs}
\label{ssec:comparison_xray}

The galaxy clusters presented in this work have a compact appearance consistent with the \qty{\sim30}{\arcsec} survey-averaged PSF of \erosita{}, leading to their misclassification as point sources. As pointed out for deeper \efeds{} observations by \citet{Bulbul2022}, some of these clusters are likely to be below the flux limit of the shallow \erass{1} survey. They may be detected in the point source catalog because an AGN boosts their flux along the line of sight or in the cluster itself. Another possible reason for misclassification could be the bright, cool cores of compact, low-redshift groups or high-redshift clusters obscuring the faint, extended X-ray emission from the outskirts, causing the source detection algorithm to classify them as point sources. Our optical identification method, which is not biased by the contamination of AGN flux, can identify the red galaxy members of the clusters in the point source catalog if they fall above the optical cut applied to the sample (see \secref{ssec:data_intro_classification_explanation} for details). In the common footprints, deep X-ray surveys with better PSF, e.g., {\it Chandra} and {\it XMM-Newton}, may detect these clusters. We, therefore, cross-matched our sample with X-ray catalogs in the literature to see which clusters are recovered, i.e., that are not part of the extent-selected catalog presented in \citetalias{Bulbul2024}.

The first large source catalogs in the X-ray band were compiled using the ROSAT All-Sky Survey \citep[RASS,][]{Voges1999}. However, the optical follow-up of these sources for confirmation had to wait until the large overlapping areas were covered with the optical surveys \citep[e.g.][]{Boehringer2000_noras, Ebeling1998, Cruddace2002, Burenin2007}. The most extensive cluster catalog was constructed from the RASS into a meta catalog of \num{1743} confirmed clusters in an area with the largest overlap with the \erosita{} survey by \citet{Piffaretti2011}. Because of their extended nature, most of these clusters are already detected in the extent-selected \erass{1} cluster catalog. Still, we find an additional \num{39} clusters with a median redshift of \num{0.27}, a higher redshift than the cross-matched sample with extent-selected clusters with a similar median offset value. The other extensive cluster catalogs, such as CODEX, are based on the X-ray detection of the RASS, identifying counterparts in the overlapping SDSS area reaching down an X-ray flux limit of \qty{e-13}{\erg\per\s\per\cm\squared}. We find \num{209} common clusters at higher redshifts in this sample with a median redshift of $z=0.33$; similarly, in this case, most of the extended sources are already matched to the extent-selected \erosita{} samples.

Right after the launch of \erosita{}, a limited area \qty{140}{\deg^2} survey in the Equatorial strip (the \erosita{} Final Equatorial Depth Survey, \efeds{}) was conducted, having $\sim10$ times deeper exposure than \erass{1}. Based on these data, an extent-selected cluster catalog was presented in \citet{Liu2022a}, while \citet{Bulbul2022} identified and classified \num{346} cluster candidates that are included in the point source catalog in \citet{Salvato2024}. While we use a classification scheme for the galaxy clusters identified at \pointlike{} X-ray positions in this work loosely resembling the one of \citet{Salvato2022}, there are some crucial differences between the underlying data, the photometric redshift estimation method and the algorithm employed to identify the clusters. The distinction between extended and point-like sources in the \efeds{} field is based on an extent-likelihood cut of $\extlike{}\geq6$; as suggested by \citet{Bulbul2022}, this cut is changed to $\extlike{}\geq3$ for the \erass{1} primary sample to facilitate the identification of galaxy groups and high redshift clusters with compact X-ray morphologies. Their clusters are classified in a scheme of four cluster classes (\clusterClass{2} to \clusterClass{5}). When cross-matching our point source sample to the \efeds{} catalogs, we find \num{54} matches to their extent-selected catalog, and \num{19} matches to their point source catalog, of which \num{2} also have a closer match to an extent-selected cluster.
The large median offset of \qty{\sim54}{\arcsec} between the \num{19} cross-matched misclassified clusters in \efeds{} and \erass{1} is likely due to low photon statistics of the \erass{1} sources, projected fainter AGN or cool-core cluster emission being missed, or AGN variability. Visual inspection of the \num{11} matches for which the offset is $\gtrsim\qty{50}{\arcsec}$ was not sufficient to distinguish between those cases. Since the underlying X-ray point sources are so far apart, the inferred cluster classes also do not match for \num{7} of them, although we note that this can also be caused by the difference in methodology. For the \num{8} close matches, the cluster classes do not match in only \num{2} cases, which can both be explained by the shallower photometry used (\lsdr{10} for this work. Another large difference is the availability of the deep Hyper Suprime Cam Subaru Strategic program \citep[\hsc{}][]{Aihara2018} data for \efeds{}, enabling more secure identification of faint counterparts). We further discuss the recovery of \efeds{} sources with respect to their X-ray luminosities in detail in \secref{sssec:4_1_1_efeds_recovery}.

Another deep limited area X-ray survey is the XXL survey performed with {\it XMM-Newton}, covering \qty{50}{\deg^{2}} \citep{Adami2018}. Other surveys based on {\it XMM-Newton} archival observations include the {\it XMM-Newton} Cluster Survey \citep[XCS][]{Mehrtens2012} and {\it XMM-Newton} Cluster Archive Super Survey \citep[X-CLASS][]{Koulouridis2021}. We find a total of 15 and 14 clusters in XXL and XCS surveys. Additionally, we found 61 cross-matched clusters in a wider area covered in the X-CLASS survey. The median redshifts range between \SIrange{0.4}{0.45}{} with $20^{\prime\prime}-30^{\prime\prime}$ offset, larger than the cross matches in the extent-selected eRASS1 sample following a similar trend overall with the other cross-matched samples. The offset observed between X-ray centroids is generally small between matched samples in the X-ray surveys. This is largely due to localization accuracy and the small PSF of the X-ray telescopes, which can measure the centroid of the ICM gas accurately.

\begin{table*}
\caption{Public Cluster Catalogs Cross-matched with Clusters in the \erass{1} Misclassified Cluster Catalog of this work}
\label{table:match}
\begin{center}
\begin{tabular}[width=0.5\textwidth]{llllllll}
\hline\hline 
External Catalog &  Common &  Total No of  & No of Clusters: & Median & Median & Reference\\
                 & Footprint & Clusters   & In the Footprint/ &   Redshift & Offset & \\               
                 &  (deg$^2$)  &  & Cross-matched &  & (arcsec) &\\
\hline 
Abell & 8784\tablefootmark{a}  & 1059 & 466/23 & 0.10 & 76.2 & \citet{Andernach1991} \\
ACO  & & 4076 & 1984/159 & 0.18 & 73.8 & \citet{Abell1989} \\
ACO~S  & & 1174 & 934/46 & 0.22 & 61.5 & \citet{Abell1989}  \\
GALWEIGHT  & & 1800 & 766/40 & 0.11 & 79.3 & \citet{Abdullah2020}\\
MaDCoWS  & 7436 & 2839 & 1309/18 & 0.90 & 34.7 & \citet{Gonzalez2019} \\
NEURALENS  & 9833 & 1312 & 697/46 & 0.46 & 19.5 & \citet{Huang2021} \\
GOGREEN-GCLASS & N/A\tablefootmark{b}  & 26 & 11/2 & 0.87 & 16.4 & \citet{Balogh2021} \\
WENHAN-HIGHZ-SPECZ & 11369 & 1959 & 24/16 & 0.72 & 48.0 & \citet{Wen2018} \\
SDSS DR8 & & 26111 & 1040 & 0.35 & 35.7 & \citet{Rykoff2014} \\
DES Y1 & 3644 & 6729 & 6075/664 & 0.42 & 28.9 & \citet{McClintock2019} \\
CODEX  &3508  & 10382 & 3106/209 & 0.33 & 57.6 & \citet{Finoguenov2020} \\
MCXC  & 13116& 1743 & 681/39 & 0.27 & 34.5 & \citet{Piffaretti2011} \\
eFEDS  & 140 & 542 & 531/54 & 0.37 & 28.9 & \citet{Liu2022a} \\
eFEDS (PS) & 140 & 346 & 345/19 & 0.42 & 54.2 & \citet{Bulbul2022} \\
XCLASS  & N/A\tablefootmark{b}  & 1559 & 800/61 & 0.45 & 23.7 &  \citet{Koulouridis2021}\\
XCS DR1 & N/A\tablefootmark{b}   & 503 & 184/14 & 0.39 & 28.1 & \citet{Mehrtens2012} \\
XXL 365  & 23 & 302 & 137/15 & 0.43 & 18.9 &  \citet{Adami2018}\\
PSZ2  & 10281 & 1653 & 633/10 & 0.42 & 76.8 & \citet{Planck2016} \\
ACT DR5  & 6877 & 4195 & 2565/418 & 0.55 & 31.9 & \citet{Hilton2021} \\
SPT~2500~deg$^2$  & 2288 & 677 & 649/66 & 0.60 & 27.1 & \citet{Bleem2015} \\
SPT~ECS  & 1798 & 470 & 381/36 & 0.58 & 31.0 & \citet{Bleem2020} \\
SPTPol~100~deg$^2$  & 78 & 89 & 89/13 & 0.46 & 32.0 & \citet{Huang2020} \\
\hline\hline
\end{tabular}
\tablefoot{
\tablefoottext{a}{The Galactic latitude cut of \citet{Abell1989} catalog is not strict, this value is calculated using $|b|>20$~deg as an approximation.}
\tablefoottext{b}{Targeted observations.}
}
\end{center}
\end{table*}
\subsection{Comparison with SZ catalogs}
\label{ssec:comparison_sz}

The Sunyaev-Zeldovich (SZ)-Effect surveys are yet another efficient way of detecting clean and complete samples of clusters of galaxies. The catalogs are compiled based on the signature generated by the inverse Compton scattering of the blackbody spectrum of the CMB by the energetic electrons in the hot ICM. Prominent cluster surveys are constructed with the Atacama Cosmology Telescope (ACT), {\it Planck} Satellite, and South Pole Telescope (SPT) data.

The most extensive cluster catalog to date is based on Planck's final All-Sky survey, which consists of \num{1653} sources \citep{Planck2016}. In the common footprint of \qty{10281}{deg^2}, we identify ten counterparts in our catalog with a median sample redshift of 0.42. {\it Planck} detected clusters show the largest offset of 78$^{\prime\prime}$ with the counterparts compared to other surveys cross-matched with \erass{1} catalogs; this is most likely due to the large beam size \citep[\qty{8.76}{\arcmin} at \qty{100}{GHz}][]{PlanckCollaboration2014}.

The SPT has performed several deep surveys in the Southern Hemisphere since its start of operations in 2007. The first SPT survey, SPT-SZ hereafter, covering a region of \qty{2500}{deg^2}, has \num{66} identified counterparts in our sample with a median redshift of \num{0.60}, a redshift much higher than counterparts in the extent-selected sample \citep{Bleem2015, Bocquet2019}. Deeper surveys were performed with an upgraded SPT-pol receiver (SPT-Pol \qty{100}{\deg^2}) and SPT-pol Extended Cluster Survey (SPT-ECS) covering a \qty{100}{\deg^2} and \qty{2700}{\deg^2} areas contains \num{13} and \num{36} common clusters with mean redshifts between 0.46 and 0.6 and centroid offsets between \qty{\sim32}{\arcsec} \citep{Huang2020, Bleem2020}.

The ACT~DR5 catalog is based on a survey performed between 2008 and 2018 in a \qty{13211}{\deg^2} region in the Equatorial region and Southern Hemisphere \citep{Hilton2021}. The wider survey area compared to SPT surveys combined with ACT's smaller beam size led to large cluster associations in both the extended source and the catalog presented in this work. We find \num{418} common clusters in the overlapping footprints with a sample's median redshift ($z_{\rm med}\,=\,0.55$) and centroid shift of \qty{31.9}{\arcsec}, consistent with our PSF size.

\section{Optical properties of the sample}\label{sec:main_sec_five}

The primary and cosmology cluster samples presented in \citetalias{Bulbul2024} are strictly selected based on their X-ray properties, i.e., the measured extent likelihood parameter in the detection chain. The optical identification and confirmation of these samples have minimal influence on the completeness as the richness limits, applied in the confirmation process, is as low as three \citepalias[for further details see][]{Kluge2024}. However, although the cluster sample examined in this work is selected via X-ray detection, the optical identification through the \eromapper{} tool plays a more critical role. As the extent likelihood does not play a significant role in the selection of this sample, the contamination rate is expected to be higher than the extent likelihood selected sample. While we have aimed to reduce optical contaminants as much as possible by employing the various cuts presented in \ref{sssec:data_intro_sample_cleaning_eromapper}, it is useful to study the multi-wavelength properties of this sample and compare them with the extent-selected sample. This section examines the redshift and color distributions of each \clusterClass{}as grouped in the previous section. 

\subsection{Redshifts and Richness Distribution of clusters}
\label{ssec:cluster_redshift_distribution}

\begin{figure}
    \centering
    \includegraphics[width=\linewidth]{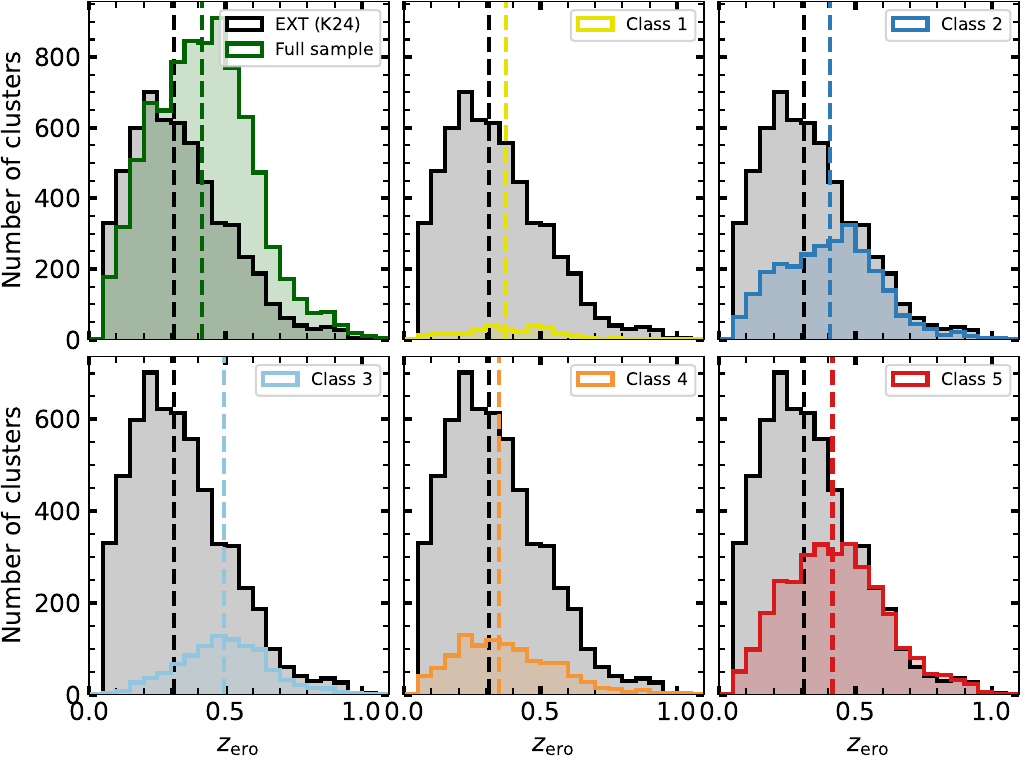}
    \caption{Distribution of the best redshifts \zEro{} assigned by \eromapper{} to the misclassified clusters, split by cluster class for the different panels, displayed in their respective colors. The upper left panel shows the distribution of the full sample combining all classes. The distribution of \NumEroExtFiltered{} extent-selected clusters \citepalias{Bulbul2024}, filtered the same way as the point source candidates, are displayed in the background in gray to guide the eye. The median of each distribution is indicated by a dashed line.}
    \label{fig:3_1_cluster_redshift_distribution}
\end{figure}

In this work, each misclassified cluster candidate is assigned at least one redshift estimate by \eromapper{}. In short, a photometric redshift estimate $z_\lambda$ reliant on the member photometry is calculated for all clusters. Clusters where a spectroscopic redshift is available for the galaxy located at the optical center of the cluster are assigned this redshift as $z_{\rm spec,cg}$. A bootstrapped bi-weight redshift estimate is calculated for clusters with at least \num{3} members having spectroscopic redshift. The final best redshift $\zEro{}$ is assigned based on a priority scheme, where $z_{\rm spec}$ is used unless the final velocity clipping during its calculation doesn't converge (which is the case for \NumEroSpecZNotConverged{} clusters). Note that we deviate from the \citetalias{Kluge2024} notation of $z_{\rm best}$ in favor of \zEro{} to better distinguish the cluster redshift from the optical counterpart redshift. Due to our selection process, we do not include any literature clusters outside the \lsdr{10} footprint cross-matched with the X-ray point source catalog. Hence, there are no clusters with only a literature redshift $z_{\rm lit}$ available. The final redshift types are summarized in \tabref{tab:redshift_type_distribution}. We present the distribution of these best redshift estimates for all \NumEroFinal{} cluster candidates, subdivided by their class in \figref{fig:3_1_cluster_redshift_distribution}. We also show the distribution of the  \NumEroExtFiltered{} extent-selected clusters that were subject to the same cuts as the PS sample \parsecrefsee{sssec:data_intro_sample_cleaning_eromapper}.

\begin{table}[ht]
    \centering
    \caption{Redshift types available for clusters in the PS cluster catalog. The first column shows the priority assigned, while the Available column denotes how many redshifts are available with the given redshift type, and the Best column reflects how many of these were used for the best redshift estimate.}
\begin{tabular}{lp{0.3\linewidth}lrr}
\toprule
Prio. & Redshift Type & Notation & Available & Best \\
\midrule
1 & Spectroscopic ($\geq 3$ members) & $z_{\rm spec}$ & 1144 & 1061 \\
2 & Spectroscopic (gal. at opt. center) & $z_{\rm spec, cg}$ & 2071 & 1137 \\
3 & Photometric & $z_\lambda$ & 8347 & 6149 \\
4 & Literature & $z_{\rm lit}$ & 2371 & 0 \\
\bottomrule
 & Best & \zEro{} &  & 8347 \\
\end{tabular}

    \label{tab:redshift_type_distribution}
\end{table}

The \NumClassOne{} \clusterClass{1} cluster redshifts follow a flat distribution with no significant features as might be expected of chance optical cluster detections close to Galactic sources.
For all distributions, notable differences to the extent-selected distribution are the generally higher median values. These are expected especially for the clusters in \clusterClass{2} and \clusterClass{3}, as the compact X-ray emission of the AGN should boost clusters at higher redshift whose extended emission would be below the detection limit. The median for \clusterClass{4} is remarkably close to the extent-selected sample, although the redshifts are more evenly distributed for the clusters in that category. The distribution of \clusterClass{5} also appears to be skewed towards higher redshifts. We anticipate detecting high-redshift cool core clusters that are missed by the extent-based selection, resulting in a sample that is skewed toward higher redshifts. This is also reflected in the higher median redshifts of the cross-matched samples with other multi-wavelength surveys presented in Sect.~\ref{sec:main_sec_four}.

A comparison of the richness distribution of the misclassified clusters in this sample to the one of the \NumEroExtFiltered{} filtered eRASS1 extent-selected sample is provided in \figref{fig:3_1_cluster_richness_distribution}. We do not distinguish between the different cluster classes as the distributions are similar. We note that the \LMax{} cut performed in \secref{sssec:data_intro_sample_cleaning_eromapper} removed mainly low richness clusters; without this cut, the distribution would peak at the minimum richness of $\lambdaNorm{}_{\rm, min}=16$. The distribution is skewed towards a lower median value ($\med{\lambdaNorm{}_{,\rm PS}}=\NumPlikeFilteredMedianRichness{}$ and $\med{\lambdaNorm{}_{,\rm EXT}}=\NumExtFilteredMedianRichness{}$) with a smaller tail. This is not surprising as clusters with higher richness values are expected to be brighter in the X-rays and thus less likely to have their extended ICM emission confused as a point source. The skewed richness distribution towards low richness indicates that this sample includes faint clusters that fall below the survey's flux limit.

\subsection{Color distribution of the \nway{} counterparts}\label{ssec:properties_color_distribution}
%

\begin{figure}
    \centering
    \includegraphics[width=\linewidth]{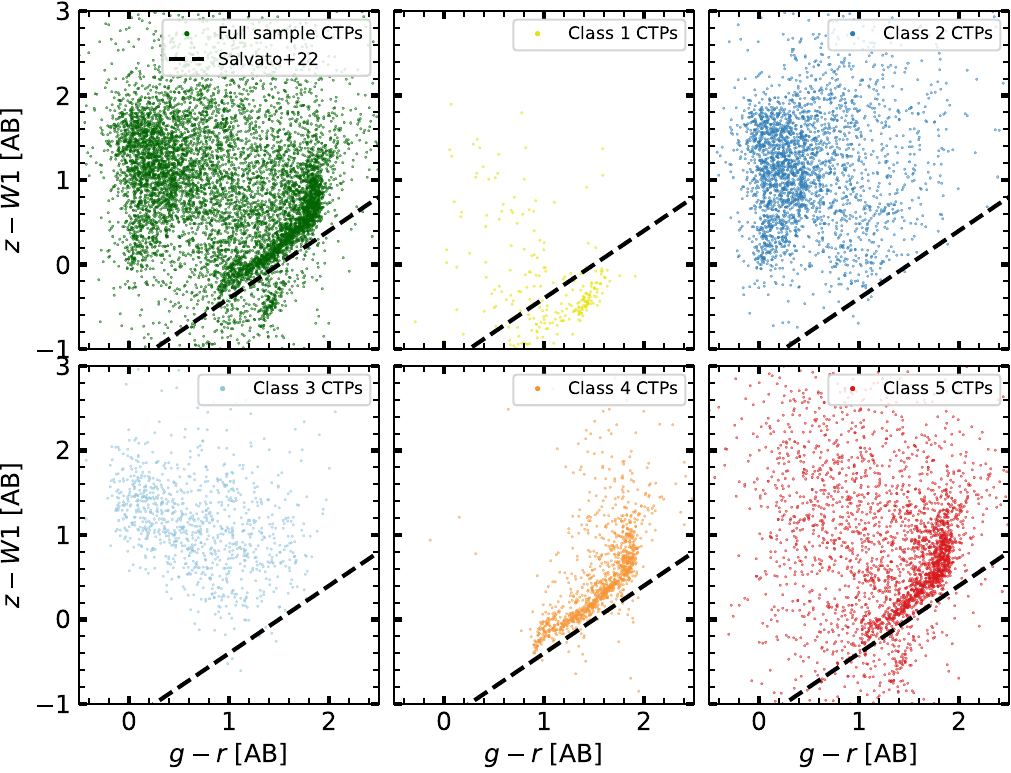}
    \caption{Distribution of the \lsdr{10} colors derived from the $grzW1$ dereddened AB magnitudes of the counterparts, first for the full sample in the upper left panel, followed by counterparts to the different cluster class objects. We show the empirical distinction line \pareqref{eq:salvato_2022_galactic_color_distinction} to distinguish between galactic and extragalactic sources.}
    \label{fig:3_1_ctp_color_distribution}
\end{figure}

\begin{figure}
    \centering
    \includegraphics[width=\linewidth]{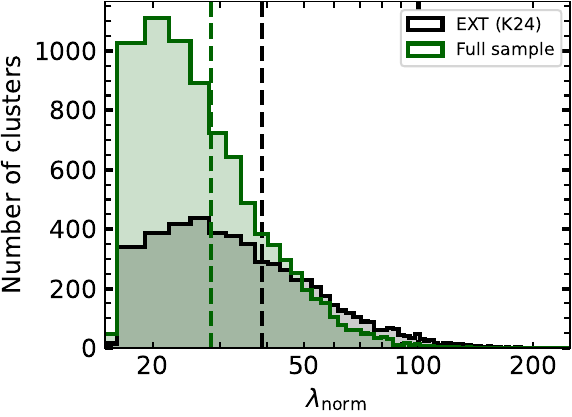}
    \caption{Distribution of the normalized richness estimate $\lambdaNorm{}$ \myeqref{eq:richness_definition} assigned by \eromapper{}. The bins are equidistant, while the $x$-axis is displayed logarithmically. The distribution of \NumEroExtFiltered{} extent-selected clusters \citepalias{Kluge2024}, filtered the same way as the misclassified clusters, is displayed in the background to guide the eye. A dashed line indicates the median of each distribution. We do not split the sample by \clusterClass{}here as the distributions are similar.}
    \label{fig:3_1_cluster_richness_distribution}
\end{figure}

Color-color diagrams are widely used in the literature to characterize the nature of the detected sources \citep[e.g.][]{Ruiz2018, Salvato2022}. We present the position of all \nway{} counterparts in the $g-r$ and $z-W1$ color-color plane of the sample of all clusters with counterparts available in \figref{fig:3_1_ctp_color_distribution} leftmost top panel. The same figure also presents the color-color diagrams of the sample split by the \clusterClass{} information. The empirical distinction line distinguishing the galactic sources from the extragalactic is used to position the counterpart in the color-color plane \citet{Salvato2022, Salvato2024}, also shown in \figref{fig:3_1_ctp_color_distribution} with a dashed line. The distinguishing line is defined as
\begin{equation}\label{eq:salvato_2022_galactic_color_distinction}
    z-W1=0.8\ (g-r)+1.2.
\end{equation}

The top left panel of Fig.~\ref{fig:3_1_ctp_color_distribution} showing the color properties of the full sample clearly shows that this diverse sample is composed of clusters with Galactic and extra Galactic counterparts assigned by \nway{}. Due to their Galactic nature, most of the counterparts in \clusterClass{1} clusters shown in the top-middle panel are below the Galactic source distinguishing line, as expected. Therefore, they are not considered for further sample investigation. There are a few \clusterClass{1} sources that are located above the distinction line. This is either due to the counterpart being associated with Gaia motions, having partially saturated photometry, or being in the range of a \hecate{} galaxy. The distribution of the counterparts of \clusterClass{2} clusters shows an apparent concentration in the upper left of the panel, displaying colors typical of AGN \citep[e.g.][]{Salvato2022}. While not as pronounced, the counterparts of \clusterClass{3} clusters are also found in this locus. This is expected as the \clusterClass{3} is designed to locate the clusters hosting bright AGN in their center. The colors of the counterparts in \clusterClass{4} are consistent with the early-type galaxies (ETG), which evolve with redshift as different spectral features enter the $grzW1$ filter bands. Notably, while \clusterClass{5} counterparts (which, by design, are only weakly associated with the X-ray point source emission) cover all areas covered by counterparts of the other classes, there is a noticeable overdensity in the locus of ETGs akin to \clusterClass{4} counterparts. This indicates that for many of these, the weakly associated counterpart is an ETG, which is reflected by \NumClassFiveStillMember{} of the \NumClassFive{} \clusterClass{5} clusters with counterparts that are associated as members galaxies by \eromapper{}. The color-color diagrams confirm that the detailed classification scheme in \secref{ssec:data_intro_classification_explanation} works well.

\section{X-ray Properties of the Sample}\label{sec:main_sec_six}

To be able to characterize the properties of the sample better and
understand the selection of this sample; we employ a reprocessing of the \erosita{} data similar to the extent-selected sample to derive the X-ray properties of the cluster candidates. For this, we perform MultiBand Projector 2D \citep[\mbproj{},][]{Sanders2018} X-ray imaging analysis using the \erass{1} data as described in \citet[\citetalias{Bulbul2024}]{Liu2023, Bulbul2022}. \mbproj{} forward fits the X-ray images of a cluster to obtain its physical models, such as the electron density profile and temperature profile. In this work, we fit the image in the soft band (\softbandRange{}) where \erosita{} has the highest sensitivity, as most of the clusters in our sample are too faint to be detected in the hard band. During the fit, the cluster center can vary the reported coordinates (R.A., Dec) in the catalog within a radius of $12\arcsec$, which accounts for the positional accuracy of the point source detection. In contrast, in the analyses of the extended sources, the centroid of the X-ray emission was allowed to vary. Due to this limitation, the offset from the initially reported X-ray position is generally smaller than for the extent-selected sample \citepalias[see Fig. 10 in][]{Bulbul2024}. 

For each cluster, we create an image and exposure map using the \erosita{} Science Analysis Software System \citep[eSASS, version c020][]{Brunner2022}. The electron density profile is described as a simplified version of the \citet{Vikhlinin2006} model:

\begin{equation}
n_{\mathrm p}n_{\mathrm e} = n_0^2 \cdot \frac{(r/r_c)^{-\alpha}}{(1+r^2/r_c^2)^{3\beta-\alpha/2}},
\end{equation}

\noindent where $n_{\rm e}$ and $n_{\rm p}$ are electron density and proton density, and we assume $n_{\rm e}=1.21n_{\rm p}$ \citep{Bulbul2010}. The central number density ($n_0$), core radius ($r_c$), and the slope of the surface brightness in the core ($\alpha$), a crucial parameter in the model, as the clusters in this sample are expected to have a peaked surface brightness profiles, and the slope of the profile at large radii ($\beta$) are free parameters in the model. We do not include the large-scale components of the model in our fit because all the clusters in our sample have very small angular sizes. The detailed 2D image fitting about the \mbproj{} can be found in Appendix A of \citetalias{Bulbul2024}. The Galactic absorption $n_{H}$ is fixed to the values obtained from the HI4PI survey \citep{HI4PI2016}. We assume the temperature is isothermal and remains constant throughout the cluster. Metal abundance is fixed at $0.3A_{sun}$ as it is hard to measure in the photon limited number survey depths \citep{Asplund2009}. C-stat is used as an estimator for the goodness-of-the-fit to avoid any biases on the best-fit parameters similar to the extent-selected sample. The physical properties of the clusters are derived at the redshifts computed by \eromapper{}.

In this section, we compare the X-ray properties, such as the distance of the X-ray centroid of the BCG, luminosity, and number density of the clusters mischaracterized in the point source sample with the extent-selected sample to understand better the reasons for the X-ray selection process and improve it for the deeper \erosita{} surveys.

\begin{figure}
    \centering
    \includegraphics[width=\linewidth]{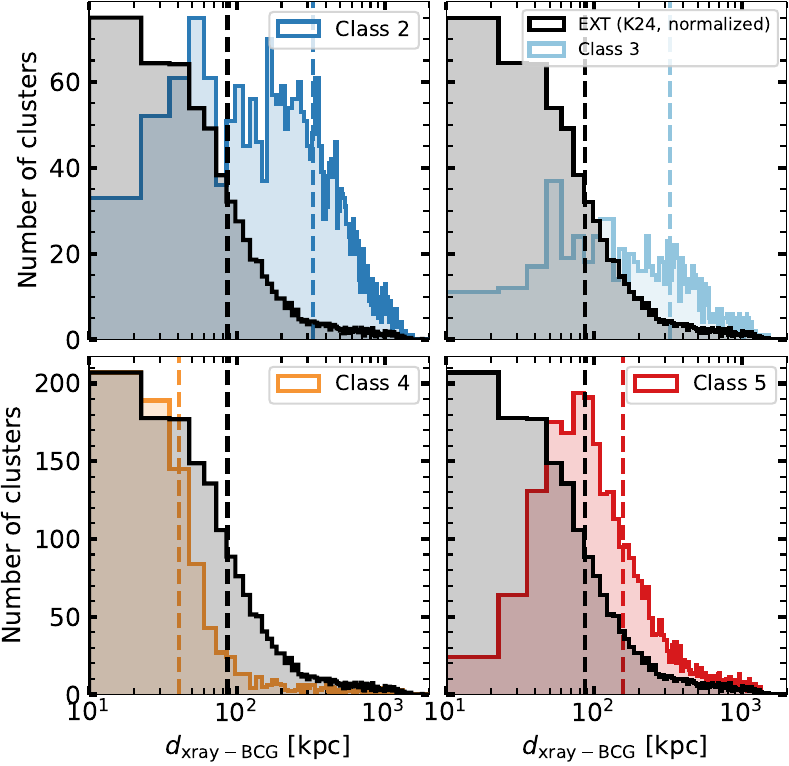}
    \caption{Distribution of the offsets of the fitted \mbproj{} X-ray centroid and the optical \eromapper{} BCG coordinates, split by cluster class. The distribution for the \NumEroExtFiltered{} filtered extent-selected clusters is shown in black, normalized for better visibility. The median offsets are indicated by the dashed lines. Compared to the median offset of \qty{\sim86}{\kpc} found for the filtered extent-selected sample, the offsets are generally higher for \clusterClass{2}, \clusterClass{3}, and \clusterClass{5} and much lower than for \clusterClass{4}. While large offsets can hint at mergers for extent-selected clusters, for the point source sample, those are more likely to indicate a chance projection of the cluster and the real point source along the line of sight.}
    \label{fig:4_1_xray_bcg_offsets}
\end{figure}

\subsection{X-ray Centroid to BCG distance}\label{ssec:4_1_xray_bcg_offsets}

The offset between the fitted X-ray center, an output by \mbproj{}, and the BCG can be used as an indicator of the dynamical state of a cluster \citep[e.g.,][]{Zenteno2020}. In the case of our sample, a large offset between the peak of the X-ray emission and the BCG can also be caused by the associated AGN being brighter in the X-rays than the extended emission of the cluster. The BCG is chosen by \eromapper{} as the brightest member galaxy in the $z$ band, which up to \qty{\sim20}{\percent} can result in an incorrect choice of the BCG, e.g., due to the true BCG being missed because of star-formation, dust absorption, masking, or unreliable photometric measurements caused by the BCG's extended outer component.
The distance between the X-ray centroid and the BCG position \bcgXrayDist{} is presented in \figref{fig:4_1_xray_bcg_offsets}. We compare those distances of classes {2} to \num{5} to the normalized distribution of the \NumEroExtFiltered{} filtered extent-selected clusters, which have a median of \qty{\sim86}{\kpc}. We omit \clusterClass{1} as for these clusters, the Galactic or nearby galaxy's X-ray emission is likely to have driven the detection, leading to large offsets.

The offset distribution of the full sample is large, with a median of $\sim$\NumFullSampleMedianXRayBcgDist{} and an extended tail. We further examine the distribution by the cluster \clusterClass{} to understand several observed trends. For \clusterClass{2} clusters, the X-ray point source emission is associated with an AGN in projection to the optical cluster. Since their position on the sky should not correlate with the cluster, some offset is expected, although especially low richness clusters with a high offset are removed by the \LMax{} cut. This explains the observed median offset of $\sim$\NumClassTwoMedianXRayBcgDist{}. Interestingly, around \qty{11}{\percent} of the \clusterClass{2} BCGs have $\bcgXrayDist{}\lesssim\qty{86}{\kpc}$. A brief visual inspection reveals that there is a chance projection of the BCG close to an AGN for around half of those. For the other half, the \nway{} counterpart seems to be either a galaxy at a different redshift than the cluster (indicating that the reported flux might stem from the cluster) or a faint object with unreliable photometry. Due to the \qty{90}{\percent} purity cut adopted to discern secure counterparts, a few contaminants are not surprising.

\begin{figure}
    \centering
    \includegraphics[width=\linewidth]{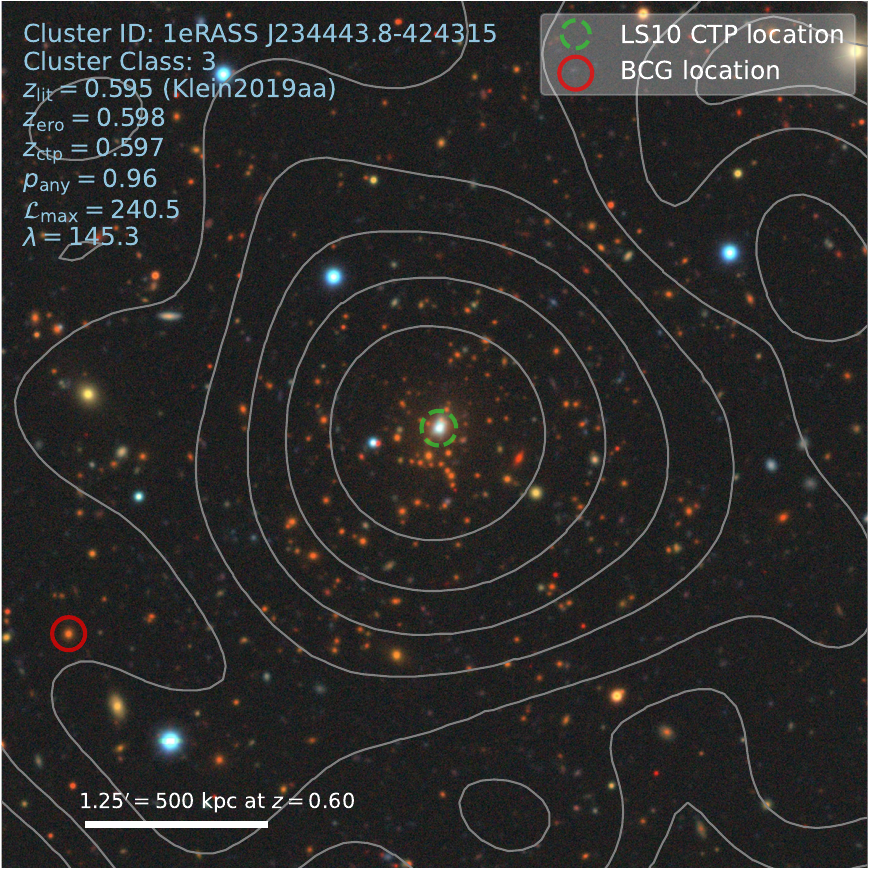}
    \caption{\lsdr{10} $grz$ image of \iauname{J234443.8-424315} overlaid with \erass{1} X-ray contours. This object, also known as the Phoenix cluster \citep{McDonald2012}, is one of the most X-ray luminous clusters and is infamous for being misclassified as a point source due to the bright AGN hosted by the BCG. The AGN is identified as the most likely \nway{} counterpart (light green circle), while the BCG reported by the \eromapper{} algorithm (red circle) is offset by $\qty{\sim870}{\kpc}$ because the true BCG's colors are too far from the red sequence. North is up; east is left, and the image is centered around the initial X-ray point source detection.}
    \label{fig:5_1_1_phoenix}
\end{figure}

The offset distribution for \clusterClass{3} clusters, shown on the top right panel of \figref{fig:4_1_xray_bcg_offsets}, appears rather flat with a high median of $\sim$\NumClassThreeMedianXRayBcgDist{}. This can be explained by multiple effects influencing \bcgXrayDist{} for these objects: the largest distances can stem from the AGN being in the outskirts of the clusters, outshining the ICM. Some might only be projected AGN since the photometric redshift estimate of the counterpart can be associated with large errors. Another reason for large offsets is misidentified BCGs, which can occur due to the true BCG not being recognized as a member. This is most prominently the case for the Phoenix cluster \parfigrefsee{fig:5_1_1_phoenix}, for which the true BCG is not identified as a member due to its blue colors, hence leading to a reported BCG at \qty{\sim870}{\kpc} away from the center.
There are also some \clusterClass{3} clusters with low offsets. As for \clusterClass{2}, the reason for those can either be that the brightest member galaxy happens to be close to the galaxy hosting the AGN (which might be the true BCG missed due to the color offset by the AGN) or in a few cases, nearby counterparts with unreliable photometry.

The low observed concentration of \bcgXrayDist{} (with the median of $\sim$\NumClassFourMedianXRayBcgDist{} being even below the one of the filtered extent-selected sample) is expected for clusters in \clusterClass{4}. In this subcategory of clusters, no AGN is found close to the reported X-ray point source emission, and the \nway{} counterpart is already a cluster member, implying that it has to be close positionally. Indeed, \NumClassFourCTPisBCG{} of the \NumClassFour{} counterparts coincide with the BCG. For \clusterClass{5} clusters, the distribution of \bcgXrayDist{} peaks at a distance of \qty{\sim100}{\kpc} as those objects with a BCG very close to the point source emission are expected to have the BCG recognized as a likely counterpart by \nway{}, in which case they are included in \clusterClass{4}. Indeed, we find the combined distribution of \clusterClass{4} and \clusterClass{5} clusters to be closer to that of the filtered extent-selected sample, albeit with a more pronounced tail towards higher offsets. Unlike for \clusterClass{2} or \clusterClass{3}, very large offsets for \clusterClass{5} are likely to indicate a disturbed dynamical state.

\subsection{X-ray luminosity measurements}
\label{sssec:4_1_1_efeds_recovery}

Akin to the analysis for the extent-selected clusters, we measure the X-ray luminosity of the misclassified clusters within \qty{300}{\kpc}. A first approximation of $R_{500}$ to estimate the \mbproj{} image size is obtained by using the best-fit luminosity-mass scaling relations of the extent-selected \efeds{} sample \citep{Chiu2022}\footnote{This is a sufficient estimation for an area of $8R_{500}\times8R_{500}$ region then used for the X-ray analysis to include both cluster and background.}. For a fair comparison, the same process is employed as in the analysis of the extent of the selected sample. In a later step, an improved estimate of the cluster's luminosity $L_{500}$, and $R_{500}$ is determined by inferring the intrinsic count rate of the clusters via background modeling plus Galactic column density ($N_{\rm H}$) and $K$-factor corrections. As cautioned by \citetalias{Bulbul2024}, the X-ray photon counts are limited due to the shallow nature of the \erass{1} survey. In the case of the sample presented here, they are typically even lower, therefore, we cannot explore the trends between various subcategories. In the catalog, we provide a measurement of $L_{300}$ for clusters in \clusterClass{3, 4, 5} with significant detection $2\sigma$ confidence level within 300~kpc. Such measurements are available for \num{401} \clusterClass{3}, \num{384} \clusterClass{4}, and \num{800}\clusterClass{5} clusters. They are biased towards the more luminous sources, as the shallow nature of \erass{1} complicates reliable flux measurements, especially for the fainter sources.


%
\begin{figure}
    \centering
    \includegraphics[width=\linewidth]{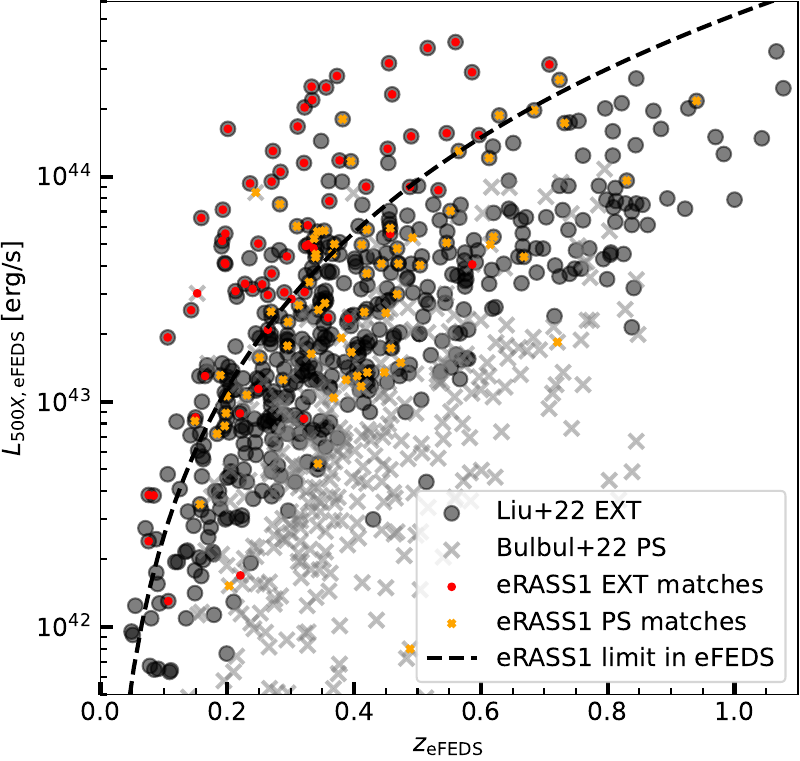}
    \caption{X-ray luminosity-redshift distribution for the \num{888} clusters reported in the \efeds{} field. The \num{346} extent-selected \efeds{} clusters \cite{Liu2022a, Bahar2022} are denoted by gray circles and the \num{542} clusters at positions of \efeds{} point sources by gray crosses. Sources with matches to the \erass{1} extent-selected catalog \citepalias{Bulbul2024} are marked by small red circles, and small yellow crosses mark sources with matches to the catalog presented in this paper. The average \erass{1} flux limit of \qty{\sim e-13}{\erg\per\s\per\cm\squared} in the \efeds{} field is indicated by the black dashed line. Most extent-selected \efeds{} clusters above the flux limit are recovered, albeit \num{54} of the \num{117} total matches to extent-selected \efeds{} clusters are found as a point source in the \erass{1} catalogs.}
    \label{fig:4_1_1_efeds_luminosity_comparison}
\end{figure}

To compare the X-ray luminosity distribution with respect to redshift with the previously published \erosita{} catalogs for all \num{888} cross-matched \efeds{} sources in \figref{fig:4_1_1_efeds_luminosity_comparison} are shown in filled circles and crosses. The redshifts and luminosities are adopted from \citet{Liu2022a} and \citet{Bahar2022} for the \num{542} extent-selected clusters and from \citet{Bulbul2022} for the \num{346} clusters at point source positions. Sources that match the extent-selected catalog are marked by small red circles, and small yellow crosses mark sources that match the catalog presented in this paper. Since \erass{1} is approximately ten times shallower than the \efeds{} area (average flux limit of \qty{\sim e-13}{\erg\per\s\per\cm\squared}), many extent-selected sources from \efeds{} are not detected. This is expected and accounted for by the selection function \citep[see][for comparisons of the selection function]{Clerc2024}; therefore, it does not influence the science analysis with the extent of the selected sample. Nonetheless, most clusters above the flux limit of the eRASS1 cluster survey are detected, shown in the black dashed line. Of the 888 clusters, the \num{113} \efeds{} clusters above the flux limit of the \erass{} cluster survey, \num{70} are detected in the \erass{1} extent-selected sample and this sample. Of the remaining \num{43} clusters, all but \num{7} can be found in the full \eromapper{} catalog on all \erass{1} point sources; they are removed due to the filtering adopted in this work. 

%
\subsection{Electron number density distribution}

\begin{figure}
    \centering
    \includegraphics[width=\linewidth]{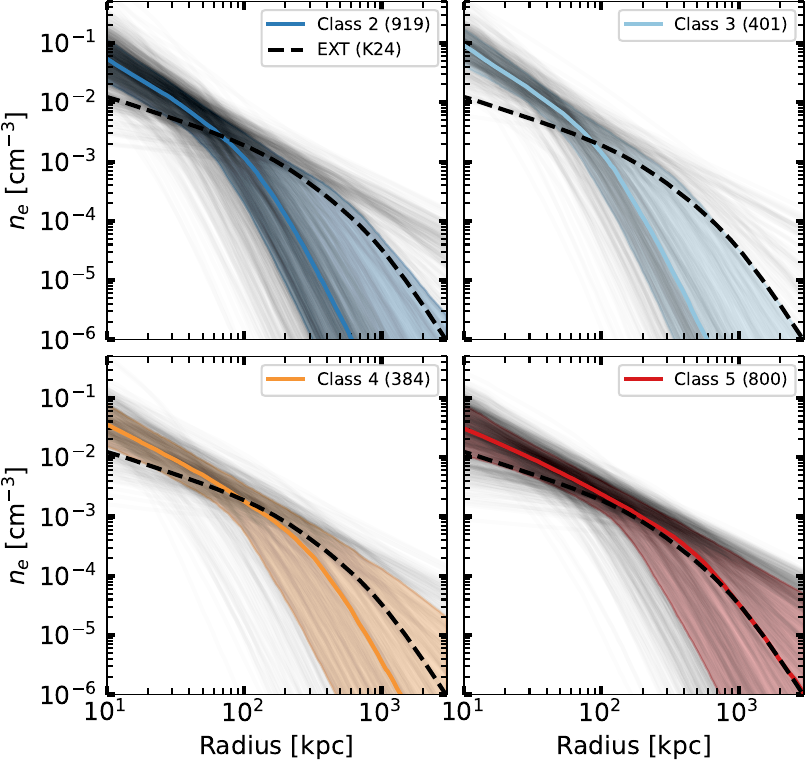}
    \caption{Electron density profiles for clusters with confident ($>2\sigma$) X-ray luminosity measurements, split by class. The median of these profiles and the $16^{\rm th}$ and $84^{\rm th}$ percentiles are denoted by the solid line and the shaded regions. We also include the median profile of the \citetalias{Bulbul2024} extent-selected sample to highlight the differences in both populations. The profiles of \clusterClass{2 and 3} clusters are cuspier, and \clusterClass{4 and 5} appear flatter and more extended; we note that we do detect extended emission, especially for the latter two classes.}
    \label{fig:3_2_xray_icm_profile}
\end{figure}

The electron number density profiles of the clusters in this sample with significant ($2\sigma$) luminosity detection are shown in \figref{fig:3_2_xray_icm_profile}, with respect to cluster class. We show the median profile and $16^{\rm th}$ and $84^{\rm th}$ percentiles for each subsample and compare it to that of the extent-selected sample of \citetalias{Bulbul2024} shown in dashed lines. Qualitatively, the median profiles of clusters in \clusterClass{2 and 3} appear to be very similar, being cuspy with high central densities and a comparatively low extent compared to the extent-selected sample. At the same time, the X-ray emission of the \clusterClass{2 and 3} candidates is most likely contaminated by the flux of the point source if it is near the cluster core defined by the X-ray detection, as the core region is not excised in this work. The clusters in these classes are likely underluminous, and their flux is below the flux limit to be detected as extended emission in the shallow first All-Sky Survey; however, they are detected in the point source catalog due to the AGN emission at the same redshift or in projection.
A small fraction of the \clusterClass{2 and 3} profiles are more extended than those of the extent-selected sample. We expect these are not recognized as extended due to being too faint compared to the nearby point source.

The median profiles of clusters in \clusterClass{4 and 5} resemble the median profile of the extent-selected sample; while being flatter and more extended than the ones for \clusterClass{2 and 3}, they still show an excess in the core in comparison to median extent-selected one, likely due to cool cores.
The similarity indicates that this subset includes a population of clusters similar to the extent-selected ones. Especially the clusters in \clusterClass{4} are likely to be clusters given their low X-ray to BCG offsets discussed in \secref{ssec:4_1_xray_bcg_offsets}; the detection algorithm {\tt ermldet} detects the core these clusters as a point source as the emission from the outskirts is faint. These clusters are expected to be detected as extended in the catalogs produced from deeper \erosita{} surveys. Clusters in \clusterClass{5} are also secure galaxy clusters identified by the optical identification of \eromapper{}. However, their electron density profiles tend to be more extended than those in \clusterClass{4}. These clusters include very extended low redshift and rich local clusters. They are not detected because their surface brightness profiles may differ from a simple $\beta$-profile, which is used in the detection chain. Including these clusters in the extent-selected samples requires another method of detection, for instance, wavelet-based algorithms \citep[\textsf{wavdetect;}][]{Kafer2020}. 

%


%
\section{Conclusion}\label{sec:main_sec_summary}

The main \erass{1} cluster catalog was compiled based on the extended morphology of the X-ray emission \citep{Bulbul2024, Kluge2024}. Galaxy cluster samples constructed using extended X-ray emission from the ICM are known to be incomplete for low flux limits and compact clusters. In this work, we examine the \erass{1} point source catalog \citep{Merloni2024} to identify clusters that are missed by the extent likelihood selection that was adopted to construct the main \erass{1} cluster catalogs. This work opens up a discovery space for finding clusters with unusually strong cool cores that the currently used detection algorithms have missed.

In order to identify the misclassified clusters as point sources, we use an analogous way to treat the sample of extended X-ray sources in \erass{1}. We apply the red sequence-based cluster-finding algorithm \eromapper{} in scanning mode at the positions of all \erosita{} detected point sources. Overdensities of passive galaxies on the red sequence are identified within a search radius $R_\lambda$, and redshifts are calculated based on their colors or archival spectroscopic data. To clean the sample further, we select secure X-ray detections with ${\rm DET\_LIKE}\geq8$ in clean sky regions (\param{IN_XGOOD}), sources with red-sequence galaxies identified with a redshift $z>0.05$ but within the local depth limit of the Legacy Surveys (\param{IN_ZVLIM}), sources with a masked area of less than \qty{30}{\percent}, and require a richness $\lambda_{\rm norm}\geq16$ and optical cluster likelihood $\LMax{}\geq20$. The catalog is then cleaned from duplicate entries by constructing associations of clusters that share more than \qty{70}{\percent} of their members with others, keeping only those with the highest \LMax{} value. Each of the point sources in the \lsdr{10} region also has an optical counterpart assigned to the X-ray emission using the \nway{} algorithm. The properties of those counterparts are quantified via the association probability, their spectroscopic (if available) and photometric redshifts, and an assessment of whether it is likely to be Galactic. The final sample comprises \NumEroFinal{} clusters in the point source catalog. In the literature, cross-matches with public optical, X-ray, and SZ surveys reveal that 5819 of the \NumEroFinal{} clusters are novel detections by \erosita{}, corresponding to \qty{70}{\percent} of the final sample. This discovery rate is similar to that of the \erass{1} extent-selected sample reported in \citetalias{Bulbul2024, Kluge2024}.

To understand the selection of \erass{1} extent-selected catalog and to identify unusual clusters, we use the optical properties provided by \nway{} and \eromapper{} to divide our sample of cluster candidates into six subclasses. The \NumClassZero{} \clusterClass{0} detections correspond to no counterpart information and, therefore, are only kept for completeness but not further analyzed. In the subsample of \NumClassOne{} cluster candidates in  \clusterClass{1}, the X-ray emission is likely linked to to and dominated by nearby identified Galactic sources or \hecate{} galaxies. As a result, we also do not conduct further analysis on these objects and only include them in the catalog for completeness. The \clusterClass{2} subsample includes \NumClassTwo{} clusters where the counterpart is securely identified by \nway{} and has colors consistent with an AGN. The \eromapper{} identified clusters for this class are located at a different redshift. These clusters' faint X-ray fluxes, below the flux limit of the \erass{1} survey, are likely to be boosted by an AGN in projection. One of the most interesting classes is \clusterClass{3}, which contains \NumClassThree{} clusters that are likely to host AGN. For this class, the secure counterpart identified by \nway{} has colors consistent with AGN, and \eromapper{} provides the same redshift for the cluster based on the red sequence. Since the faint extended X-ray emission from the cluster itself would be dominated by the bright AGN, these clusters are misclassified by the detection algorithm as point sources. The clusters in this subsample are intriguing for AGN feedback studies and are ideal targets for follow-up observations. A prime example of a typical \clusterClass{3} cluster is the Phoenix cluster \citep{McDonald2012}. Although the Phoenix cluster is correctly identified and classified in our sample, regrettably, we do not find any similar extreme case in the Western Galactic Hemisphere. Regardless, we successfully identify several strong cool-core clusters hosting a BCG with significant star formation or AGN activity. The in-depth studies of the X-ray and optical properties of this subsample will be presented in subsequent work (Kluge et al., in prep.). \clusterClass{4} contains \NumClassFour{} clusters where the counterpart has colors typical of early-type red galaxies and is identified as a member galaxy by \eromapper{}. The X-ray morphology of ICM emission from clusters in this class is likely to be more compact than for the extent-selected ones, either because they are located at higher redshifts or have a bright cool core. \clusterClass{5} includes \NumClassFive{} clusters where no secure \nway{} counterpart can be identified, which can either be caused by the optical data being too shallow, or the absence of such a single source counterpart, implying that the cluster identified by \eromapper{} is the only correct association. Therefore, this class should contain a large number of true clusters.

To understand the selection further, we investigate the electron number density profiles and X-ray centroid offsets obtained from fitting \erosita{} X-ray imaging data. Excluding the clusters in the \clusterClass{5} subsample, the electron density profiles of most clusters show a peaked X-ray morphology in the sample, confirming compact that X-ray emission, either due to a central point source or a compact core, is responsible for the misclassification by the source detection algorithm. The clusters of \clusterClass{2} and \clusterClass{3} are associated with AGN along the line of sight, with a large (possibly random) X-ray to BCG offset and PSF-like cuspy electron density profiles. On the other hand, \clusterClass{4} clusters have similar BCG-to-X-ray offsets to the extent-selected clusters, indicating a similar population. 

This work sheds light on the X-ray selection of shallow surveys such as the first \erosita{} All-Sky Survey. We show that the incompleteness in the X-ray detection process of galaxy clusters is induced mostly due to the dense point source population in the surveys. A large fraction of high redshift clusters presented in this sample will be detected as extended sources in deeper \efeds{} or \erass{:5} data. As those surveys become deeper, the faint ICM emission outshone by nearby point sources can be more consistently disentangled. Expanding the redshift range to $z>0.8-1.1$ in forthcoming cosmological analyses using clusters identified in this work will open the discovery space to the transition from the matter-dominated epoch to the dark-energy-dominated cosmic epoch.

\section*{Data Availability}
We provide the X-ray and optical properties for the \NumEroFinal{} misclassified clusters presented in this paper, along with the \nway{} counterpart information and our classification flags. Descriptions of the column names are provided in \tabref{tab:column_names}. Reasonable requests for the member galaxy catalogs can be made to the corresponding author.\\

{\tiny
\noindent\textit{Acknowledgements.}
This work is based on data from eROSITA, the soft X-ray instrument aboard SRG, a joint Russian-German science mission supported by the Russian Space Agency (Roskosmos), in the interests of the Russian Academy of Sciences represented by its Space Research Institute (IKI), and the Deutsches Zentrum für Luft- und Raumfahrt (DLR). The SRG spacecraft was built by Lavochkin Association (NPOL) and its subcontractors, and is operated by NPOL with support from the Max Planck Institute for Extraterrestrial Physics (MPE).\\
The development and construction of the eROSITA X-ray instrument was led by MPE, with contributions from the Dr. Karl Remeis Observatory Bamberg \& ECAP (FAU Erlangen-Nuernberg), the University of Hamburg Observatory, the Leibniz Institute for Astrophysics Potsdam (AIP), and the Institute for Astronomy and Astrophysics of the University of Tübingen, with the support of DLR and the Max Planck Society. The Argelander Institute for Astronomy of the University of Bonn and the Ludwig Maximilians Universität Munich also participated in the science preparation for eROSITA.\\
The eROSITA data shown here were processed using the eSASS/NRTA software system developed by the German eROSITA consortium.\\
E. Bulbul, A. Liu, S. Zelmer, and X. Zhang acknowledge
financial support from the European Research Council (ERC) Consolidator
Grant under the European Union’s Horizon 2020 research and innovation
program (grant agreement CoG DarkQuest No 101002585). N. Clerc was financially
supported by CNES.\\
The Legacy Surveys consist of three individual and complementary projects: the Dark Energy Camera Legacy Survey (DECaLS; Proposal ID \#2014B-0404; PIs: David Schlegel and Arjun Dey), the Beijing-Arizona Sky Survey (BASS; NOAO Prop. ID \#2015A-0801; PIs: Zhou Xu and Xiaohui Fan), and the Mayall z-band Legacy Survey (MzLS; Prop. ID \#2016A-0453; PI: Arjun Dey). DECaLS, BASS and MzLS together include data obtained, respectively, at the Blanco telescope, Cerro Tololo Inter-American Observatory, NSF's NOIRLab; the Bok telescope, Steward Observatory, University of Arizona; and the Mayall telescope, Kitt Peak National Observatory, NOIRLab. Pipeline processing and analyses of the data were supported by NOIRLab and the Lawrence Berkeley National Laboratory (LBNL). The Legacy Surveys project is honored to be permitted to conduct astronomical research on Iolkam Du'ag (Kitt Peak), a mountain with particular significance to the Tohono O'odham Nation.
NOIRLab is operated by the Association of Universities for Research in Astronomy (AURA) under a cooperative agreement with the National Science Foundation. LBNL is managed by the Regents of the University of California under contract to the U.S. Department of Energy.
This project used data obtained with the Dark Energy Camera (DECam), which was constructed by the Dark Energy Survey (DES) collaboration. Funding for the DES Projects has been provided by the U.S. Department of Energy, the U.S. National Science Foundation, the Ministry of Science and Education of Spain, the Science and Technology Facilities Council of the United Kingdom, the Higher Education Funding Council for England, the National Center for Supercomputing Applications at the University of Illinois at Urbana-Champaign, the Kavli Institute of Cosmological Physics at the University of Chicago, Center for Cosmology and Astro-Particle Physics at the Ohio State University, the Mitchell Institute for Fundamental Physics and Astronomy at Texas A\&M University, Financiadora de Estudos e Projetos, Fundacao Carlos Chagas Filho de Amparo, Financiadora de Estudos e Projetos, Fundacao Carlos Chagas Filho de Amparo a Pesquisa do Estado do Rio de Janeiro, Conselho Nacional de Desenvolvimento Cientifico e Tecnologico and the Ministerio da Ciencia, Tecnologia e Inovacao, the Deutsche Forschungsgemeinschaft and the Collaborating Institutions in the Dark Energy Survey. The Collaborating Institutions are Argonne National Laboratory, the University of California at Santa Cruz, the University of Cambridge, Centro de Investigaciones Energeticas, Medioambientales y Tecnologicas-Madrid, the University of Chicago, University College London, the DES-Brazil Consortium, the University of Edinburgh, the Eidgenossische Technische Hochschule (ETH) Zurich, Fermi National Accelerator Laboratory, the University of Illinois at Urbana-Champaign, the Institut de Ciencies de l'Espai (IEEC/CSIC), the Institut de Fisica d'Altes Energies, Lawrence Berkeley National Laboratory, the Ludwig Maximilians Universitat Munchen and the associated Excellence Cluster Universe, the University of Michigan, NSF's NOIRLab, the University of Nottingham, the Ohio State University, the University of Pennsylvania, the University of Portsmouth, SLAC National Accelerator Laboratory, Stanford University, the University of Sussex, and Texas A\&M University.
BASS is a key project of the Telescope Access Program (TAP), which has been funded by the National Astronomical Observatories of China, the Chinese Academy of Sciences (the Strategic Priority Research Program “The Emergence of Cosmological Structures” Grant \# XDB09000000), and the Special Fund for Astronomy from the Ministry of Finance. The BASS is also supported by the External Cooperation Program of Chinese Academy of Sciences (Grant \# 114A11KYSB20160057), and Chinese National Natural Science Foundation (Grant \# 12120101003, \# 11433005).
The Legacy Survey team makes use of data products from the Near-Earth Object Wide-field Infrared Survey Explorer (NEOWISE), which is a project of the Jet Propulsion Laboratory/California Institute of Technology. NEOWISE is funded by the National Aeronautics and Space Administration.
The Legacy Surveys imaging of the DESI footprint is supported by the Director, Office of Science, Office of High Energy Physics of the U.S. Department of Energy under Contract No. DE-AC02-05CH1123, by the National Energy Research Scientific Computing Center, a DOE Office of Science User Facility under the same contract; and by the U.S. National Science Foundation, Division of Astronomical Sciences under Contract No. AST-0950945 to NOAO.\\
This work made use of the following Python libraries for data processing: \texttt{SciPy} \citep{2020SciPy-NMeth}, \texttt{matplotlib} \citep{Hunter2007}, \texttt{NumPy} \citep{harris2020array}, adn \texttt{Astropy} \citep{astropy:2022}.\\
This research made use of the \texttt{VizieR} catalog access tool, CDS,
Strasbourg, France \citep{10.26093/cds/vizier}. The original description 
of the VizieR service was published in \citet{vizier2000}.}

\begin{table*}[ht]
\caption{Column names, symbols adopted to represent them, units, data types and brief descriptions of the columns provided for the catalog of the \NumEroFinal{} misidentified clusters. We abbreviate the counterpart assigned by \nway{} as CTP for brevity. We provide variants of column names via square brackets with the variants separated by semicolons; for example, \param{BEST_Z[;_ERR]} indicates the presence of a \param{BEST_Z} and a \param{BEST_Z_ERR} column. The suffixes \param{_ERR}, \param{_L}, and \param{_U} denote the symmetrical, lower, and upper $1\sigma$ errors of the quantity preceding them, respectively.}
\label{tab:column_names}\footnotesize
\begin{tabular}{lllll}
\toprule
Column Name & Symbol & Unit & Type & Description \\
\midrule
\param{DETUID} &  &  & 32A & Unique eSASS detection ID \\
\param{IAUNAME} &  &  & 23A & Point source detection name in the IAU format \\
\param{RA} &  & $\deg$ & E4 & Right Ascension J2000 of the point source detection \\
\param{DEC} &  & $\deg$ & E4 & Declination J2000 of the point source detection \\
\param{RA_OPT} &  & $\deg$ & E4 & Right Ascension J2000 of cluster optical center \\
\param{DEC_OPT} &  & $\deg$ & E4 & Declination J2000 of cluster optical center \\
\param{RA_BCG} &  & $\deg$ & E4 & Right Ascension J2000 of the brightest member \\
\param{DEC_BCG} &  & $\deg$ & E4 & Declination J2000 of the brightest member \\
\param{RA_XFIT} &  & $\deg$ & E8 & Central Right Ascension J2000 of the \mbproj{} X-ray fit \\
\param{DEC_XFIT} &  & $\deg$ & E8 & Central Declination J2000 of the \mbproj{} X-ray fit \\
\param{DET_LIKE_0} & \detlike{} &  & E4 & X-ray detection likelihood \\
\param{REFMAG_BCG} &  & AB mag & E4 & Galactic extinction-corrected $z$-band mag. of the brightest member \\
\param{BEST_Z[;_ERR]} & $\zEro$ &  & E4 & Best available cluster redshift \\
\param{BEST_Z_TYPE} &  &  & 16A & Type of best available cluster redshift \\
\param{Z_LAMBDA[;_ERR]} & $z_\lambda$ &  & E4 & Uncorrected cluster photometric redshift \\
\param{LIT_Z[;_ERR]} & $z_{\rm lit}$ &  & E4 & Literature cluster redshift \\
\param{LIT_Z_SRC} &  &  & 128A & Reference for the literature redshift \\
\param{SPEC_Z_BOOT[;_ERR]} & $z_{\rm spec}$ &  & E4 & Bootstrap estimate of biweight spectroscopic redshift \\
\param{N_MEMBERS} & $N_{\rm spec}$ &  & i4 & Number of members used in biweight $z_{\rm spec}$ and \veldisp{} estimate \\
\param{CG_SPEC_Z[;_ERR]} & $z_{\rm spec,cg}$ &  & E4 & Spectroscopic redshift of the central galaxy \\
\param{BCG_SPEC_Z} &  &  & E4 & Spectroscopic redshift of the brightest member \\
\param{VDISP_BOOT[;_ERR]} & \veldisp & km/s & E4 & Bootstrap estimate of velocity dispersion \\
\param{VDISP_BOOT_FLAG} &  & & E4 & Mean \veldisp{} flag for bootstrap estimate \\
\param{VDISP_TYPE} &  &  & 8A & Estimate used to determine \veldisp{}. Gapper or Biweight depending on $N_{\rm spec}$ \\
\param{LAMBDA_NORM[;_ERR]} & $\lambda_{\rm norm}$ &  & E4 & Richness normalized to the the definition of the $grz$ run \\
\param{LAMBDA_OPT_NORM[;_ERR]} &  &  & E4 & Richness (optical center) normalized to the definition of the $grz$ run \\
\param{SCALEVAL} & $S$ &  & E4 & Richness scale factor, Eq. 2 of \citet{Rykoff2016} \\
\param{MASKFRAC} & $f_{\rm mask}$ &  & E4 & Fraction of cluster area which is masked \\
\param{RUN} &  &  & 128A & \redmapper{} calibration: \param{survey_bands_refband_iteration} \\
\param{LMAX} & \LMax{} &  & E4 & Optical maximum likelihood \\
\param{NHI} & $N_{\rm HI}$ & $10^{21}$ cm$^{-2}$ & E4 & HI Column density from HI4PI. Nside = 1024 \\
\param{LIMMAG_[G;R;I;Z;W1]} &  & AB mag & E4 & Limiting $grizW1$ galaxy magnitude from \citet{Rykoff2015} depth model \\
\param{MATCH_NAME} & & & 128A & The names of cross-matched clusters \parsecrefsee{sec:main_sec_four}\\
\param{X_F_300[;_L;_U]} & $F_X$ & \unit{\erg\per\square\cm\per\second}  & E8 & X-ray flux in \softbandRange{} within \qty{300}{\kpc} \\
\param{X_CR_300[;_L;_U]} & ${\rm CR}_X$ & \unit{Counts\per\s} & E8 & X-ray count rate in \softbandRange{} within \qty{300}{\kpc} \\
\param{X_CTS_300[;_L;_U]} & ${\rm CTS}_X$ & \unit{Counts} & E8 & Total X-ray Counts in \softbandRange{} within \qty{300}{\kpc} \\
\param{X_L_300[;_L;_U]} & $L_X$ & \unit{\erg\per\second}  & E8 & X-ray luminosity in \softbandRange{} within \qty{300}{\kpc} \\
\param{SPLIT_NEIGHBOR_NUM} &  &  & i4 & Amount of neighboring X-ray point sources pointing to same cluster \\
\param{HAS_CLOSE_EXT_NEIGHBOR} &  &  & L & Whether the cluster is in the vicinity of an extent-selected cluster \\
\param{RA_CTP_LS10} &  & $\deg$ & E8 & Right Ascension J2000 of the counterpart in \lsdr{10} \\
\param{DEC_CTP_LS10} &  & $\deg$ & E8 &  Declination J2000 of the counterpart in \lsdr{10}  \\
\param{CTP_LS10_FULLID} &  &  & 18A & LS10 ID, consisting of \param{releaseID_brickID_objID} \\
\param{CTP_NWAY_P_ANY} & \pAny{} &  & E4 & The \nway{} probability of the counterpart being correct \\
\param{CTP_CLASS_GAL_EXGAL} &  &  & i4 & The Galactic/extragalactic CTP class $\in\{-5,-1,1,2,3,4\}$ \citep{Salvato2024} \\
\param{CTP_NWAY_COMPUR8} &  &  & E8 & \pAny{} completeness/purity crossover at $\detlike{}=8$ \citep[see][]{Salvato2024} \\
\param{CTP_LS10_TYPE} &  &  & 3A & CTP source type in LS10 (PSF, EXP, DEV, REX, SER) \\
\param{CTP_BEST_Z[;_ERR]} & \zCtp{} &  & E8 & Best available counterpart redshift \\
\param{CTP_BEST_Z_TYPE} &  &  & 9A & Type of best available counterpart redshift (SPEC or PHOT) \\
\param{CTP_FLAG_BEST_Z_UNRELIABLE} &  &  & L & The best redshift estimate has a high relative error \\
\param{CTP_NWAY_HAS_OTHER_MATCH} &  &  & L & There exists a secondary (less likely) CTP that was discarded \\
\param{CTP_IS_GALACTIC} &  &  & L & CTP is considered Galactic (derived from \param{CTP_CLASS_GAL_EXGAL}) \\
\param{CTP_IN_HECATE_RANGE} &  &  & L & CTP is in range of a \hecate{} galaxy \\
\param{CTP_IS_SECURE} &  &  & L & CTP is considered secure (derived from \param{CTP_NWAY_COMPUR8}) \\
\param{CTP_Z_MATCHES_CLUSTER_Z} &  &  & L & CTP and cluster redshift match \pareqref{eq:same_redshift_criterion} \\
\param{CTP_IS_CLUSTER_MEMBER} &  &  & L & CTP is member galaxy of the cluster \\
\param{CLUSTER_CLASS} &  &  & i4 & The \clusterClass{}$\in\{0,1,2,3,4,5\}$ assigned to this source \\
\bottomrule
\end{tabular}

\end{table*}

\bibliographystyle{aa}
\bibliography{technicalities/library}

\begin{thebibliography}{104}
\expandafter\ifx\csname natexlab\endcsname\relax\def\natexlab#1{#1}\fi

\bibitem[{Abbott {et~al.}(2020)Abbott, Aguena, Alarcon, Allam, Allen, Annis, Avila, Bacon, Bechtol, Bermeo, Bernstein, Bertin, Bhargava, Bocquet, Brooks, Brout, Buckley-Geer, Burke, Carnero~Rosell, Carrasco~Kind, Carretero, Castander, Cawthon, Chang, Chen, Choi, Costanzi, Crocce, da~Costa, Davis, De~Vicente, DeRose, Desai, Diehl, Dietrich, Dodelson, Doel, Drlica-Wagner, Eckert, Eifler, Elvin-Poole, Estrada, Everett, Evrard, Farahi, Ferrero, Flaugher, Fosalba, Frieman, Garc{\'\i}a-Bellido, Gatti, Gaztanaga, Gerdes, Giannantonio, Giles, Grandis, Gruen, Gruendl, Gschwend, Gutierrez, Hartley, Hinton, Hollowood, Honscheid, Hoyle, Huterer, James, Jarvis, Jeltema, Johnson, Johnson, Kent, Krause, Kron, Kuehn, Kuropatkin, Lahav, Li, Lidman, Lima, Lin, MacCrann, Maia, Mantz, Marshall, Martini, Mayers, Melchior, Mena-Fern{\'a}ndez, Menanteau, Miquel, Mohr, Nichol, Nord, Ogando, Palmese, Paz-Chinch{\'o}n, Plazas, Prat, Rau, Romer, Roodman, Rooney, Rozo, Rykoff, Sako, Samuroff, S{\'a}nchez, Sanchez, Saro, Scarpine,
  Schubnell, Scolnic, Serrano, Sevilla-Noarbe, Sheldon, Smith, Smith, Suchyta, Swanson, Tarle, Thomas, To, Troxel, Tucker, Varga, von~der Linden, Walker, Wechsler, Weller, Wilkinson, Wu, Yanny, Zhang, Zhang, Zuntz, \& Collaboration}]{Abbott2020}
Abbott, T.~M.~C., Aguena, M., Alarcon, A., {et~al.} 2020, \prd, 102, 023509

\bibitem[{{Abdullah} {et~al.}(2020){Abdullah}, {Klypin}, \& {Wilson}}]{Abdullah2020}
{Abdullah}, M.~H., {Klypin}, A., \& {Wilson}, G. 2020, \apj, 901, 90

\bibitem[{{Abell} {et~al.}(1989){Abell}, {Corwin}, \& {Olowin}}]{Abell1989}
{Abell}, G.~O., {Corwin}, Harold~G., J., \& {Olowin}, R.~P. 1989, \apjs, 70, 1

\bibitem[{Adami {et~al.}(2018)Adami, Giles, Koulouridis, Pacaud, Caretta, Pierre, Eckert, Ramos-Ceja, Gastaldello, Fotopoulou, Guglielmo, Lidman, Sadibekova, Iovino, Maughan, Chiappetti, Alis, Altieri, Baldry, Bottini, Birkinshaw, Bremer, Brown, Cucciati, Driver, Elmer, Ettori, Evrard, Faccioli, Granett, Grootes, Guzzo, Hopkins, Horellou, Lef{\`e}vre, Liske, Malek, Marulli, Maurogordato, Owers, Paltani, Poggianti, Polletta, Plionis, Pollo, Pompei, Ponman, Rapetti, Ricci, Robotham, Tuffs, Tasca, Valtchanov, Vergani, Wagner, Willis, \& Consortium}]{Adami2018}
Adami, C., Giles, P., Koulouridis, E., {et~al.} 2018, \aap, 620, A5

\bibitem[{{Aihara} {et~al.}(2018){Aihara}, {Arimoto}, {Armstrong}, {Arnouts}, {Bahcall}, {Bickerton}, {Bosch}, {Bundy}, {Capak}, {Chan}, {Chiba}, {Coupon}, {Egami}, {Enoki}, {Finet}, {Fujimori}, {Fujimoto}, {Furusawa}, {Furusawa}, {Goto}, {Goulding}, {Greco}, {Greene}, {Gunn}, {Hamana}, {Harikane}, {Hashimoto}, {Hattori}, {Hayashi}, {Hayashi}, {He{\l}miniak}, {Higuchi}, {Hikage}, {Ho}, {Hsieh}, {Huang}, {Huang}, {Ikeda}, {Imanishi}, {Inoue}, {Iwasawa}, {Iwata}, {Jaelani}, {Jian}, {Kamata}, {Karoji}, {Kashikawa}, {Katayama}, {Kawanomoto}, {Kayo}, {Koda}, {Koike}, {Kojima}, {Komiyama}, {Konno}, {Koshida}, {Koyama}, {Kusakabe}, {Leauthaud}, {Lee}, {Lin}, {Lin}, {Lupton}, {Mandelbaum}, {Matsuoka}, {Medezinski}, {Mineo}, {Miyama}, {Miyatake}, {Miyazaki}, {Momose}, {More}, {More}, {Moritani}, {Moriya}, {Morokuma}, {Mukae}, {Murata}, {Murayama}, {Nagao}, {Nakata}, {Niida}, {Niikura}, {Nishizawa}, {Obuchi}, {Oguri}, {Oishi}, {Okabe}, {Okamoto}, {Okura}, {Ono}, {Onodera}, {Onoue}, {Osato}, {Ouchi}, {Price}, {Pyo},
  {Sako}, {Sawicki}, {Shibuya}, {Shimasaku}, {Shimono}, {Shirasaki}, {Silverman}, {Simet}, {Speagle}, {Spergel}, {Strauss}, {Sugahara}, {Sugiyama}, {Suto}, {Suyu}, {Suzuki}, {Tait}, {Takada}, {Takata}, {Tamura}, {Tanaka}, {Tanaka}, {Tanaka}, {Tanaka}, {Terai}, {Terashima}, {Toba}, {Tominaga}, {Toshikawa}, {Turner}, {Uchida}, {Uchiyama}, {Umetsu}, {Uraguchi}, {Urata}, {Usuda}, {Utsumi}, {Wang}, {Wang}, {Wong}, {Yabe}, {Yamada}, {Yamanoi}, {Yasuda}, {Yeh}, {Yonehara}, \& {Yuma}}]{Aihara2018}
{Aihara}, H., {Arimoto}, N., {Armstrong}, R., {et~al.} 2018, \pasj, 70, S4

\bibitem[{{Andernach}(1991)}]{Andernach1991}
{Andernach}, H. 1991, in Astronomical Society of the Pacific Conference Series, Vol.~15, Large-scale Structures and Peculiar Motions in the Universe, ed. D.~W. {Latham} \& L.~A.~N. {da Costa}, 279--284

\bibitem[{{Artis} {et~al.}(2024){Artis}, {Bulbul}, {Grandis}, {Ghirardini}, {Clerc}, {Seppi}, {Comparat}, {Cataneo}, {von der Linden}, {Bahar}, {Balzer}, {Chiu}, {Gruen}, {Kleinebreil}, {Kluge}, {Krippendorf}, {Li}, {Liu}, {Malavasi}, {Merloni}, {Miyatake}, {Miyazaki}, {Nandra}, {Okabe}, {Pacaud}, {Predehl}, {Ramos-Ceja}, {Reiprich}, {Sanders}, {Schrabback}, {Zelmer}, \& {Zhang}}]{Artis2024b}
{Artis}, E., {Bulbul}, E., {Grandis}, S., {et~al.} 2024, arXiv e-prints, arXiv:2410.09499

\bibitem[{Artis {et~al.}(2024)Artis, Ghirardini, Bulbul, Grandis, Garrel, Clerc, Seppi, Comparat, Cataneo, Bahar, Balzer, Chiu, Gruen, Kleinebreil, Kluge, Krippendorf, Li, Liu, Merloni, Miyatake, Miyazaki, Nandra, Okabe, Pacaud, Predehl, Ramos-Ceja, Reiprich, Sanders, Schrabback, Zelmer, \& Zhang}]{Artis2024}
Artis, E., Ghirardini, V., Bulbul, E., {et~al.} 2024, arXiv e-prints, arXiv:2402.08459

\bibitem[{{Asplund} {et~al.}(2009){Asplund}, {Grevesse}, {Sauval}, \& {Scott}}]{Asplund2009}
{Asplund}, M., {Grevesse}, N., {Sauval}, A.~J., \& {Scott}, P. 2009, \araa, 47, 481

\bibitem[{{Astropy Collaboration} {et~al.}(2022){Astropy Collaboration}, {Price-Whelan}, {Lim}, {Earl}, {Starkman}, {Bradley}, {Shupe}, {Patil}, {Corrales}, {Brasseur}, {N{"o}the}, {Donath}, {Tollerud}, {Morris}, {Ginsburg}, {Vaher}, {Weaver}, {Tocknell}, {Jamieson}, {van Kerkwijk}, {Robitaille}, {Merry}, {Bachetti}, {G{"u}nther}, {Aldcroft}, {Alvarado-Montes}, {Archibald}, {B{'o}di}, {Bapat}, {Barentsen}, {Baz{'a}n}, {Biswas}, {Boquien}, {Burke}, {Cara}, {Cara}, {Conroy}, {Conseil}, {Craig}, {Cross}, {Cruz}, {D'Eugenio}, {Dencheva}, {Devillepoix}, {Dietrich}, {Eigenbrot}, {Erben}, {Ferreira}, {Foreman-Mackey}, {Fox}, {Freij}, {Garg}, {Geda}, {Glattly}, {Gondhalekar}, {Gordon}, {Grant}, {Greenfield}, {Groener}, {Guest}, {Gurovich}, {Handberg}, {Hart}, {Hatfield-Dodds}, {Homeier}, {Hosseinzadeh}, {Jenness}, {Jones}, {Joseph}, {Kalmbach}, {Karamehmetoglu}, {Ka{l}uszy{'n}ski}, {Kelley}, {Kern}, {Kerzendorf}, {Koch}, {Kulumani}, {Lee}, {Ly}, {Ma}, {MacBride}, {Maljaars}, {Muna}, {Murphy}, {Norman}, {O'Steen},
  {Oman}, {Pacifici}, {Pascual}, {Pascual-Granado}, {Patil}, {Perren}, {Pickering}, {Rastogi}, {Roulston}, {Ryan}, {Rykoff}, {Sabater}, {Sakurikar}, {Salgado}, {Sanghi}, {Saunders}, {Savchenko}, {Schwardt}, {Seifert-Eckert}, {Shih}, {Jain}, {Shukla}, {Sick}, {Simpson}, {Singanamalla}, {Singer}, {Singhal}, {Sinha}, {Sip{H{o}}cz}, {Spitler}, {Stansby}, {Streicher}, {{{S}}umak}, {Swinbank}, {Taranu}, {Tewary}, {Tremblay}, {Val-Borro}, {Van Kooten}, {Vasovi{'c}}, {Verma}, {de Miranda Cardoso}, {Williams}, {Wilson}, {Winkel}, {Wood-Vasey}, {Xue}, {Yoachim}, {Zhang}, {Zonca}, \& {Astropy Project Contributors}}]{astropy:2022}
{Astropy Collaboration}, {Price-Whelan}, A.~M., {Lim}, P.~L., {et~al.} 2022, \apj, 935, 167

\bibitem[{{Bahar} {et~al.}(2022){Bahar}, {Bulbul}, {Clerc}, {Ghirardini}, {Liu}, {Nandra}, {Pacaud}, {Chiu}, {Comparat}, {Ider-Chitham}, {Klein}, {Liu}, {Merloni}, {Migkas}, {Okabe}, {Ramos-Ceja}, {Reiprich}, {Sanders}, \& {Schrabback}}]{Bahar2022}
{Bahar}, Y.~E., {Bulbul}, E., {Clerc}, N., {et~al.} 2022, \aap, 661, A7

\bibitem[{Bahar {et~al.}(2024)Bahar, Bulbul, Ghirardini, Sanders, Zhang, Liu, Clerc, Artis, Balzer, Biffi, Bose, Comparat, Dolag, Garrel, Hadzhiyska, Hern{\'a}ndez-Aguayo, Hernquist, Kluge, Krippendorf, Merloni, Nandra, Pakmor, Popesso, Ramos-Ceja, Seppi, Springel, Weller, \& Zelmer}]{Bahar2024}
Bahar, Y.~E., Bulbul, E., Ghirardini, V., {et~al.} 2024, arXiv e-prints, arXiv:2401.17276

\bibitem[{{Balogh} {et~al.}(2021){Balogh}, {van der Burg}, {Muzzin}, {Rudnick}, {Wilson}, {Webb}, {Biviano}, {Boak}, {Cerulo}, {Chan}, {Cooper}, {Gilbank}, {Gwyn}, {Lidman}, {Matharu}, {McGee}, {Old}, {Pintos-Castro}, {Reeves}, {Shipley}, {Vulcani}, {Yee}, {Alonso}, {Bellhouse}, {Cooke}, {Davidson}, {De Lucia}, {Demarco}, {Drakos}, {Fillingham}, {Finoguenov}, {Forrest}, {Golledge}, {Jablonka}, {Lambas Garcia}, {McNab}, {Muriel}, {Nantais}, {Noble}, {Parker}, {Petter}, {Poggianti}, {Townsend}, {Valotto}, {Webb}, \& {Zaritsky}}]{Balogh2021}
{Balogh}, M.~L., {van der Burg}, R. F.~J., {Muzzin}, A., {et~al.} 2021, \mnras, 500, 358

\bibitem[{{Bleem} {et~al.}(2020){Bleem}, {Bocquet}, {Stalder}, {Gladders}, {Ade}, {Allen}, {Anderson}, {Annis}, {Ashby}, {Austermann}, {Avila}, {Avva}, {Bayliss}, {Beall}, {Bechtol}, {Bender}, {Benson}, {Bertin}, {Bianchini}, {Blake}, {Brodwin}, {Brooks}, {Buckley-Geer}, {Burke}, {Carlstrom}, {Rosell}, {Carrasco Kind}, {Carretero}, {Chang}, {Chiang}, {Citron}, {Moran}, {Costanzi}, {Crawford}, {Crites}, {da Costa}, {de Haan}, {De Vicente}, {Desai}, {Diehl}, {Dietrich}, {Dobbs}, {Eifler}, {Everett}, {Flaugher}, {Floyd}, {Frieman}, {Gallicchio}, {Garc{\'\i}a-Bellido}, {George}, {Gerdes}, {Gilbert}, {Gruen}, {Gruendl}, {Gschwend}, {Gupta}, {Gutierrez}, {Halverson}, {Harrington}, {Henning}, {Heymans}, {Holder}, {Hollowood}, {Holzapfel}, {Honscheid}, {Hrubes}, {Huang}, {Hubmayr}, {Irwin}, {James}, {Jeltema}, {Joudaki}, {Khullar}, {Klein}, {Knox}, {Kuropatkin}, {Lee}, {Li}, {Lidman}, {Lowitz}, {MacCrann}, {Mahler}, {Maia}, {Marshall}, {McDonald}, {McMahon}, {Melchior}, {Menanteau}, {Meyer}, {Miquel}, {Mocanu},
  {Mohr}, {Montgomery}, {Nadolski}, {Natoli}, {Nibarger}, {Noble}, {Novosad}, {Padin}, {Palmese}, {Parkinson}, {Patil}, {Paz-Chinch{\'o}n}, {Plazas}, {Pryke}, {Ramachandra}, {Reichardt}, {Remolina Gonz{\'a}lez}, {Romer}, {Roodman}, {Ruhl}, {Rykoff}, {Saliwanchik}, {Sanchez}, {Saro}, {Sayre}, {Schaffer}, {Schrabback}, {Serrano}, {Sharon}, {Sievers}, {Smecher}, {Smith}, {Soares-Santos}, {Stark}, {Story}, {Suchyta}, {Tarle}, {Tucker}, {Vanderlinde}, {Veach}, {Vieira}, {Wang}, {Weller}, {Whitehorn}, {Wu}, {Yefremenko}, \& {Zhang}}]{Bleem2020}
{Bleem}, L.~E., {Bocquet}, S., {Stalder}, B., {et~al.} 2020, \apjs, 247, 25

\bibitem[{Bleem {et~al.}(2015)Bleem, Stalder, de~Haan, Aird, Allen, Applegate, Ashby, Bautz, Bayliss, Benson, Bocquet, Brodwin, Carlstrom, Chang, Chiu, Cho, Clocchiatti, Crawford, Crites, Desai, Dietrich, Dobbs, Foley, Forman, George, Gladders, Gonzalez, Halverson, Hennig, Hoekstra, Holder, Holzapfel, Hrubes, Jones, Keisler, Knox, Lee, Leitch, Liu, Lueker, Luong-Van, Mantz, Marrone, McDonald, McMahon, Meyer, Mocanu, Mohr, Murray, Padin, Pryke, Reichardt, Rest, Ruel, Ruhl, Saliwanchik, Saro, Sayre, Schaffer, Schrabback, Shirokoff, Song, Spieler, Stanford, Staniszewski, Stark, Story, Stubbs, Vanderlinde, Vieira, Vikhlinin, Williamson, Zahn, \& Zenteno}]{Bleem2015}
Bleem, L.~E., Stalder, B., de~Haan, T., {et~al.} 2015, \apjs, 216, 27

\bibitem[{{Bocquet} {et~al.}(2019){Bocquet}, {Dietrich}, {Schrabback}, {Bleem}, {Klein}, {Allen}, {Applegate}, {Ashby}, {Bautz}, {Bayliss}, {Benson}, {Brodwin}, {Bulbul}, {Canning}, {Capasso}, {Carlstrom}, {Chang}, {Chiu}, {Cho}, {Clocchiatti}, {Crawford}, {Crites}, {de Haan}, {Desai}, {Dobbs}, {Foley}, {Forman}, {Garmire}, {George}, {Gladders}, {Gonzalez}, {Grandis}, {Gupta}, {Halverson}, {Hlavacek-Larrondo}, {Hoekstra}, {Holder}, {Holzapfel}, {Hou}, {Hrubes}, {Huang}, {Jones}, {Khullar}, {Knox}, {Kraft}, {Lee}, {von der Linden}, {Luong-Van}, {Mantz}, {Marrone}, {McDonald}, {McMahon}, {Meyer}, {Mocanu}, {Mohr}, {Morris}, {Padin}, {Patil}, {Pryke}, {Rapetti}, {Reichardt}, {Rest}, {Ruhl}, {Saliwanchik}, {Saro}, {Sayre}, {Schaffer}, {Shirokoff}, {Stalder}, {Stanford}, {Staniszewski}, {Stark}, {Story}, {Strazzullo}, {Stubbs}, {Vanderlinde}, {Vieira}, {Vikhlinin}, {Williamson}, \& {Zenteno}}]{Bocquet2019}
{Bocquet}, S., {Dietrich}, J.~P., {Schrabback}, T., {et~al.} 2019, \apj, 878, 55

\bibitem[{B{\"o}hringer {et~al.}(2004)B{\"o}hringer, Schuecker, Guzzo, Collins, Voges, Cruddace, Ortiz-Gil, Chincarini, De~Grandi, Edge, MacGillivray, Neumann, Schindler, \& Shaver}]{Boehringer2004}
B{\"o}hringer, H., Schuecker, P., Guzzo, L., {et~al.} 2004, \aap, 425, 367

\bibitem[{B{\"o}hringer {et~al.}(2000)B{\"o}hringer, Voges, Huchra, McLean, Giacconi, Rosati, Burg, Mader, Schuecker, Simi{\c{c}}, Komossa, Reiprich, Retzlaff, \& Tr{\"u}mper}]{Boehringer2000}
B{\"o}hringer, H., Voges, W., Huchra, J.~P., {et~al.} 2000, \apjs, 129, 435

\bibitem[{{B{\"o}hringer} {et~al.}(2000){B{\"o}hringer}, {Voges}, {Huchra}, {McLean}, {Giacconi}, {Rosati}, {Burg}, {Mader}, {Schuecker}, {Simi{\c{c}}}, {Komossa}, {Reiprich}, {Retzlaff}, \& {Tr{\"u}mper}}]{Boehringer2000_noras}
{B{\"o}hringer}, H., {Voges}, W., {Huchra}, J.~P., {et~al.} 2000, \apjs, 129, 435

\bibitem[{{Boller} {et~al.}(2016){Boller}, {Freyberg}, {Tr{\"u}mper}, {Haberl}, {Voges}, \& {Nandra}}]{Boller2016}
{Boller}, T., {Freyberg}, M.~J., {Tr{\"u}mper}, J., {et~al.} 2016, \aap, 588, A103

\bibitem[{Bower {et~al.}(1992)Bower, Lucey, \& Ellis}]{Bower1992}
Bower, R.~G., Lucey, J.~R., \& Ellis, R.~S. 1992, \mnras, 254, 589

\bibitem[{Brunner {et~al.}(2022)Brunner, Liu, Lamer, Georgakakis, Merloni, Brusa, Bulbul, Dennerl, Friedrich, Liu, Maitra, Nandra, Ramos-Ceja, Sanders, Stewart, Boller, Buchner, Clerc, Comparat, Dwelly, Eckert, Finoguenov, Freyberg, Ghirardini, Gueguen, Haberl, Kreykenbohm, Krumpe, Osterhage, Pacaud, Predehl, Reiprich, Robrade, Salvato, Santangelo, Schrabback, Schwope, \& Wilms}]{Brunner2022}
Brunner, H., Liu, T., Lamer, G., {et~al.} 2022, \aap, 661, A1

\bibitem[{Bulbul {et~al.}(2024)Bulbul, Liu, Kluge, Zhang, Sanders, Bahar, Ghirardini, Artis, Seppi, Garrel, Ramos-Ceja, Comparat, Balzer, B{\"o}ckmann, Br{\"u}ggen, Clerc, Dennerl, Dolag, Freyberg, Grandis, Gruen, Kleinebreil, Krippendorf, Lamer, Merloni, Migkas, Nandra, Pacaud, Predehl, Reiprich, Schrabback, Veronica, Weller, \& Zelmer}]{Bulbul2024}
Bulbul, E., Liu, A., Kluge, M., {et~al.} 2024, arXiv e-prints, arXiv:2402.08452

\bibitem[{Bulbul {et~al.}(2022)Bulbul, Liu, Pasini, Comparat, Hoang, Klein, Ghirardini, Salvato, Merloni, Seppi, Wolf, Anderson, Bahar, Brusa, Br{\"u}ggen, Buchner, Dwelly, Ibarra-Medel, Ider~Chitham, Liu, Nandra, Ramos-Ceja, Sanders, \& Shen}]{Bulbul2022}
Bulbul, E., Liu, A., Pasini, T., {et~al.} 2022, \aap, 661, A10

\bibitem[{{Bulbul} {et~al.}(2010){Bulbul}, {Hasler}, {Bonamente}, \& {Joy}}]{Bulbul2010}
{Bulbul}, G.~E., {Hasler}, N., {Bonamente}, M., \& {Joy}, M. 2010, \apj, 720, 1038

\bibitem[{{Burenin} {et~al.}(2007){Burenin}, {Vikhlinin}, {Hornstrup}, {Ebeling}, {Quintana}, \& {Mescheryakov}}]{Burenin2007}
{Burenin}, R.~A., {Vikhlinin}, A., {Hornstrup}, A., {et~al.} 2007, \apjs, 172, 561

\bibitem[{{Busch} \& {White}(2017)}]{Busch2017}
{Busch}, P. \& {White}, S. D.~M. 2017, \mnras, 470, 4767

\bibitem[{Carlstr{\"o}m {et~al.}(2011)Carlstr{\"o}m, Garaud, \& Babaev}]{Carlstroem2011}
Carlstr{\"o}m, J., Garaud, J., \& Babaev, E. 2011, \prb, 84, 134518

\bibitem[{Chiu {et~al.}(2022)Chiu, Ghirardini, Liu, Grandis, Bulbul, Bahar, Comparat, Bocquet, Clerc, Klein, Liu, Li, Miyatake, Mohr, More, Oguri, Okabe, Pacaud, Ramos-Ceja, Reiprich, Schrabback, \& Umetsu}]{Chiu2022}
Chiu, I.~N., Ghirardini, V., Liu, A., {et~al.} 2022, \aap, 661, A11

\bibitem[{Clerc {et~al.}(2024)Clerc, Comparat, Seppi, Artis, Bahar, Balzer, Bulbul, Dauser, Garrel, Ghirardini, Grandis, Kirsch, Kluge, Liu, Pacaud, Ramos-Ceja, Reiprich, Sanders, Wilms, \& Zhang}]{Clerc2024}
Clerc, N., Comparat, J., Seppi, R., {et~al.} 2024, arXiv e-prints, arXiv:2402.08457

\bibitem[{{Clerc} \& {Finoguenov}(2023)}]{Clerc2022}
{Clerc}, N. \& {Finoguenov}, A. 2023, in Handbook of X-ray and Gamma-ray Astrophysics, 123

\bibitem[{Clerc {et~al.}(2018)Clerc, Ramos-Ceja, Ridl, Lamer, Brunner, Hofmann, Comparat, Pacaud, K{\"a}fer, Reiprich, Merloni, Schmid, Brand, Wilms, Friedrich, Finoguenov, Dauser, \& Kreykenbohm}]{Clerc2018}
Clerc, N., Ramos-Ceja, M.~E., Ridl, J., {et~al.} 2018, \aap, 617, A92

\bibitem[{{Cruddace} {et~al.}(2002){Cruddace}, {Voges}, {B{\"o}hringer}, {Collins}, {Romer}, {MacGillivray}, {Yentis}, {Schuecker}, {Ebeling}, \& {De Grandi}}]{Cruddace2002}
{Cruddace}, R., {Voges}, W., {B{\"o}hringer}, H., {et~al.} 2002, \apjs, 140, 239

\bibitem[{Dey {et~al.}(2019)Dey, Schlegel, Lang, Blum, Burleigh, Fan, Findlay, Finkbeiner, Herrera, Juneau, Landriau, Levi, McGreer, Meisner, Myers, Moustakas, Nugent, Patej, Schlafly, Walker, Valdes, Weaver, Y{\`e}che, Zou, Zhou, Abareshi, Abbott, Abolfathi, Aguilera, Alam, Allen, Alvarez, Annis, Ansarinejad, Aubert, Beechert, Bell, BenZvi, Beutler, Bielby, Bolton, Brice{\~n}o, Buckley-Geer, Butler, Calamida, Carlberg, Carter, Casas, Castander, Choi, Comparat, Cukanovaite, Delubac, DeVries, Dey, Dhungana, Dickinson, Ding, Donaldson, Duan, Duckworth, Eftekharzadeh, Eisenstein, Etourneau, Fagrelius, Farihi, Fitzpatrick, Font-Ribera, Fulmer, G{\"a}nsicke, Gaztanaga, George, Gerdes, Gontcho, Gorgoni, Green, Guy, Harmer, Hernandez, Honscheid, Huang, James, Jannuzi, Jiang, Joyce, Karcher, Karkar, Kehoe, Kneib, Kueter-Young, Lan, Lauer, Le~Guillou, Le~Van~Suu, Lee, Lesser, Perreault~Levasseur, Li, Mann, Marshall, Mart{\'\i}nez-V{\'a}zquez, Martini, du~Mas~des Bourboux, McManus, Meier, M{\'e}nard, Metcalfe,
  Mu{\~n}oz-Guti{\'e}rrez, Najita, Napier, Narayan, Newman, Nie, Nord, Norman, Olsen, Paat, Palanque-Delabrouille, Peng, Poppett, Poremba, Prakash, Rabinowitz, Raichoor, Rezaie, Robertson, Roe, Ross, Ross, Rudnick, Safonova, Saha, S{\'a}nchez, Savary, Schweiker, Scott, Seo, Shan, Silva, Slepian, Soto, Sprayberry, Staten, Stillman, Stupak, Summers, Sien~Tie, Tirado, Vargas-Maga{\~n}a, Vivas, Wechsler, Williams, Yang, Yang, Yapici, Zaritsky, Zenteno, Zhang, Zhang, Zhou, \& Zhou}]{Dey2019}
Dey, A., Schlegel, D.~J., Lang, D., {et~al.} 2019, \aj, 157, 168

\bibitem[{Donahue {et~al.}(2020)Donahue, Funkhouser, Koeppe, Frisbie, \& Voit}]{Donahue2020}
Donahue, M., Funkhouser, K., Koeppe, D., Frisbie, R. L.~S., \& Voit, G.~M. 2020, \apj, 889, 121

\bibitem[{Dressler(1980)}]{Dressler1980}
Dressler, A. 1980, \apj, 236, 351

\bibitem[{{Ebeling} {et~al.}(1998){Ebeling}, {Edge}, {Bohringer}, {Allen}, {Crawford}, {Fabian}, {Voges}, \& {Huchra}}]{Ebeling1998}
{Ebeling}, H., {Edge}, A.~C., {Bohringer}, H., {et~al.} 1998, \mnras, 301, 881

\bibitem[{{Finoguenov} {et~al.}(2020){Finoguenov}, {Rykoff}, {Clerc}, {Costanzi}, {Hagstotz}, {Ider Chitham}, {Kiiveri}, {Kirkpatrick}, {Capasso}, {Comparat}, {Damsted}, {Dupke}, {Erfanianfar}, {Patrick Henry}, {Kaefer}, {Kneib}, {Lindholm}, {Rozo}, {van Waerbeke}, \& {Weller}}]{Finoguenov2020}
{Finoguenov}, A., {Rykoff}, E., {Clerc}, N., {et~al.} 2020, \aap, 638, A114

\bibitem[{{Gaia Collaboration} {et~al.}(2016){Gaia Collaboration}, {Prusti}, {de Bruijne}, {Brown}, {Vallenari}, {Babusiaux}, {Bailer-Jones}, {Bastian}, {Biermann}, {Evans}, {Eyer}, {Jansen}, {Jordi}, {Klioner}, {Lammers}, {Lindegren}, {Luri}, {Mignard}, {Milligan}, {Panem}, {Poinsignon}, {Pourbaix}, {Randich}, {Sarri}, {Sartoretti}, {Siddiqui}, {Soubiran}, {Valette}, {van Leeuwen}, {Walton}, {Aerts}, {Arenou}, {Cropper}, {Drimmel}, {H{\o}g}, {Katz}, {Lattanzi}, {O'Mullane}, {Grebel}, {Holland}, {Huc}, {Passot}, {Bramante}, {Cacciari}, {Casta{\~n}eda}, {Chaoul}, {Cheek}, {De Angeli}, {Fabricius}, {Guerra}, {Hern{\'a}ndez}, {Jean-Antoine-Piccolo}, {Masana}, {Messineo}, {Mowlavi}, {Nienartowicz}, {Ord{\'o}{\~n}ez-Blanco}, {Panuzzo}, {Portell}, {Richards}, {Riello}, {Seabroke}, {Tanga}, {Th{\'e}venin}, {Torra}, {Els}, {Gracia-Abril}, {Comoretto}, {Garcia-Reinaldos}, {Lock}, {Mercier}, {Altmann}, {Andrae}, {Astraatmadja}, {Bellas-Velidis}, {Benson}, {Berthier}, {Blomme}, {Busso}, {Carry}, {Cellino}, {Clementini},
  {Cowell}, {Creevey}, {Cuypers}, {Davidson}, {De Ridder}, {de Torres}, {Delchambre}, {Dell'Oro}, {Ducourant}, {Fr{\'e}mat}, {Garc{\'\i}a-Torres}, {Gosset}, {Halbwachs}, {Hambly}, {Harrison}, {Hauser}, {Hestroffer}, {Hodgkin}, {Huckle}, {Hutton}, {Jasniewicz}, {Jordan}, {Kontizas}, {Korn}, {Lanzafame}, {Manteiga}, {Moitinho}, {Muinonen}, {Osinde}, {Pancino}, {Pauwels}, {Petit}, {Recio-Blanco}, {Robin}, {Sarro}, {Siopis}, {Smith}, {Smith}, {Sozzetti}, {Thuillot}, {van Reeven}, {Viala}, {Abbas}, {Abreu Aramburu}, {Accart}, {Aguado}, {Allan}, {Allasia}, {Altavilla}, {{\'A}lvarez}, {Alves}, {Anderson}, {Andrei}, {Anglada Varela}, {Antiche}, {Antoja}, {Ant{\'o}n}, {Arcay}, {Atzei}, {Ayache}, {Bach}, {Baker}, {Balaguer-N{\'u}{\~n}ez}, {Barache}, {Barata}, {Barbier}, {Barblan}, {Baroni}, {Barrado y Navascu{\'e}s}, {Barros}, {Barstow}, {Becciani}, {Bellazzini}, {Bellei}, {Bello Garc{\'\i}a}, {Belokurov}, {Bendjoya}, {Berihuete}, {Bianchi}, {Bienaym{\'e}}, {Billebaud}, {Blagorodnova}, {Blanco-Cuaresma}, {Boch},
  {Bombrun}, {Borrachero}, {Bouquillon}, {Bourda}, {Bouy}, {Bragaglia}, {Breddels}, {Brouillet}, {Br{\"u}semeister}, {Bucciarelli}, {Budnik}, {Burgess}, {Burgon}, {Burlacu}, {Busonero}, {Buzzi}, {Caffau}, {Cambras}, {Campbell}, {Cancelliere}, {Cantat-Gaudin}, {Carlucci}, {Carrasco}, {Castellani}, {Charlot}, {Charnas}, {Charvet}, {Chassat}, {Chiavassa}, {Clotet}, {Cocozza}, {Collins}, {Collins}, \& {Costigan}}]{Gaia2016}
{Gaia Collaboration}, {Prusti}, T., {de Bruijne}, J.~H.~J., {et~al.} 2016, \aap, 595, A1

\bibitem[{{Gaia Collaboration} {et~al.}(2023){Gaia Collaboration}, {Vallenari}, {Brown}, {Prusti}, {de Bruijne}, {Arenou}, {Babusiaux}, {Biermann}, {Creevey}, {Ducourant}, {Evans}, {Eyer}, {Guerra}, {Hutton}, {Jordi}, {Klioner}, {Lammers}, {Lindegren}, {Luri}, {Mignard}, {Panem}, {Pourbaix}, {Randich}, {Sartoretti}, {Soubiran}, {Tanga}, {Walton}, {Bailer-Jones}, {Bastian}, {Drimmel}, {Jansen}, {Katz}, {Lattanzi}, {van Leeuwen}, {Bakker}, {Cacciari}, {Casta{\~n}eda}, {De Angeli}, {Fabricius}, {Fouesneau}, {Fr{\'e}mat}, {Galluccio}, {Guerrier}, {Heiter}, {Masana}, {Messineo}, {Mowlavi}, {Nicolas}, {Nienartowicz}, {Pailler}, {Panuzzo}, {Riclet}, {Roux}, {Seabroke}, {Sordo}, {Th{\'e}venin}, {Gracia-Abril}, {Portell}, {Teyssier}, {Altmann}, {Andrae}, {Audard}, {Bellas-Velidis}, {Benson}, {Berthier}, {Blomme}, {Burgess}, {Busonero}, {Busso}, {C{\'a}novas}, {Carry}, {Cellino}, {Cheek}, {Clementini}, {Damerdji}, {Davidson}, {de Teodoro}, {Nu{\~n}ez Campos}, {Delchambre}, {Dell'Oro}, {Esquej},
  {Fern{\'a}ndez-Hern{\'a}ndez}, {Fraile}, {Garabato}, {Garc{\'\i}a-Lario}, {Gosset}, {Haigron}, {Halbwachs}, {Hambly}, {Harrison}, {Hern{\'a}ndez}, {Hestroffer}, {Hodgkin}, {Holl}, {Jan{\ss}en}, {Jevardat de Fombelle}, {Jordan}, {Krone-Martins}, {Lanzafame}, {L{\"o}ffler}, {Marchal}, {Marrese}, {Moitinho}, {Muinonen}, {Osborne}, {Pancino}, {Pauwels}, {Recio-Blanco}, {Reyl{\'e}}, {Riello}, {Rimoldini}, {Roegiers}, {Rybizki}, {Sarro}, {Siopis}, {Smith}, {Sozzetti}, {Utrilla}, {van Leeuwen}, {Abbas}, {{\'A}brah{\'a}m}, {Abreu Aramburu}, {Aerts}, {Aguado}, {Ajaj}, {Aldea-Montero}, {Altavilla}, {{\'A}lvarez}, {Alves}, {Anders}, {Anderson}, {Anglada Varela}, {Antoja}, {Baines}, {Baker}, {Balaguer-N{\'u}{\~n}ez}, {Balbinot}, {Balog}, {Barache}, {Barbato}, {Barros}, {Barstow}, {Bartolom{\'e}}, {Bassilana}, {Bauchet}, {Becciani}, {Bellazzini}, {Berihuete}, {Bernet}, {Bertone}, {Bianchi}, {Binnenfeld}, {Blanco-Cuaresma}, {Blazere}, {Boch}, {Bombrun}, {Bossini}, {Bouquillon}, {Bragaglia}, {Bramante}, {Breedt},
  {Bressan}, {Brouillet}, {Brugaletta}, {Bucciarelli}, {Burlacu}, {Butkevich}, {Buzzi}, {Caffau}, {Cancelliere}, {Cantat-Gaudin}, {Carballo}, {Carlucci}, {Carnerero}, {Carrasco}, {Casamiquela}, {Castellani}, {Castro-Ginard}, {Chaoul}, {Charlot}, {Chemin}, {Chiaramida}, {Chiavassa}, {Chornay}, {Comoretto}, {Contursi}, {Cooper}, {Cornez}, {Cowell}, {Crifo}, {Cropper}, {Crosta}, {Crowley}, {Dafonte}, {Dapergolas}, {David}, {David}, {de Laverny}, {De Luise}, {De March}, {De Ridder}, {de Souza}, {de Torres}, {del Peloso}, {del Pozo}, {Delbo}, {Delgado}, {Delisle}, {Demouchy}, {Dharmawardena}, {Di Matteo}, {Diakite}, {Diener}, {Distefano}, {Dolding}, {Edvardsson}, {Enke}, {Fabre}, {Fabrizio}, {Faigler}, {Fedorets}, {Fernique}, {Fienga}, {Figueras}, {Fournier}, {Fouron}, {Fragkoudi}, {Gai}, {Garcia-Gutierrez}, {Garcia-Reinaldos}, {Garc{\'\i}a-Torres}, {Garofalo}, {Gavel}, {Gavras}, {Gerlach}, {Geyer}, {Giacobbe}, {Gilmore}, {Girona}, {Giuffrida}, {Gomel}, {Gomez}, {Gonz{\'a}lez-N{\'u}{\~n}ez},
  {Gonz{\'a}lez-Santamar{\'\i}a}, {Gonz{\'a}lez-Vidal}, {Granvik}, {Guillout}, {Guiraud}, {Guti{\'e}rrez-S{\'a}nchez}, {Guy}, {Hatzidimitriou}, {Hauser}, {Haywood}, {Helmer}, {Helmi}, {Sarmiento}, {Hidalgo}, {Hilger}, {H{\l}adczuk}, {Hobbs}, {Holland}, {Huckle}, {Jardine}, {Jasniewicz}, {Jean-Antoine Piccolo}, {Jim{\'e}nez-Arranz}, {Jorissen}, {Juaristi Campillo}, {Julbe}, {Karbevska}, {Kervella}, {Khanna}, {Kontizas}, {Kordopatis}, {Korn}, {K{\'o}sp{\'a}l}, {Kostrzewa-Rutkowska}, {Kruszy{\'n}ska}, {Kun}, {Laizeau}, {Lambert}, {Lanza}, {Lasne}, {Le Campion}, {Lebreton}, {Lebzelter}, {Leccia}, {Leclerc}, {Lecoeur-Taibi}, {Liao}, {Licata}, {Lindstr{\o}m}, {Lister}, {Livanou}, {Lobel}, {Lorca}, {Loup}, {Madrero Pardo}, {Magdaleno Romeo}, {Managau}, {Mann}, {Manteiga}, {Marchant}, {Marconi}, {Marcos}, {Marcos Santos}, {Mar{\'\i}n Pina}, {Marinoni}, {Marocco}, {Marshall}, {Martin Polo}, {Mart{\'\i}n-Fleitas}, {Marton}, {Mary}, {Masip}, {Massari}, {Mastrobuono-Battisti}, {Mazeh}, {McMillan}, {Messina}, {Michalik},
  {Millar}, {Mints}, {Molina}, {Molinaro}, {Moln{\'a}r}, {Monari}, {Mongui{\'o}}, {Montegriffo}, {Montero}, {Mor}, {Mora}, {Morbidelli}, {Morel}, {Morris}, {Muraveva}, {Murphy}, {Musella}, {Nagy}, {Noval}, {Oca{\~n}a}, {Ogden}, {Ordenovic}, {Osinde}, {Pagani}, {Pagano}, {Palaversa}, {Palicio}, {Pallas-Quintela}, {Panahi}, {Payne-Wardenaar}, {Pe{\~n}alosa Esteller}, {Penttil{\"a}}, {Pichon}, {Piersimoni}, {Pineau}, {Plachy}, {Plum}, {Poggio}, {Pr{\v{s}}a}, {Pulone}, {Racero}, {Ragaini}, {Rainer}, {Raiteri}, {Rambaux}, {Ramos}, {Ramos-Lerate}, {Re Fiorentin}, {Regibo}, {Richards}, {Rios Diaz}, {Ripepi}, {Riva}, {Rix}, {Rixon}, {Robichon}, {Robin}, {Robin}, {Roelens}, {Rogues}, {Rohrbasser}, {Romero-G{\'o}mez}, {Rowell}, {Royer}, {Ruz Mieres}, {Rybicki}, {Sadowski}, {S{\'a}ez N{\'u}{\~n}ez}, {Sagrist{\`a} Sell{\'e}s}, {Sahlmann}, {Salguero}, {Samaras}, {Sanchez Gimenez}, {Sanna}, {Santove{\~n}a}, {Sarasso}, {Schultheis}, {Sciacca}, {Segol}, {Segovia}, {S{\'e}gransan}, {Semeux}, {Shahaf}, {Siddiqui}, {Siebert},
  {Siltala}, {Silvelo}, {Slezak}, {Slezak}, {Smart}, {Snaith}, {Solano}, {Solitro}, {Souami}, {Souchay}, {Spagna}, {Spina}, {Spoto}, {Steele}, {Steidelm{\"u}ller}, {Stephenson}, {S{\"u}veges}, {Surdej}, {Szabados}, {Szegedi-Elek}, {Taris}, {Taylor}, {Teixeira}, {Tolomei}, {Tonello}, {Torra}, {Torra}, {Torralba Elipe}, {Trabucchi}, {Tsounis}, {Turon}, {Ulla}, {Unger}, {Vaillant}, {van Dillen}, {van Reeven}, {Vanel}, {Vecchiato}, {Viala}, {Vicente}, {Voutsinas}, {Weiler}, {Wevers}, {Wyrzykowski}, {Yoldas}, {Yvard}, {Zhao}, {Zorec}, {Zucker}, \& {Zwitter}}]{Gaia2023}
{Gaia Collaboration}, {Vallenari}, A., {Brown}, A.~G.~A., {et~al.} 2023, \aap, 674, A1

\bibitem[{Ghirardini {et~al.}(2024)Ghirardini, Bulbul, Artis, Clerc, Garrel, Grandis, Kluge, Liu, Bahar, Balzer, Chiu, Comparat, Gruen, Kleinebreil, Krippendorf, Merloni, Nandra, Okabe, Pacaud, Predehl, Ramos-Ceja, Reiprich, Sanders, Schrabback, Seppi, Zelmer, Zhang, Bornemann, Brunner, Burwitz, Coutinho, Dennerl, Freyberg, Friedrich, Gaida, Gueguen, Haberl, Kink, Lamer, Li, Liu, Maitra, Meidinger, Mueller, Miyatake, Miyazaki, Robrade, Schwope, \& Stewart}]{Ghirardini2024}
Ghirardini, V., Bulbul, E., Artis, E., {et~al.} 2024, arXiv e-prints, arXiv:2402.08458

\bibitem[{Gladders \& Yee(2000)}]{Gladders2000}
Gladders, M.~D. \& Yee, H.~K.~C. 2000, \aj, 120, 2148

\bibitem[{{Gonzalez} {et~al.}(2019){Gonzalez}, {Gettings}, {Brodwin}, {Eisenhardt}, {Stanford}, {Wylezalek}, {Decker}, {Marrone}, {Moravec}, {O'Donnell}, {Stalder}, {Stern}, {Abdulla}, {Brown}, {Carlstrom}, {Chambers}, {Hayden}, {Lin}, {Magnier}, {Masci}, {Mantz}, {McDonald}, {Mo}, {Perlmutter}, {Wright}, \& {Zeimann}}]{Gonzalez2019}
{Gonzalez}, A.~H., {Gettings}, D.~P., {Brodwin}, M., {et~al.} 2019, \apjs, 240, 33

\bibitem[{Grandis {et~al.}(2024)Grandis, Ghirardini, Bocquet, Garrel, Mohr, Liu, Kluge, Kimmig, Reiprich, Alarcon, Amon, Artis, Bahar, Balzer, Bechtol, Becker, Bernstein, Bulbul, Campos, Carnero~Rosell, Carrasco~Kind, Cawthon, Chang, Chen, Chiu, Choi, Clerc, Comparat, Cordero, Davis, Derose, Diehl, Dodelson, Doux, Drlica-Wagner, Eckert, Elvin-Poole, Everett, Ferte, Gatt, Giannini, Giles, Gruen, Gruendl, Harrison, Hartley, Herner, Huf, Kleinebreil, Kuropatkin, Leget, Maccrann, Mccullough, Merloni, Myles, Nandra, Navarro-Alsina, Okabe, Pacaud, Pandey, Prat, Predehl, Ramos, Raveri, Rollins, Roodman, Ross, Rykoff, Sanchez, Sanders, Schrabback, Secco, Seppi, Sevilla-Noarbe, Sheldon, Shin, Troxel, Tutusaus, Varga, Wu, Yanny, Yin, Zhang, Zhang, Alves, Bhargava, Brooks, Burke, Carretero, Costanzi, da~Costa, Pereira, De~Vicente, Desai, Doel, Ferrero, Flaugher, Friedel, Frieman, Garc{\'\i}a-Bellido, Gutierrez, Hinton, Hollowood, Honscheid, James, Jeffrey, Lahav, Lee, Marshall, Menanteau, Ogando, Pieres,
  Plazas~Malag{\'o}n, Romer, Sanchez, Schubnell, Smith, Suchyta, Swanson, Tarle, Weaverdyck, \& Weller}]{Grandis2024}
Grandis, S., Ghirardini, V., Bocquet, S., {et~al.} 2024, arXiv e-prints, arXiv:2402.08455

\bibitem[{{Green} {et~al.}(2017){Green}, {Edge}, {Ebeling}, {Burgett}, {Draper}, {Kaiser}, {Kudritzki}, {Magnier}, {Metcalfe}, {Wainscoat}, \& {Waters}}]{Green2017}
{Green}, T.~S., {Edge}, A.~C., {Ebeling}, H., {et~al.} 2017, \mnras, 465, 4872

\bibitem[{Harris {et~al.}(2020)Harris, Millman, van~der Walt, Gommers, Virtanen, Cournapeau, Wieser, Taylor, Berg, Smith, Kern, Picus, Hoyer, van Kerkwijk, Brett, Haldane, del R{\'{i}}o, Wiebe, Peterson, G{\'{e}}rard-Marchant, Sheppard, Reddy, Weckesser, Abbasi, Gohlke, \& Oliphant}]{harris2020array}
Harris, C.~R., Millman, K.~J., van~der Walt, S.~J., {et~al.} 2020, Nature, 585, 357

\bibitem[{{HI4PI Collaboration} {et~al.}(2016){HI4PI Collaboration}, {Ben Bekhti}, {Fl{\"o}er}, {Keller}, {Kerp}, {Lenz}, {Winkel}, {Bailin}, {Calabretta}, {Dedes}, {Ford}, {Gibson}, {Haud}, {Janowiecki}, {Kalberla}, {Lockman}, {McClure-Griffiths}, {Murphy}, {Nakanishi}, {Pisano}, \& {Staveley-Smith}}]{HI4PI2016}
{HI4PI Collaboration}, {Ben Bekhti}, N., {Fl{\"o}er}, L., {et~al.} 2016, \aap, 594, A116

\bibitem[{{Hilton} {et~al.}(2021){Hilton}, {Sif{\'o}n}, {Naess}, {Madhavacheril}, {Oguri}, {Rozo}, {Rykoff}, {Abbott}, {Adhikari}, {Aguena}, {Aiola}, {Allam}, {Amodeo}, {Amon}, {Annis}, {Ansarinejad}, {Aros-Bunster}, {Austermann}, {Avila}, {Bacon}, {Battaglia}, {Beall}, {Becker}, {Bernstein}, {Bertin}, {Bhandarkar}, {Bhargava}, {Bond}, {Brooks}, {Burke}, {Calabrese}, {Carrasco Kind}, {Carretero}, {Choi}, {Choi}, {Conselice}, {da Costa}, {Costanzi}, {Crichton}, {Crowley}, {D{\"u}nner}, {Denison}, {Devlin}, {Dicker}, {Diehl}, {Dietrich}, {Doel}, {Duff}, {Duivenvoorden}, {Dunkley}, {Everett}, {Ferraro}, {Ferrero}, {Fert{\'e}}, {Flaugher}, {Frieman}, {Gallardo}, {Garc{\'\i}a-Bellido}, {Gaztanaga}, {Gerdes}, {Giles}, {Golec}, {Gralla}, {Grandis}, {Gruen}, {Gruendl}, {Gschwend}, {Gutierrez}, {Han}, {Hartley}, {Hasselfield}, {Hill}, {Hilton}, {Hincks}, {Hinton}, {Ho}, {Honscheid}, {Hoyle}, {Hubmayr}, {Huffenberger}, {Hughes}, {Jaelani}, {Jain}, {James}, {Jeltema}, {Kent}, {Knowles}, {Koopman}, {Kuehn}, {Lahav},
  {Lima}, {Lin}, {Lokken}, {Loubser}, {MacCrann}, {Maia}, {Marriage}, {Martin}, {McMahon}, {Melchior}, {Menanteau}, {Miquel}, {Miyatake}, {Moodley}, {Morgan}, {Mroczkowski}, {Nati}, {Newburgh}, {Niemack}, {Nishizawa}, {Ogando}, {Orlowski-Scherer}, {Page}, {Palmese}, {Partridge}, {Paz-Chinch{\'o}n}, {Phakathi}, {Plazas}, {Robertson}, {Romer}, {Carnero Rosell}, {Salatino}, {Sanchez}, {Schaan}, {Schillaci}, {Sehgal}, {Serrano}, {Shin}, {Simon}, {Smith}, {Soares-Santos}, {Spergel}, {Staggs}, {Storer}, {Suchyta}, {Swanson}, {Tarle}, {Thomas}, {To}, {Trac}, {Ullom}, {Vale}, {Van Lanen}, {Vavagiakis}, {De Vicente}, {Wilkinson}, {Wollack}, {Xu}, \& {Zhang}}]{Hilton2021}
{Hilton}, M., {Sif{\'o}n}, C., {Naess}, S., {et~al.} 2021, \apjs, 253, 3

\bibitem[{{Huang} {et~al.}(2020){Huang}, {Bleem}, {Stalder}, {Ade}, {Allen}, {Anderson}, {Austermann}, {Avva}, {Beall}, {Bender}, {Benson}, {Bianchini}, {Bocquet}, {Brodwin}, {Carlstrom}, {Chang}, {Chiang}, {Citron}, {Moran}, {Crawford}, {Crites}, {Haan}, {Dobbs}, {Everett}, {Floyd}, {Gallicchio}, {George}, {Gilbert}, {Gladders}, {Guns}, {Gupta}, {Halverson}, {Harrington}, {Henning}, {Hilton}, {Holder}, {Holzapfel}, {Hrubes}, {Hubmayr}, {Irwin}, {Khullar}, {Knox}, {Lee}, {Li}, {Lowitz}, {McDonald}, {McMahon}, {Meyer}, {Mocanu}, {Montgomery}, {Nadolski}, {Natoli}, {Nibarger}, {Noble}, {Novosad}, {Padin}, {Patil}, {Pryke}, {Reichardt}, {Ruhl}, {Saliwanchik}, {Saro}, {Sayre}, {Schaffer}, {Sharon}, {Sievers}, {Smecher}, {Stark}, {Story}, {Tucker}, {Vanderlinde}, {Veach}, {Vieira}, {Wang}, {Whitehorn}, {Wu}, \& {Yefremenko}}]{Huang2020}
{Huang}, N., {Bleem}, L.~E., {Stalder}, B., {et~al.} 2020, \aj, 159, 110

\bibitem[{{Huang} {et~al.}(2021){Huang}, {Storfer}, {Gu}, {Ravi}, {Pilon}, {Sheu}, {Venguswamy}, {Banka}, {Dey}, {Landriau}, {Lang}, {Meisner}, {Moustakas}, {Myers}, {Sajith}, {Schlafly}, \& {Schlegel}}]{Huang2021}
{Huang}, X., {Storfer}, C., {Gu}, A., {et~al.} 2021, \apj, 909, 27

\bibitem[{Hunter(2007)}]{Hunter2007}
Hunter, J.~D. 2007, Computing in Science \& Engineering, 9, 90

\bibitem[{Ider~Chitham {et~al.}(2020)Ider~Chitham, Comparat, Finoguenov, Clerc, Kirkpatrick, Damsted, Kukkola, Capasso, Nandra, Merloni, Bulbul, Rykoff, Schneider, \& Brownstein}]{IderChitham2020}
Ider~Chitham, J., Comparat, J., Finoguenov, A., {et~al.} 2020, \mnras, 499, 4768

\bibitem[{{K{\"a}fer} {et~al.}(2020){K{\"a}fer}, {Finoguenov}, {Eckert}, {Clerc}, {Ramos-Ceja}, {Sanders}, \& {Ghirardini}}]{Kafer2020}
{K{\"a}fer}, F., {Finoguenov}, A., {Eckert}, D., {et~al.} 2020, \aap, 634, A8

\bibitem[{Kleinebreil {et~al.}(2024)Kleinebreil, Grandis, Schrabback, Ghirardini, Chiu, Liu, Kluge, Reiprich, Artis, Bahar, Balzer, Bulbul, Clerc, Comparat, Garrel, Gruen, Li, Miyatake, Miyazaki, Ramos-Ceja, Sanders, Seppi, Okabe, \& Zhang}]{Kleinebreil2024}
Kleinebreil, F., Grandis, S., Schrabback, T., {et~al.} 2024, arXiv e-prints, arXiv:2402.08456

\bibitem[{Kluge {et~al.}(2024)Kluge, Comparat, Liu, Balzer, Bulbul, Ider~Chitham, Ghirardini, Garrel, Bahar, Artis, Bender, Clerc, Dwelly, Fabricius, Grandis, Hern{\'a}ndez-Lang, Hill, Joshi, Lamer, Merloni, Nandra, Pacaud, Predehl, Ramos-Ceja, Reiprich, Salvato, Sanders, Schrabback, Seppi, Zelmer, Zenteno, \& Zhang}]{Kluge2024}
Kluge, M., Comparat, J., Liu, A., {et~al.} 2024, arXiv e-prints, arXiv:2402.08453

\bibitem[{{Kluge} {et~al.}(2020){Kluge}, {Neureiter}, {Riffeser}, {Bender}, {Goessl}, {Hopp}, {Schmidt}, {Ries}, \& {Brosch}}]{Kluge2020}
{Kluge}, M., {Neureiter}, B., {Riffeser}, A., {et~al.} 2020, \apjs, 247, 43

\bibitem[{{Koulouridis} {et~al.}(2021){Koulouridis}, {Clerc}, {Sadibekova}, {Chira}, {Drigga}, {Faccioli}, {Le F{\`e}vre}, {Garrel}, {Gaynullina}, {Gkini}, {Kosiba}, {Pacaud}, {Pierre}, {Ridl}, {Tazhenova}, {Adami}, {Altieri}, {Baguley}, {Cabanac}, {Cucchetti}, {Khalikova}, {Lieu}, {Melin}, {Molham}, {Ramos-Ceja}, {Soucail}, {Takey}, \& {Valtchanov}}]{Koulouridis2021}
{Koulouridis}, E., {Clerc}, N., {Sadibekova}, T., {et~al.} 2021, \aap, 652, A12

\bibitem[{{Kovlakas} {et~al.}(2021){Kovlakas}, {Zezas}, {Andrews}, {Basu-Zych}, {Fragos}, {Hornschemeier}, {Kouroumpatzakis}, {Lehmer}, \& {Ptak}}]{Kovlakas2021}
{Kovlakas}, K., {Zezas}, A., {Andrews}, J.~J., {et~al.} 2021, \mnras, 506, 1896

\bibitem[{Lang(2014)}]{Lang2014}
Lang, D. 2014, The Astronomical Journal, 147, 108

\bibitem[{{Lang} {et~al.}(2016){Lang}, {Hogg}, \& {Mykytyn}}]{Lang2016}
{Lang}, D., {Hogg}, D.~W., \& {Mykytyn}, D. 2016, {The Tractor: Probabilistic astronomical source detection and measurement}, Astrophysics Source Code Library, record ascl:1604.008

\bibitem[{Liu {et~al.}(2022)Liu, Bulbul, Ghirardini, Liu, Klein, Clerc, {\"O}zsoy, Ramos-Ceja, Pacaud, Comparat, Okabe, Bahar, Biffi, Brunner, Br{\"u}ggen, Buchner, Ider~Chitham, Chiu, Dolag, Gatuzz, Gonzalez, Hoang, Lamer, Merloni, Nandra, Oguri, Ota, Predehl, Reiprich, Salvato, Schrabback, Sanders, Seppi, \& Thibaud}]{Liu2022a}
Liu, A., Bulbul, E., Ghirardini, V., {et~al.} 2022, \aap, 661, A2

\bibitem[{Liu {et~al.}(2024)Liu, Bulbul, Kluge, Ghirardini, Zhang, Sanders, Artis, Bahar, Balzer, Br{\"u}ggen, Clerc, Comparat, Garrel, Gatuzz, Grandis, Lamer, Merloni, Migkas, Nandra, Predehl, Ramos-Ceja, Reiprich, Seppi, \& Zelmer}]{Liu2024}
Liu, A., Bulbul, E., Kluge, M., {et~al.} 2024, \aap, 683, A130

\bibitem[{{Liu} {et~al.}(2023){Liu}, {Bulbul}, {Ramos-Ceja}, {Sanders}, {Ghirardini}, {Bahar}, {Yeung}, {Gatuzz}, {Freyberg}, {Garrel}, {Zhang}, {Merloni}, \& {Nandra}}]{Liu2023}
{Liu}, A., {Bulbul}, E., {Ramos-Ceja}, M.~E., {et~al.} 2023, \aap, 670, A96

\bibitem[{{Liu} {et~al.}(2024){Liu}, {Bulbul}, {Shin}, {von der Linden}, {Ghirardini}, {Kluge}, {Sanders}, {Grandis}, {Zhang}, {Artis}, {Bahar}, {Balzer}, {Clerc}, {Malavasi}, {Merloni}, {Nandra}, {Ramos-Ceja}, \& {Zelmer}}]{Liu2024b}
{Liu}, A., {Bulbul}, E., {Shin}, T., {et~al.} 2024, arXiv e-prints, arXiv:2404.17345

\bibitem[{Mainzer {et~al.}(2014)Mainzer, Bauer, Cutri, Grav, Masiero, Beck, Clarkson, Conrow, Dailey, Eisenhardt, Fabinsky, Fajardo-Acosta, Fowler, Gelino, Grillmair, Heinrichsen, Kendall, Kirkpatrick, Liu, Masci, McCallon, Nugent, Papin, Rice, Royer, Ryan, Sevilla, Sonnett, Stevenson, Thompson, Wheelock, Wiemer, Wittman, Wright, \& Yan}]{Mainzer2014}
Mainzer, A., Bauer, J., Cutri, R.~M., {et~al.} 2014, The Astrophysical Journal, 792, 30

\bibitem[{{McClintock} {et~al.}(2019){McClintock}, {Varga}, {Gruen}, {Rozo}, {Rykoff}, {Shin}, {Melchior}, {DeRose}, {Seitz}, {Dietrich}, {Sheldon}, {Zhang}, {von der Linden}, {Jeltema}, {Mantz}, {Romer}, {Allen}, {Becker}, {Bermeo}, {Bhargava}, {Costanzi}, {Everett}, {Farahi}, {Hamaus}, {Hartley}, {Hollowood}, {Hoyle}, {Israel}, {Li}, {MacCrann}, {Morris}, {Palmese}, {Plazas}, {Pollina}, {Rau}, {Simet}, {Soares-Santos}, {Troxel}, {Vergara Cervantes}, {Wechsler}, {Zuntz}, {Abbott}, {Abdalla}, {Allam}, {Annis}, {Avila}, {Bridle}, {Brooks}, {Burke}, {Carnero Rosell}, {Carrasco Kind}, {Carretero}, {Castander}, {Crocce}, {Cunha}, {D'Andrea}, {da Costa}, {Davis}, {De Vicente}, {Diehl}, {Doel}, {Drlica-Wagner}, {Evrard}, {Flaugher}, {Fosalba}, {Frieman}, {Garc{\'\i}a-Bellido}, {Gaztanaga}, {Gerdes}, {Giannantonio}, {Gruendl}, {Gutierrez}, {Honscheid}, {James}, {Kirk}, {Krause}, {Kuehn}, {Lahav}, {Li}, {Lima}, {March}, {Marshall}, {Menanteau}, {Miquel}, {Mohr}, {Nord}, {Ogando}, {Roodman}, {Sanchez}, {Scarpine},
  {Schindler}, {Sevilla-Noarbe}, {Smith}, {Smith}, {Sobreira}, {Suchyta}, {Swanson}, {Tarle}, {Tucker}, {Vikram}, {Walker}, {Weller}, \& {DES Collaboration}}]{McClintock2019}
{McClintock}, T., {Varga}, T.~N., {Gruen}, D., {et~al.} 2019, \mnras, 482, 1352

\bibitem[{McDonald {et~al.}(2012)McDonald, Bayliss, Benson, Foley, Ruel, Sullivan, Veilleux, Aird, Ashby, Bautz, Bazin, Bleem, Brodwin, Carlstrom, Chang, Cho, Clocchiatti, Crawford, Crites, de~Haan, Desai, Dobbs, Dudley, Egami, Forman, Garmire, George, Gladders, Gonzalez, Halverson, Harrington, High, Holder, Holzapfel, Hoover, Hrubes, Jones, Joy, Keisler, Knox, Lee, Leitch, Liu, Lueker, Luong-van, Mantz, Marrone, McMahon, Mehl, Meyer, Miller, Mocanu, Mohr, Montroy, Murray, Natoli, Padin, Plagge, Pryke, Rawle, Reichardt, Rest, Rex, Ruhl, Saliwanchik, Saro, Sayre, Schaffer, Shaw, Shirokoff, Simcoe, Song, Spieler, Stalder, Staniszewski, Stark, Story, Stubbs, {\v{S}}uhada, van Engelen, Vanderlinde, Vieira, Vikhlinin, Williamson, Zahn, \& Zenteno}]{McDonald2012}
McDonald, M., Bayliss, M., Benson, B.~A., {et~al.} 2012, \nat, 488, 349

\bibitem[{{Mehrtens} {et~al.}(2012){Mehrtens}, {Romer}, {Hilton}, {Lloyd-Davies}, {Miller}, {Stanford}, {Hosmer}, {Hoyle}, {Collins}, {Liddle}, {Viana}, {Nichol}, {Stott}, {Dubois}, {Kay}, {Sahl{\'e}n}, {Young}, {Short}, {Christodoulou}, {Watson}, {Davidson}, {Harrison}, {Baruah}, {Smith}, {Burke}, {Mayers}, {Deadman}, {Rooney}, {Edmondson}, {West}, {Campbell}, {Edge}, {Mann}, {Sabirli}, {Wake}, {Benoist}, {da Costa}, {Maia}, \& {Ogando}}]{Mehrtens2012}
{Mehrtens}, N., {Romer}, A.~K., {Hilton}, M., {et~al.} 2012, \mnras, 423, 1024

\bibitem[{Meisner {et~al.}(2017{\natexlab{a}})Meisner, Lang, \& Schlegel}]{Meisner2017a}
Meisner, A.~M., Lang, D., \& Schlegel, D.~J. 2017{\natexlab{a}}, The Astronomical Journal, 154, 161

\bibitem[{Meisner {et~al.}(2017{\natexlab{b}})Meisner, Lang, \& Schlegel}]{Meisner2017b}
Meisner, A.~M., Lang, D., \& Schlegel, D.~J. 2017{\natexlab{b}}, The Astronomical Journal, 153, 38

\bibitem[{{Merloni} {et~al.}(2024){Merloni}, {Lamer}, {Liu}, {Ramos-Ceja}, {Brunner}, {Bulbul}, {Dennerl}, {Doroshenko}, {Freyberg}, {Friedrich}, {Gatuzz}, {Georgakakis}, {Haberl}, {Igo}, {Kreykenbohm}, {Liu}, {Maitra}, {Malyali}, {Mayer}, {Nandra}, {Predehl}, {Robrade}, {Salvato}, {Sanders}, {Stewart}, {Tub{\'\i}n-Arenas}, {Weber}, {Wilms}, {Arcodia}, {Artis}, {Aschersleben}, {Avakyan}, {Aydar}, {Bahar}, {Balzer}, {Becker}, {Berger}, {Boller}, {Bornemann}, {Br{\"u}ggen}, {Brusa}, {Buchner}, {Burwitz}, {Camilloni}, {Clerc}, {Comparat}, {Coutinho}, {Czesla}, {Dannhauer}, {Dauner}, {Dauser}, {Dietl}, {Dolag}, {Dwelly}, {Egg}, {Ehl}, {Freund}, {Friedrich}, {Gaida}, {Garrel}, {Ghirardini}, {Gokus}, {Gr{\"u}nwald}, {Grandis}, {Grotova}, {Gruen}, {Gueguen}, {H{\"a}mmerich}, {Hamaus}, {Hasinger}, {Haubner}, {Homan}, {Ider Chitham}, {Joseph}, {Joyce}, {K{\"o}nig}, {Kaltenbrunner}, {Khokhriakova}, {Kink}, {Kirsch}, {Kluge}, {Knies}, {Krippendorf}, {Krumpe}, {Kurpas}, {Li}, {Liu}, {Locatelli}, {Lorenz}, {M{\"u}ller},
  {Magaudda}, {Mannes}, {McCall}, {Meidinger}, {Michailidis}, {Migkas}, {Mu{\~n}oz-Giraldo}, {Musiimenta}, {Nguyen-Dang}, {Ni}, {Olechowska}, {Ota}, {Pacaud}, {Pasini}, {Perinati}, {Pires}, {Pommranz}, {Ponti}, {Poppenhaeger}, {P{\"u}hlhofer}, {Rau}, {Reh}, {Reiprich}, {Roster}, {Saeedi}, {Santangelo}, {Sasaki}, {Schmitt}, {Schneider}, {Schrabback}, {Schuster}, {Schwope}, {Seppi}, {Serim}, {Shreeram}, {Sokolova-Lapa}, {Starck}, {Stelzer}, {Stierhof}, {Suleimanov}, {Tenzer}, {Traulsen}, {Tr{\"u}mper}, {Tsuge}, {Urrutia}, {Veronica}, {Waddell}, {Willer}, {Wolf}, {Yeung}, {Zainab}, {Zangrandi}, {Zhang}, {Zhang}, \& {Zheng}}]{Merloni2024}
{Merloni}, A., {Lamer}, G., {Liu}, T., {et~al.} 2024, \aap, 682, A34

\bibitem[{Ochsenbein(1996)}]{10.26093/cds/vizier}
Ochsenbein, F. 1996, The VizieR database of astronomical catalogues

\bibitem[{{Ochsenbein} {et~al.}(2000){Ochsenbein}, {Bauer}, \& {Marcout}}]{vizier2000}
{Ochsenbein}, F., {Bauer}, P., \& {Marcout}, J. 2000, \aaps, 143, 23

\bibitem[{{Oguri}(2014)}]{Oguri2014}
{Oguri}, M. 2014, \mnras, 444, 147

\bibitem[{{Piffaretti} {et~al.}(2011){Piffaretti}, {Arnaud}, {Pratt}, {Pointecouteau}, \& {Melin}}]{Piffaretti2011}
{Piffaretti}, R., {Arnaud}, M., {Pratt}, G.~W., {Pointecouteau}, E., \& {Melin}, J.~B. 2011, \aap, 534, A109

\bibitem[{{Planck Collaboration} {et~al.}(2014){Planck Collaboration}, Ade, Aghanim, Armitage-Caplan, Arnaud, Ashdown, Atrio-Barandela, Aumont, Aussel, Baccigalupi, Banday, Barreiro, Barrena, Bartelmann, Bartlett, Battaner, Benabed, Beno{\^\i}t, Benoit-L{\'e}vy, Bernard, Bersanelli, Bielewicz, Bikmaev, Bobin, Bock, B{\"o}hringer, Bonaldi, Bond, Borrill, Bouchet, Bridges, Bucher, Burenin, Burigana, Butler, Cardoso, Carvalho, Catalano, Challinor, Chamballu, Chary, Chen, Chiang, Chiang, Chon, Christensen, Churazov, Church, Clements, Colombi, Colombo, Comis, Couchot, Coulais, Crill, Curto, Cuttaia, Da~Silva, Dahle, Danese, Davies, Davis, de~Bernardis, de~Rosa, de~Zotti, Delabrouille, Delouis, D{\'e}mocl{\`e}s, D{\'e}sert, Dickinson, Diego, Dolag, Dole, Donzelli, Dor{\'e}, Douspis, Dupac, Efstathiou, Eisenhardt, En{\ss}lin, Eriksen, Feroz, Finelli, Flores-Cacho, Forni, Frailis, Franceschi, Fromenteau, Galeotta, Ganga, G{\'e}nova-Santos, Giard, Giardino, Gilfanov, Giraud-H{\'e}raud, Gonz{\'a}lez-Nuevo, G{\'o}rski,
  Grainge, Gratton, Gregorio, Groeneboom, Gruppuso, Hansen, Hanson, Harrison, Hempel, Henrot-Versill{\'e}, Hern{\'a}ndez-Monteagudo, Herranz, Hildebrandt, Hivon, Hobson, Holmes, Hornstrup, Hovest, Huffenberger, Hurier, Hurley-Walker, Jaffe, Jaffe, Jones, Juvela, Keih{\"a}nen, Keskitalo, Khamitov, Kisner, Kneissl, Knoche, Knox, Kunz, Kurki-Suonio, Lagache, L{\"a}hteenm{\"a}ki, Lamarre, Lasenby, Laureijs, Lawrence, Leahy, Leonardi, Le{\'o}n-Tavares, Lesgourgues, Li, Liddle, Liguori, Lilje, Linden-V{\o}rnle, L{\'o}pez-Caniego, Lubin, Mac{\'\i}as-P{\'e}rez, MacTavish, Maffei, Maino, Mandolesi, Maris, Marshall, Martin, Mart{\'\i}nez-Gonz{\'a}lez, Masi, Massardi, Matarrese, Matthai, Mazzotta, Mei, Meinhold, Melchiorri, Melin, Mendes, Mennella, Migliaccio, Mikkelsen, Mitra, Miville-Desch{\^e}nes, Moneti, Montier, Morgante, Mortlock, Munshi, Murphy, Naselsky, Nati, Natoli, Nesvadba, Netterfield, N{\o}rgaard-Nielsen, Noviello, Novikov, Novikov, O'Dwyer, Olamaie, Osborne, Oxborrow, Paci, Pagano, Pajot, Paoletti,
  Pasian, Patanchon, Pearson, Perdereau, Perotto, Perrott, Perrotta, Piacentini, Piat, Pierpaoli, Pietrobon, Plaszczynski, Pointecouteau, Polenta, Ponthieu, Popa, Poutanen, Pratt, Pr{\'e}zeau, Prunet, Puget, Rachen, Reach, Rebolo, Reinecke, Remazeilles, Renault, Ricciardi, Riller, Ristorcelli, Rocha, Rosset, Roudier, Rowan-Robinson, Rubi{\~n}o-Mart{\'\i}n, Rumsey, Rusholme, Sandri, Santos, Saunders, Savini, Schammel, Scott, Seiffert, Shellard, Shimwell, Spencer, Stanford, Starck, Stolyarov, Stompor, Sudiwala, Sunyaev, Sureau, Sutton, Suur-Uski, Sygnet, Tauber, Tavagnacco, Terenzi, Toffolatti, Tomasi, Tristram, Tucci, Tuovinen, T{\"u}rler, Umana, Valenziano, Valiviita, Van~Tent, Vibert, Vielva, Villa, Vittorio, Wade, Wandelt, White, White, Yvon, Zacchei, \& Zonca}]{PlanckCollaboration2014}
{Planck Collaboration}, Ade, P.~A.~R., Aghanim, N., {et~al.} 2014, \aap, 571, A29

\bibitem[{{Planck Collaboration} {et~al.}(2016){Planck Collaboration}, {Ade}, {Aghanim}, {Arnaud}, {Ashdown}, {Aumont}, {Baccigalupi}, {Banday}, {Barreiro}, {Barrena}, {Bartlett}, {Bartolo}, {Battaner}, {Battye}, {Benabed}, {Beno{\^\i}t}, {Benoit-L{\'e}vy}, {Bernard}, {Bersanelli}, {Bielewicz}, {Bikmaev}, {B{\"o}hringer}, {Bonaldi}, {Bonavera}, {Bond}, {Borrill}, {Bouchet}, {Bucher}, {Burenin}, {Burigana}, {Butler}, {Calabrese}, {Cardoso}, {Carvalho}, {Catalano}, {Challinor}, {Chamballu}, {Chary}, {Chiang}, {Chon}, {Christensen}, {Clements}, {Colombi}, {Colombo}, {Combet}, {Comis}, {Couchot}, {Coulais}, {Crill}, {Curto}, {Cuttaia}, {Dahle}, {Danese}, {Davies}, {Davis}, {de Bernardis}, {de Rosa}, {de Zotti}, {Delabrouille}, {D{\'e}sert}, {Dickinson}, {Diego}, {Dolag}, {Dole}, {Donzelli}, {Dor{\'e}}, {Douspis}, {Ducout}, {Dupac}, {Efstathiou}, {Eisenhardt}, {Elsner}, {En{\ss}lin}, {Eriksen}, {Falgarone}, {Fergusson}, {Feroz}, {Ferragamo}, {Finelli}, {Forni}, {Frailis}, {Fraisse}, {Franceschi}, {Frejsel},
  {Galeotta}, {Galli}, {Ganga}, {G{\'e}nova-Santos}, {Giard}, {Giraud-H{\'e}raud}, {Gjerl{\o}w}, {Gonz{\'a}lez-Nuevo}, {G{\'o}rski}, {Grainge}, {Gratton}, {Gregorio}, {Gruppuso}, {Gudmundsson}, {Hansen}, {Hanson}, {Harrison}, {Hempel}, {Henrot-Versill{\'e}}, {Hern{\'a}ndez-Monteagudo}, {Herranz}, {Hildebrandt}, {Hivon}, {Hobson}, {Holmes}, {Hornstrup}, {Hovest}, {Huffenberger}, {Hurier}, {Jaffe}, {Jaffe}, {Jin}, {Jones}, {Juvela}, {Keih{\"a}nen}, {Keskitalo}, {Khamitov}, {Kisner}, {Kneissl}, {Knoche}, {Kunz}, {Kurki-Suonio}, {Lagache}, {Lamarre}, {Lasenby}, {Lattanzi}, {Lawrence}, {Leonardi}, {Lesgourgues}, {Levrier}, {Liguori}, {Lilje}, {Linden-V{\o}rnle}, {L{\'o}pez-Caniego}, {Lubin}, {Mac{\'\i}as-P{\'e}rez}, {Maggio}, {Maino}, {Mak}, {Mandolesi}, {Mangilli}, {Martin}, {Mart{\'\i}nez-Gonz{\'a}lez}, {Masi}, {Matarrese}, {Mazzotta}, {McGehee}, {Mei}, {Melchiorri}, {Melin}, {Mendes}, {Mennella}, {Migliaccio}, {Mitra}, {Miville-Desch{\^e}nes}, {Moneti}, {Montier}, {Morgante}, {Mortlock}, {Moss}, {Munshi},
  {Murphy}, {Naselsky}, {Nastasi}, {Nati}, {Natoli}, {Netterfield}, {N{\o}rgaard-Nielsen}, {Noviello}, {Novikov}, {Novikov}, {Olamaie}, {Oxborrow}, {Paci}, {Pagano}, {Pajot}, {Paoletti}, {Pasian}, {Patanchon}, {Pearson}, {Perdereau}, {Perotto}, {Perrott}, {Perrotta}, {Pettorino}, {Piacentini}, {Piat}, {Pierpaoli}, {Pietrobon}, {Plaszczynski}, {Pointecouteau}, {Polenta}, {Pratt}, {Pr{\'e}zeau}, {Prunet}, {Puget}, {Rachen}, {Reach}, {Rebolo}, {Reinecke}, {Remazeilles}, {Renault}, {Renzi}, {Ristorcelli}, {Rocha}, {Rosset}, {Rossetti}, {Roudier}, {Rozo}, {Rubi{\~n}o-Mart{\'\i}n}, {Rumsey}, {Rusholme}, {Rykoff}, {Sandri}, {Santos}, {Saunders}, {Savelainen}, {Savini}, {Schammel}, {Scott}, {Seiffert}, {Shellard}, {Shimwell}, {Spencer}, {Stanford}, {Stern}, {Stolyarov}, {Stompor}, {Streblyanska}, {Sudiwala}, {Sunyaev}, {Sutton}, {Suur-Uski}, {Sygnet}, {Tauber}, {Terenzi}, {Toffolatti}, {Tomasi}, {Tramonte}, {Tristram}, {Tucci}, {Tuovinen}, {Umana}, {Valenziano}, {Valiviita}, {Van Tent}, {Vielva}, {Villa}, {Wade},
  {Wandelt}, {Wehus}, {White}, {Wright}, {Yvon}, {Zacchei}, \& {Zonca}}]{Planck2016}
{Planck Collaboration}, {Ade}, P.~A.~R., {Aghanim}, N., {et~al.} 2016, \aap, 594, A27

\bibitem[{Predehl {et~al.}(2021)Predehl, Andritschke, Arefiev, Babyshkin, Batanov, Becker, B{\"o}hringer, Bogomolov, Boller, Borm, Bornemann, Br{\"a}uninger, Br{\"u}ggen, Brunner, Brusa, Bulbul, Buntov, Burwitz, Burkert, Clerc, Churazov, Coutinho, Dauser, Dennerl, Doroshenko, Eder, Emberger, Eraerds, Finoguenov, Freyberg, Friedrich, Friedrich, F{\"u}rmetz, Georgakakis, Gilfanov, Granato, Grossberger, Gueguen, Gureev, Haberl, H{\"a}lker, Hartner, Hasinger, Huber, Ji, Kienlin, Kink, Korotkov, Kreykenbohm, Lamer, Lomakin, Lapshov, Liu, Maitra, Meidinger, Menz, Merloni, Mernik, Mican, Mohr, M{\"u}ller, Nandra, Nazarov, Pacaud, Pavlinsky, Perinati, Pfeffermann, Pietschner, Ramos-Ceja, Rau, Reiffers, Reiprich, Robrade, Salvato, Sanders, Santangelo, Sasaki, Scheuerle, Schmid, Schmitt, Schwope, Shirshakov, Steinmetz, Stewart, Str{\"u}der, Sunyaev, Tenzer, Tiedemann, Tr{\"u}mper, Voron, Weber, Wilms, \& Yaroshenko}]{Predehl2021}
Predehl, P., Andritschke, R., Arefiev, V., {et~al.} 2021, \aap, 647, A1

\bibitem[{{Rozo} {et~al.}(2011){Rozo}, {Rykoff}, {Koester}, {Nord}, {Wu}, {Evrard}, \& {Wechsler}}]{Rozo2011}
{Rozo}, E., {Rykoff}, E., {Koester}, B., {et~al.} 2011, \apj, 740, 53

\bibitem[{Ruiz {et~al.}(2018)Ruiz, Corral, Mountrichas, \& Georgantopoulos}]{Ruiz2018}
Ruiz, A., Corral, A., Mountrichas, G., \& Georgantopoulos, I. 2018, \aap, 618, A52

\bibitem[{{Rykoff} {et~al.}(2012){Rykoff}, {Koester}, {Rozo}, {Annis}, {Evrard}, {Hansen}, {Hao}, {Johnston}, {McKay}, \& {Wechsler}}]{Rykoff2012}
{Rykoff}, E.~S., {Koester}, B.~P., {Rozo}, E., {et~al.} 2012, \apj, 746, 178

\bibitem[{Rykoff {et~al.}(2014)Rykoff, Rozo, Busha, Cunha, Finoguenov, Evrard, Hao, Koester, Leauthaud, Nord, Pierre, Reddick, Sadibekova, Sheldon, \& Wechsler}]{Rykoff2014}
Rykoff, E.~S., Rozo, E., Busha, M.~T., {et~al.} 2014, \apj, 785, 104

\bibitem[{{Rykoff} {et~al.}(2016){Rykoff}, {Rozo}, {Hollowood}, {Bermeo-Hernandez}, {Jeltema}, {Mayers}, {Romer}, {Rooney}, {Saro}, {Vergara Cervantes}, {Wechsler}, {Wilcox}, {Abbott}, {Abdalla}, {Allam}, {Annis}, {Benoit-L{\'e}vy}, {Bernstein}, {Bertin}, {Brooks}, {Burke}, {Capozzi}, {Carnero Rosell}, {Carrasco Kind}, {Castander}, {Childress}, {Collins}, {Cunha}, {D'Andrea}, {da Costa}, {Davis}, {Desai}, {Diehl}, {Dietrich}, {Doel}, {Evrard}, {Finley}, {Flaugher}, {Fosalba}, {Frieman}, {Glazebrook}, {Goldstein}, {Gruen}, {Gruendl}, {Gutierrez}, {Hilton}, {Honscheid}, {Hoyle}, {James}, {Kay}, {Kuehn}, {Kuropatkin}, {Lahav}, {Lewis}, {Lidman}, {Lima}, {Maia}, {Mann}, {Marshall}, {Martini}, {Melchior}, {Miller}, {Miquel}, {Mohr}, {Nichol}, {Nord}, {Ogando}, {Plazas}, {Reil}, {Sahl{\'e}n}, {Sanchez}, {Santiago}, {Scarpine}, {Schubnell}, {Sevilla-Noarbe}, {Smith}, {Soares-Santos}, {Sobreira}, {Stott}, {Suchyta}, {Swanson}, {Tarle}, {Thomas}, {Tucker}, {Uddin}, {Viana}, {Vikram}, {Walker}, {Zhang}, \& {DES
  Collaboration}}]{Rykoff2016}
{Rykoff}, E.~S., {Rozo}, E., {Hollowood}, D., {et~al.} 2016, \apjs, 224, 1

\bibitem[{{Rykoff} {et~al.}(2015){Rykoff}, {Rozo}, \& {Keisler}}]{Rykoff2015}
{Rykoff}, E.~S., {Rozo}, E., \& {Keisler}, R. 2015, arXiv e-prints, arXiv:1509.00870

\bibitem[{Salvato(2024)}]{Salvato2024}
Salvato, M. 2024, TODO

\bibitem[{Salvato {et~al.}(2018)Salvato, Buchner, Budav{\'a}ri, Dwelly, Merloni, Brusa, Rau, Fotopoulou, \& Nandra}]{Salvato2018}
Salvato, M., Buchner, J., Budav{\'a}ri, T., {et~al.} 2018, \mnras, 473, 4937

\bibitem[{Salvato {et~al.}(2022)Salvato, Wolf, Dwelly, Georgakakis, Brusa, Merloni, Liu, Toba, Nandra, Lamer, Buchner, Schneider, Freund, Rau, Schwope, Nishizawa, Klein, Arcodia, Comparat, Musiimenta, Nagao, Brunner, Malyali, Finoguenov, Anderson, Shen, Ibarra-Medel, Trump, Brandt, Urry, Rivera, Krumpe, Urrutia, Miyaji, Ichikawa, Schneider, Fresco, Boller, Haase, Brownstein, Lane, Bizyaev, \& Nitschelm}]{Salvato2022}
Salvato, M., Wolf, J., Dwelly, T., {et~al.} 2022, \aap, 661, A3

\bibitem[{{Sanders} {et~al.}(2018){Sanders}, {Fabian}, {Russell}, \& {Walker}}]{Sanders2018}
{Sanders}, J.~S., {Fabian}, A.~C., {Russell}, H.~R., \& {Walker}, S.~A. 2018, \mnras, 474, 1065

\bibitem[{{Saxena} {et~al.}(2024){Saxena}, {Salvato}, {Roster}, {Shirley}, {Buchner}, {Wolf}, {Kohl}, {Starck}, {Dwelly}, {Comparat}, {Malyali}, {Krippendorf}, {Zenteno}, {Lang}, {Schlegel}, {Zhou}, {Dey}, {Valdes}, {Myers}, {Assef}, {Ricci}, {Temple}, {Merloni}, {Koekemoer}, {Anderson}, {Morrison}, {Liu}, \& {Nandra}}]{Saxena2024}
{Saxena}, A., {Salvato}, M., {Roster}, W., {et~al.} 2024, \aap, 690, A365

\bibitem[{{Schechter}(1976)}]{Schechter1976}
{Schechter}, P. 1976, \apj, 203, 297

\bibitem[{{Seppi} {et~al.}(2022){Seppi}, {Comparat}, {Bulbul}, {Nandra}, {Merloni}, {Clerc}, {Liu}, {Ghirardini}, {Liu}, {Salvato}, {Sanders}, {Wilms}, {Dwelly}, {Dauser}, {K{\"o}nig}, {Ramos-Ceja}, {Garrel}, \& {Reiprich}}]{Seppi2022}
{Seppi}, R., {Comparat}, J., {Bulbul}, E., {et~al.} 2022, \aap, 665, A78

\bibitem[{{Seppi} {et~al.}(2024){Seppi}, {Comparat}, {Ghirardini}, {Garrel}, {Artis}, {S{\'a}nchez}, {Liu}, {Clerc}, {Bulbul}, {Grandis}, {Kluge}, {Reiprich}, {Merloni}, {Zhang}, {Bahar}, {Shreeram}, {Sanders}, {Ramos-Ceja}, \& {Krumpe}}]{Seppi2024}
{Seppi}, R., {Comparat}, J., {Ghirardini}, V., {et~al.} 2024, \aap, 686, A196

\bibitem[{{Silva} {et~al.}(2016){Silva}, {Blum}, {Allen}, {Dey}, {Schlegel}, {Lang}, {Moustakas}, {Meisner}, {Valdes}, {Patej}, {Myers}, {Sprayberry}, {Saha}, {Olsen}, {Safonova}, {Yang}, {Burleigh}, \& {MzLS Team}}]{Silva2016aa}
{Silva}, D.~R., {Blum}, R.~D., {Allen}, L., {et~al.} 2016, in American Astronomical Society Meeting Abstracts, Vol. 228, American Astronomical Society Meeting Abstracts \#228, 317.02

\bibitem[{Somboonpanyakul {et~al.}(2021)Somboonpanyakul, McDonald, Gaspari, Stalder, \& Stark}]{Somboonpanyakul2021b}
Somboonpanyakul, T., McDonald, M., Gaspari, M., Stalder, B., \& Stark, A.~A. 2021, \apj, 910, 60

\bibitem[{{Sunyaev} {et~al.}(2021){Sunyaev}, {Arefiev}, {Babyshkin}, {Bogomolov}, {Borisov}, {Buntov}, {Brunner}, {Burenin}, {Churazov}, {Coutinho}, {Eder}, {Eismont}, {Freyberg}, {Gilfanov}, {Gureyev}, {Hasinger}, {Khabibullin}, {Kolmykov}, {Komovkin}, {Krivonos}, {Lapshov}, {Levin}, {Lomakin}, {Lutovinov}, {Medvedev}, {Merloni}, {Mernik}, {Mikhailov}, {Molodtsov}, {Mzhelsky}, {M{\"u}ller}, {Nandra}, {Nazarov}, {Pavlinsky}, {Poghodin}, {Predehl}, {Robrade}, {Sazonov}, {Scheuerle}, {Shirshakov}, {Tkachenko}, \& {Voron}}]{Sunyaeve2021}
{Sunyaev}, R., {Arefiev}, V., {Babyshkin}, V., {et~al.} 2021, \aap, 656, A132

\bibitem[{{Sunyaev} \& {Zeldovich}(1972)}]{Sunyaev1972}
{Sunyaev}, R.~A. \& {Zeldovich}, Y.~B. 1972, Comments on Astrophysics and Space Physics, 4, 173

\bibitem[{{Vikhlinin} {et~al.}(2006){Vikhlinin}, {Kravtsov}, {Forman}, {Jones}, {Markevitch}, {Murray}, \& {Van Speybroeck}}]{Vikhlinin2006}
{Vikhlinin}, A., {Kravtsov}, A., {Forman}, W., {et~al.} 2006, \apj, 640, 691

\bibitem[{Virtanen {et~al.}(2020)Virtanen, Gommers, Oliphant, Haberland, Reddy, Cournapeau, Burovski, Peterson, Weckesser, Bright, {van der Walt}, Brett, Wilson, Millman, Mayorov, Nelson, Jones, Kern, Larson, Carey, Polat, Feng, Moore, {VanderPlas}, Laxalde, Perktold, Cimrman, Henriksen, Quintero, Harris, Archibald, Ribeiro, Pedregosa, {van Mulbregt}, \& {SciPy 1.0 Contributors}}]{2020SciPy-NMeth}
Virtanen, P., Gommers, R., Oliphant, T.~E., {et~al.} 2020, Nature Methods, 17, 261

\bibitem[{Voges {et~al.}(1999)Voges, Aschenbach, Boller, Br{\"a}uninger, Briel, Burkert, Dennerl, Englhauser, Gruber, Haberl, Hartner, Hasinger, K{\"u}rster, Pfeffermann, Pietsch, Predehl, Rosso, Schmitt, Tr{\"u}mper, \& Zimmermann}]{Voges1999}
Voges, W., Aschenbach, B., Boller, T., {et~al.} 1999, \aap, 349, 389

\bibitem[{{Von Der Linden} {et~al.}(2007){Von Der Linden}, {Best}, {Kauffmann}, \& {White}}]{VonDerLinden2007}
{Von Der Linden}, A., {Best}, P.~N., {Kauffmann}, G., \& {White}, S. D.~M. 2007, \mnras, 379, 867

\bibitem[{{Wen} \& {Han}(2018)}]{Wen2018}
{Wen}, Z.~L. \& {Han}, J.~L. 2018, \mnras, 481, 4158

\bibitem[{{Zenteno} {et~al.}(2020){Zenteno}, {Hern{\'a}ndez-Lang}, {Klein}, {Vergara Cervantes}, {Hollowood}, {Bhargava}, {Palmese}, {Strazzullo}, {Romer}, {Mohr}, {Jeltema}, {Saro}, {Lidman}, {Gruen}, {Ojeda}, {Katzenberger}, {Aguena}, {Allam}, {Avila}, {Bayliss}, {Bertin}, {Brooks}, {Buckley-Geer}, {Burke}, {Capasso}, {Carnero Rosell}, {Carrasco Kind}, {Carretero}, {Castander}, {Costanzi}, {da Costa}, {De Vicente}, {Desai}, {Diehl}, {Doel}, {Eifler}, {Evrard}, {Flaugher}, {Floyd}, {Fosalba}, {Frieman}, {Garc{\'\i}a-Bellido}, {Gerdes}, {Gonzalez}, {Gruendl}, {Gschwend}, {Gutierrez}, {Hartley}, {Hinton}, {Honscheid}, {James}, {Kuehn}, {Lahav}, {Lima}, {McDonald}, {Maia}, {March}, {Melchior}, {Menanteau}, {Miquel}, {Ogando}, {Paz-Chinch{\'o}n}, {Plazas}, {Roodman}, {Rykoff}, {Sanchez}, {Scarpine}, {Schubnell}, {Serrano}, {Sevilla-Noarbe}, {Smith}, {Soares-Santos}, {Suchyta}, {Swanson}, {Tarle}, {Thomas}, {Varga}, {Walker}, {Wilkinson}, \& {DES Collaboration}}]{Zenteno2020}
{Zenteno}, A., {Hern{\'a}ndez-Lang}, D., {Klein}, M., {et~al.} 2020, \mnras, 495, 705

\bibitem[{{Zou} {et~al.}(2017){Zou}, {Zhou}, {Fan}, {Zhang}, {Zhou}, {Nie}, {Peng}, {McGreer}, {Jiang}, {Dey}, {Fan}, {He}, {Jiang}, {Lang}, {Lesser}, {Ma}, {Mao}, {Schlegel}, \& {Wang}}]{Zou2017aa}
{Zou}, H., {Zhou}, X., {Fan}, X., {et~al.} 2017, \pasp, 129, 064101

\bibitem[{{Zu} {et~al.}(2017){Zu}, {Mandelbaum}, {Simet}, {Rozo}, \& {Rykoff}}]{Zu2017}
{Zu}, Y., {Mandelbaum}, R., {Simet}, M., {Rozo}, E., \& {Rykoff}, E.~S. 2017, \mnras, 470, 551

\end{thebibliography}

\appendix

\end{document}